\documentclass[]{aa}
\usepackage{microtype}
\usepackage{booktabs}
\usepackage{placeins}
\usepackage{verbatim}
\usepackage{color}
\usepackage{graphicx}
\usepackage{amsmath}
\usepackage{xspace}
\usepackage[parfill]{parskip}
\usepackage{txfonts}
\usepackage[]{hyperref}
\newcommand{\therepository}{https://bitbucket.org/radboudradiolab/gwtoolbox}
\newcommand{\thewebsite}{https:gw-universe.org}
\newcommand{\eptabh}{$1.4\times10^{-15}$}
\newcommand{\eptacs}{$2.5\times10^{-10}$}
\newcommand{\eptarl}{$4.7\times10^{-10}$}
\newcommand{\pptabh}{$2.45\times10^{-15}$}
\newcommand{\pptacs}{$6.4\times10^{-10}$}
\newcommand{\pptarl}{$1.2\times10^{-9}$}
\newcommand{\nanobh}{$1.6\times10^{-15}$}
\newcommand{\nanocs}{$3.6\times10^{-10}$}
\newcommand{\nanorl}{$6.3\times10^{-10}$}
\newcommand{\iptabh}{$9.9\times10^{-16}$}
\newcommand{\iptacs}{$1.0\times10^{-10}$}
\newcommand{\iptarl}{$2.0\times10^{-10}$}
\newcommand{\GWT}{\texttt{GW-Toolbox}\xspace}

\newcommand{\msun}{$M_\odot$}
\hypersetup{
    colorlinks=true,
    linkcolor=blue,
    filecolor=blue,
    citecolor=blue,
    urlcolor=cyan,
    pdfpagemode=FullScreen,
    }

\begin{document} 
   \title{The Gravitational Wave Universe Toolbox: }

   \subtitle{A software package to simulate observation of the Gravitational Wave Universe with different detectors}

   \author{Shu-Xu Yi\inst{1,5}\thanks{sxyi@ihep.ac.cn}
          \and
          Gijs Nelemans\inst{1,2,3}
          \and
          Christiaan Brinkerink\inst{1}
          \and
          Zuzanna Kostrzewa-Rutkowska\inst{4,1,2}
          \and
          Sjoerd T. Timmer\inst{1}
          \and 
          Fiorenzo Stoppa\inst{1}
          \and
          Elena M. Rossi\inst{4}
          \and
          Simon F. Portegies Zwart\inst{4}
          }

\institute{
   Department of Astrophysics/IMAPP, Radboud University, P O Box 9010, NL-6500 GL Nijmegen, The Netherlands
         \and
          SRON, Netherlands Institute for Space Research, Sorbonnelaan 2, NL-3584 CA Utrecht, The Netherlands
         \and
         Institute of Astronomy, KU Leuven, Celestijnenlaan 200D, B-3001 Leuven, Belgium
         \and
         Leiden Observatory, Leiden University, PO Box 9513, NL-2300 RA Leiden, the Netherlands
         \and 
         Key Laboratory of Particle Astrophysics, Institute of High Energy Physics, Chinese Academy of Sciences, 19B Yuquan Road, Beijing 100049, People’s Republic of China
   }

   \date{}

 
  \abstract
   {As the importance of Gravitational Wave (GW)  astrophysics increases rapidly, astronomers interested in GW who are not experts can have the need to get a quick idea of what GW sources can be detected by certain detectors and the accuracy of the measured parameters.}
   {The \GWT is a set of easy-to-use, flexible tools to simulate observations on the GW universe with different detectors, including ground-based interferometers (advanced LIGO, advanced VIRGO, KAGRA, Einstein Telescope, Cosmic Explorer and also customised interferometers), space-borne interferometers (LISA and a customised design), pulsar timing arrays mimicking the current working ones (EPTA, PPTA, NANOGrav, IPTA) and future ones. We include a broad range of sources such as mergers of stellar mass compact objects, namely black holes, neutron stars and black hole-neutron stars, supermassive black hole binaries mergers and inspirals, Galactic double white dwarfs in ultra-compact orbit, extreme mass ratio inspirals and Stochastic GW backgrounds.}
   {We collect methods to simulate source populations and determine their detectability with various detectors. The paper aims at giving a comprehensive description of the methodology and functionality of the \GWT. }
   {The \GWT produces results that are consistent with previous work in literature, and the tools can be accessed with a website interface (\href{\thewebsite}{gw-universe.org}) or as a python package (\href{\therepository}{https://bitbucket.org/radboudradiolab/gwtoolbox}). In the future, it will be upgraded with more functions.}
   {}


   \maketitle
%
\section{Introduction}
Gravitational Wave (GW) astrophysics rises quickly into a pivotal branch of astronomy. A growing number of researchers from different fields find GW interesting and relevant to their various scientific goals \citep[e.g.][]{2013PhRvD..87f4036P,2016PhRvX...6a1035L,2018CQGra..35p3001C,2018arXiv180701060N,2019arXiv190304592M,2020GReGr..52...81B}. The first opened frequency window covers $\sim10-1000$ Hz (a.k.a., the kHz band), which corresponds to the working frequency range of ground-based interferometers, e.g., the operating 2nd generation detectors advanced Laser Interferometer Gravitational-Wave Observatory (LIGO)/Virgo interferometer (Virgo)/Kamioka Gravitational Wave Detector (KAGRA)  \citep{2010CQGra..27h4006H,2015CQGra..32b4001A,2019NatAs...3...35K} and the planned 3rd generation detectors e.g., Einstein Telescope (ET, \citealt{2010CQGra..27s4002P}) and Cosmic Explorer (CE, \citealt{2019BAAS...51g..35R}). Space-borne interferometers, {\it e.g.,} Laser Interferometer Space Antenna (LISA, \citealt{2017arXiv170200786A}) and other similar projects, e.g., DECIGO \citep{2006CQGra..23S.125K}, Taiji \citep{2020arXiv200810332M} and Tianqin \citep{2016CQGra..33c5010L}, will cover the GW spectrum in the frequency range $\sim10^{-3}-1$\,Hz (a.k.a., mHz band). At even lower frequencies, pulsar timing arrays (PTAs, \citealt[e.g.][]{2017arXiv170701615H,2020JApA...41....8D}) are used to probe GWs with frequencies around $10^{-8}-10^{-5}$\,Hz ((a.k.a., nHz band).  There are also attempts to search for GW at frequencies higher than the kHz regime, i.e., MHz/GHz \citep{2020arXiv201112414A}. Although this frequency range is another important part of the GW Universe, which is full of opportunities to discovery new physics and phenomena, we do not include detectors and sources of this frequency range in the \GWT at this stage, due to its larger uncertainties.

Together, these detectors are sensitive to a very broad range of GW sources, where higher frequency detectors are sensitive to smaller objects. In the frequency range $\sim10-1000$ Hz, there are inspiral and mergers of stellar mass compact object binaries, spinning neutron stars and supernovae explosions (see e.g., \citealt{1979A&A....72..120C,1987thyg.book..330T,1989CQGra...6.1761S,1997NewA....2...43L,2000ApJ...528L..17P, 2002ApJ...572..407B,2009CQGra..26t4015O,2011CQGra..28u5006G,2014LRR....17....3P}). On the other hand, the early inspiral phase of stellar mass compact object binaries occupy the frequency range $\sim10^{-3}-1$\,Hz \citep{2016PhRvL.116w1102S}. Also white dwarfs binaries, which are not compact enough to be detected by kHz detectors, are prominent sources for mHz detectors \citep[e.g.,][]{2001A&A...375..890N,2010ApJ...717.1006R,2016PhRvL.116w1102S,2018MNRAS.480.2704L,2018PhRvL.121m1105T,2020MNRAS.494L..75S,2020MNRAS.492.3061L}. Since BH sizes increase with mass, the mHZ detectors are sensitive to mergers of supermassive BHs (SMBHs $10^3-10^8 \, M_\odot$ \citep[e.g.,][]{2005ApJ...623...23S,2006PhRvD..73f4030B} and extreme mass ratio inspirals (EMRIs, see \citealt{2004CQGra..21S1595G,2007CQGra..24R.113A}). Again, in the early inspiral phase, SMBH binaries occupy the low frequency end of the spectrum $\sim10^{-8}-10^{-5}$ Hz, either as individual sources or as a stochastic background \citep{1978SvA....22...36S,1979ApJ...234.1100D,1983ApJ...265L..39H,2005ApJ...625L.123J,2008MNRAS.390..192S}. 

The first detection of GW was made by the LIGO/Virgo Collaboration (LVC) in 2015 \citep{2016PhRvL.116f1102A}. The event GW150914 originates from the merger of a binary black hole (BBH) and already had interesting astrophysical implications \citep[see][]{2016ApJ...818L..22A,2016Natur.534..512B}. Since then, there have been 90 GW events detected (as reported by LVC \citealt{2019PhRvX...9c1040A,2020arXiv201014527A,GWTC-3}, and there are more candidates reported by other groups from the public strain data). These events include mergers of BBH, double neutron stars (DNS, e.g., GW170817 and GW190425) and Black Hole-Neutron stars (BHNS, e.g. GW190426, GW200105, GW200115 \citep{2021ApJ...915L...5A}). The population of detected BBH mergers provides important clues on stellar formation and evolution history \citep[e.g.,][]{2020arXiv201014533T}. Among the detected sources, there are several unique systems which provide input to and even challenge current stellar evolution theory (e.g., GW190412, GW190814) and provide insight into the nature of neutron star matter (GW170817, GW190425). The multi-messenger observations of the DNS merger event GW170817/GRB 170817A/AT 2017gfo brought huge progress on astrophysics, fundamental physics and cosmology \citep[e.g.][]{2017PhRvL.119p1101A,2017ApJ...848L..12A,2017ApJ...848L..13A,2017Natur.551...85A,2017ApJ...850L..39A,2017ApJ...850L..40A,2017Sci...358.1556C,2017Natur.551...67P,2018PhRvL.120i1101A,2018PhRvL.121p1101A,2019PhRvX...9a1001A,2019PhRvL.122f1104A,2019PhRvL.123a1102A,2020CQGra..37d5006A}. Future detectors like ET will vastly increase the detection reach and thus allow even broader science investigations \citep[e.g.][]{2020JCAP...03..050M}.

 Also PTA observations are routinely happening, and although there has no conclusive evidence on GW detection with PTA\footnote{NanoGrav found evidence for a common stochastic signal across pulsars, but there is no significant evidence of that being a GW \citep{2020ApJ...905L..34A}.}, the communities are putting more and more stringent upper limits on the nHz GW signals \citep{2011MNRAS.414.3117V,2013ApJ...762...94D,2015MNRAS.453.2576L,2018ApJ...859...47A,2016MNRAS.458.1267V,2019ApJ...880..116A,2020ApJ...889...38A} and already ruled out some theories of galaxy evolution \citep{2013Sci...342..334S,2015Sci...349.1522S}.  Future studies, in particular including SKA will make significant steps in the science that can be done \citep[e.g.][]{2015aska.confE..37J}.

 In the following sections we describe the \GWT, while in section~\ref{Results} we show the results and compare these to the literature. In sections~\ref{sec:discussion} and \ref{sec:summary} we discuss the caveats, further plans and summarize the paper.

\section{The Gravitational Wave Universe Toolbox}
\begin{figure}
    \centering
    \includegraphics[width=0.5\textwidth]{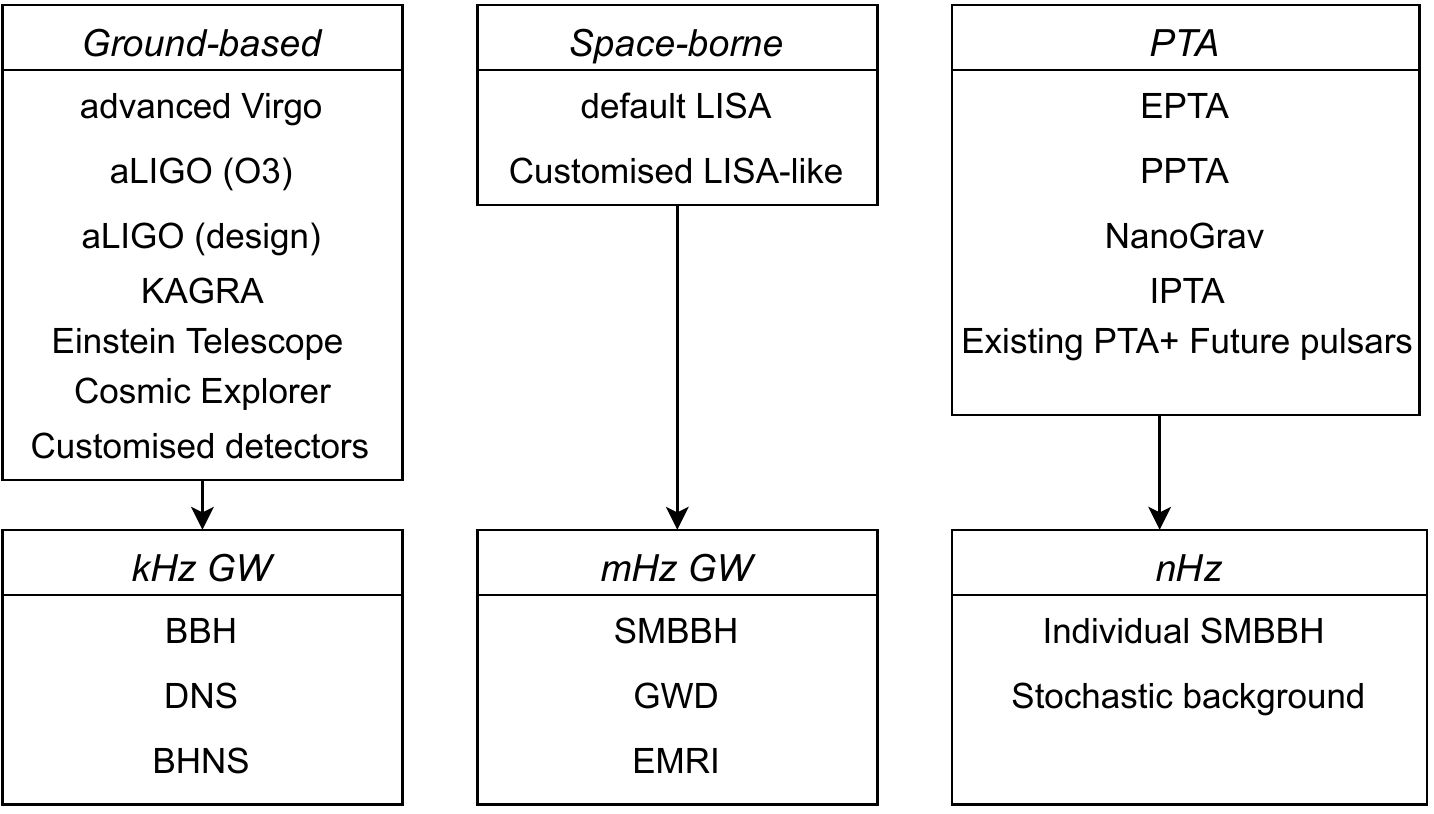}
    \caption{A summary of detectors and sources included in the \GWT}
    \label{fig:setup}
\end{figure}

The \GWT (website: \href{\thewebsite}{gw-universe.org})\footnote{Python package repository:\\ \href{\therepository}{bitbucket.org/radboudradiolab/gwtoolbox}.} is a set of tools for a broad audience to quickly simulate the observation on the GW universe with different detectors. The results include the expected number of detections, synthetic catalogues and parameter uncertainties. We include three classes of GW detectors, namely Earth-based interferometers, space-borne interferometers, and PTAs. In each of these classes, the \GWT has the following detectors with default and customised settings:
\begin{itemize}
\item {\it Earth-based interferometers:}
\begin{itemize}
    \item Advanced LIGO in O3
    \item Advanced LIGO at final design sensitivity
    \item Advanced Virgo at final design sensitivity
    \item KAGRA
    \item Einstein telescope 
    \item Cosmic Explorer
    \item A LIGO/Virgo-Like interferometer with customised parameters
\end{itemize}
\item {\it Space-borne interferometers:}
\begin{itemize}
    \item default LISA
    \item LISA-like spacecraft with customised parameters
\end{itemize}
\item {\it Pulsar Timing Arrays: }
\begin{itemize}
    \item Existing EPTA
    \item Existing PPTA
    \item Existing NANOGrav
    \item Existing IPTA
    \item any of the existing PTAs plus simulated new pulsars to mimick future PTAs
\end{itemize}
\end{itemize}
The three classes of detectors correspond to the three main frequency regimes, namely kHz, mHz and nHz. In each of these regimes, we include the following sources.
\begin{itemize}
\item {\it kHz GW:}
\begin{itemize}
    \item Binary BH (BBH) mergers
    \item Double Neutron Star (DNS) mergers
    \item BH-Neutron Star (BHNS) mergers
\end{itemize}
\item {\it mHz GW:}
\begin{itemize}
    \item Supermassive BH Binary Mergers (SMBBH)
    \item Close Galactic White Dwarf binaries insprials (GWD)
    \item Extreme Mass Ratio Inspirals (EMRIs, Stellar mass BHs inspiraling into supermassive BHs)
\end{itemize}

\item {\it nHz GW:}
\begin{itemize}
    \item Individual SMBBH inspiral
    \item A stochastic GW background
\end{itemize}
\end{itemize}
See figure ~\ref{fig:setup} for a summary of the detectors and sources. 
\begin{figure*}[ht!]
    \centering
    \includegraphics[width=0.48\textwidth, height=7cm]{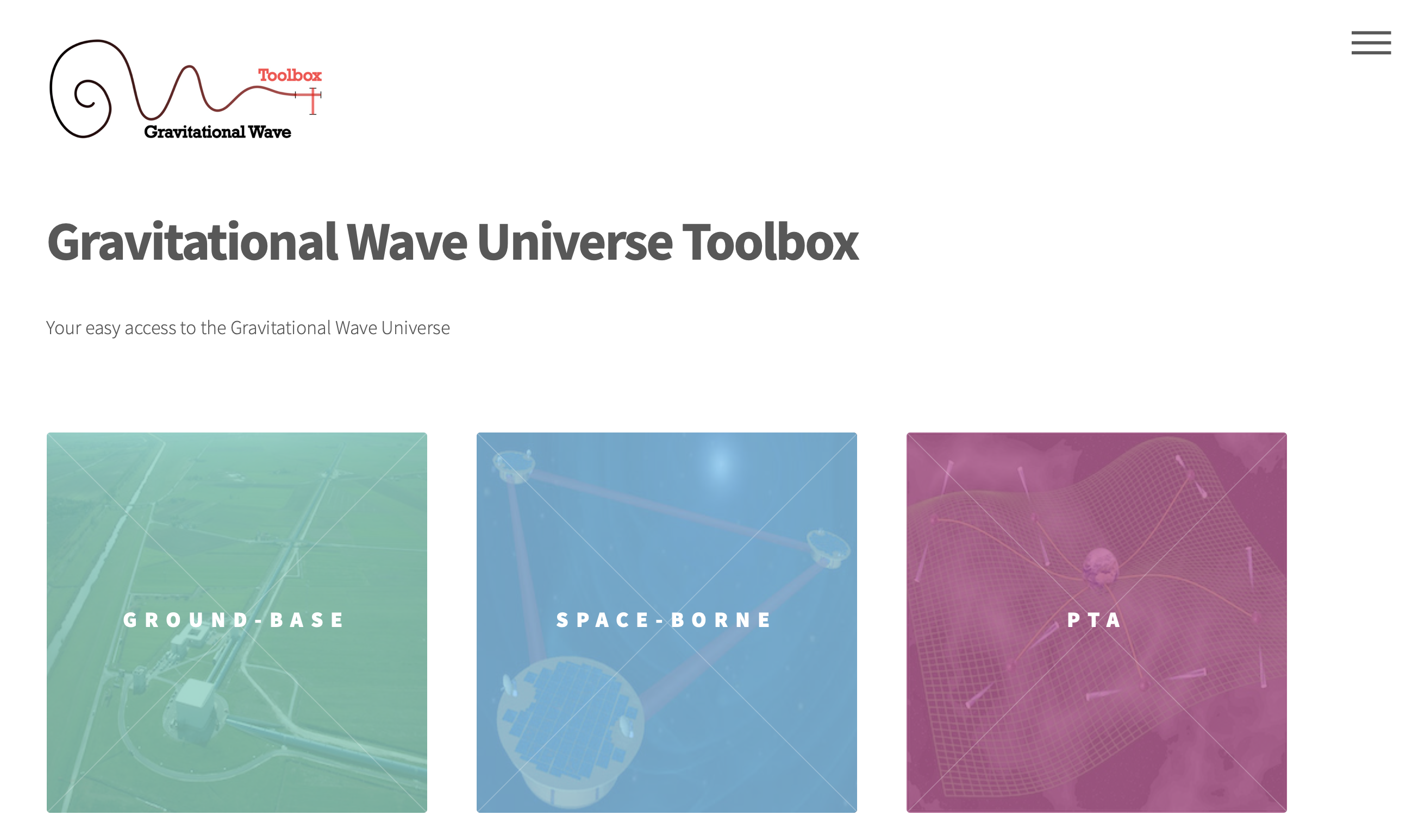}
    \includegraphics[width=0.5\textwidth, height=8cm]{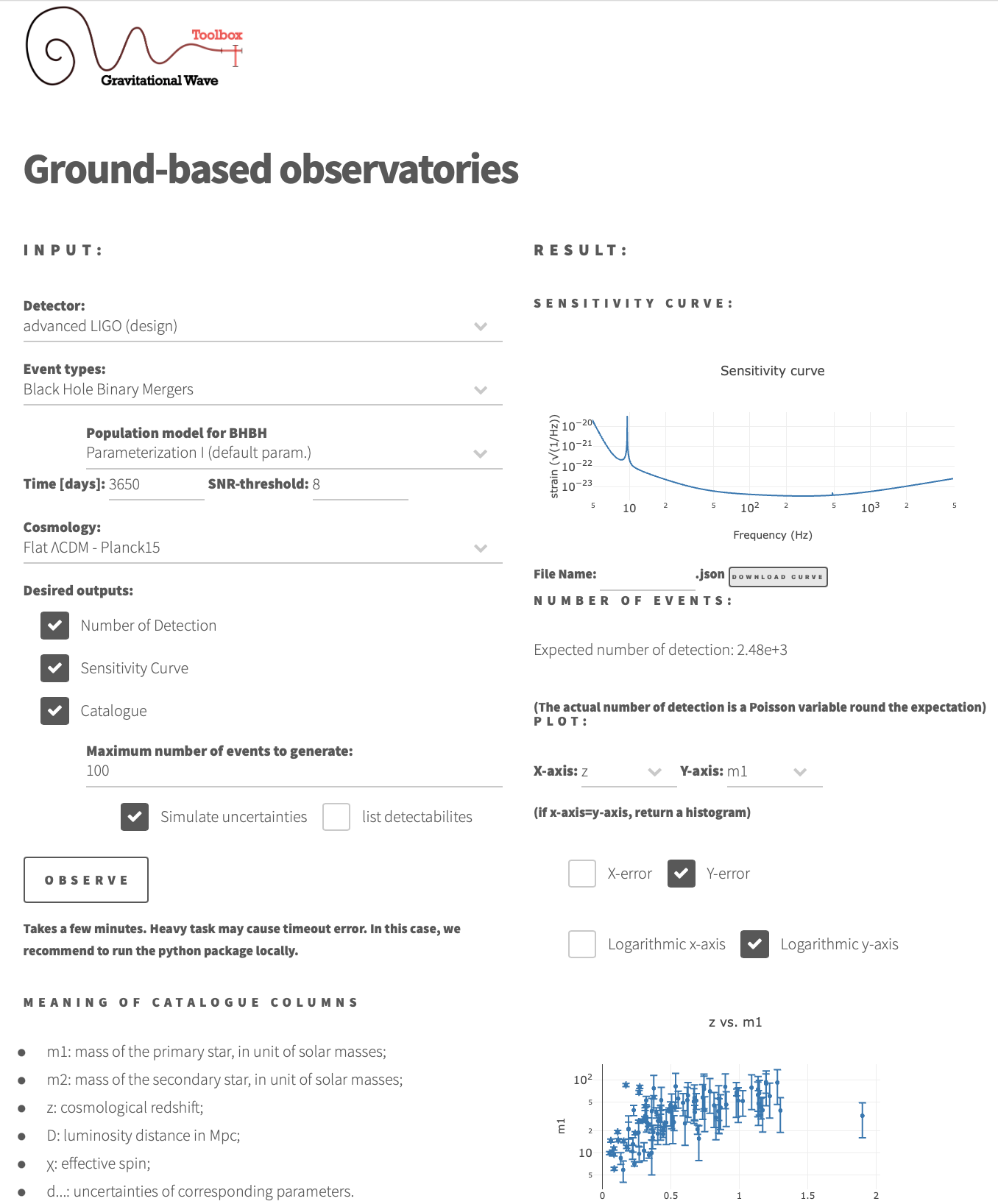}
    \caption{Screen shot of the GWT start page (left) and ground-based observatories page with results for advanced LIGO (right).}
    \label{fig:GWT}
\end{figure*}
In practice, the logical flow of the \GWT is to first select the detector class and the detector parameters, then choose the source class, select its parameters and run. For some of the sources, the underlying cosmological model is also relevant to the simulation, where users are able to select a certain build-in cosmology or to input parameters for a self-defined $\Lambda$-CDM cosmological model. Examples in this paper are simulated with the cosmology model that corresponds to Planck DR15 \citep{2016A&A...594A...1P} We use the \texttt{astropy} package \citep{2013A&A...558A..33A,2018AJ....156..123A} with the cosmology model ``Planck15". Changing to more up-to-date ``Planck18" cosmology model does not significantly change the simulation catalogues.

The results in general are the expected number of detections of the selected sources, and a synthetic catalogue with/without uncertainties on parameters. Plots and histograms of parameters of the catalogue can be made in-browser, and the figures and catalogue can also be exported. In figure~\ref{fig:GWT}, a screen shot of the \GWT website with the top-level selection is shown (left panel), in addition to an example of the kHz detector selection page (right panel).

This paper aims at giving a comprehensive description of the methodology and the functionality of the \GWT. The paper is organised with a similar structure as the set-up of the \GWT. For each class of sources, a model for the Universe is made in which the sources are distributed in space, time and other relevant parameters (such as mass, spins etc) according to a pre-defined population model. The user is allowed to change the population model, according to a parametrised formalism. In section 3, we give description on each of the population models.

The resulting GW source population is then simulated to be observed against a selected GW detector from the above list. After source and detector are determined, the \emph{detected} sources are selected using a signal-to-noise ratio (SNR) criterion. Using the Fisher Matrix formalism, the uncertainties of parameters of the detected sources are estimated. In section 4, we describe how the response of the detector is represented, and the algorithm to generate the catalogue of detected sources. For PTAs, we describe our simplified representation of the pulsar array, the properties of the timing noises and the observation campaign.





\section{Implementation 1: The Universe model}
We first turn to the implementation of the Universe model, in which we select source population models from the literature and in some cases, allow users to change or adapt the source population. There are many reviews of GW sources that discuss these in more detail, such as \cite{2009LRR....12....2S,2013FrPhy...8..771G} and below we discuss the relevant ones. We start with sources for ground-based detectors, where we concentrate on compact binary mergers detectable with ground-based detectors, and then move to space-borne detectors, with supermassive binary black hole mergers, white dwarf compact binaries and extreme mass-ratio inspirals (EMRIs) as sources, and finish with pulsar timing arrays for which we discuss individual supermassive black hole binaries as well as stochastic background. 

\subsection{Sources for Ground Based detectors}
There have been many studies of the formation and population of sources for ground based detectors \citep[e.g.][]{1991ApJ...380L..17P,2001MNRAS.324..797S,2001ApJ...554..548F,2002ApJ...572..407B,2015ApJ...806..263D,2015ApJ...814...58D} and the population depends in a complicated way on many uncertain aspects of binary evolution, the formation of binary stars and even the metallicity evolution of the cosmic star formation history. We take a simple, yet flexible approach by using a parametrised description of the source populations (but see e.g. \citealt{2017ApJS..230...15M,2019MNRAS.482.5012C,2020A&A...636A..10C} for discussions of some of the complications).

\subsubsection{Binary Black Hole Mergers}\label{sec:BBH_model}

The population of BBH mergers is the most prominent in current GW detectors \citep{2020arXiv201014527A}. The source population is characterised by the merger rate as a function of redshift and the distribution of masses and spins \citep[e.g.][]{2017MNRAS.472.2422M,2011ApJ...741..103F,2017PhRvD..95j3010K,2018ApJ...856..173T,2019MNRAS.483.3288P,2019ApJ...882L..24A}. 

In the population model for BBH, the merger rate density is expressed as:
\begin{equation}
    \dot{n}(z, m_1,q,\chi)=\mathcal{R}(z)f(m_1)\pi(q)P(\chi),
\end{equation}
where $f(m_1)$ is the mass function of the primary (heavier) black hole, $\pi(q)$ and $P(\chi)$ are the probability distributions of the mass ratio $q\equiv m_2/m_1$ and the effective spin $\chi$ respectively. $\mathcal{R}$ as function of $z$ is often refer to as the cosmic merger rate density. We take the parameterisation as in \citep{2019ApJ...886L...1V}:
\begin{equation}\label{eq:Rz}
    \mathcal{R}(z_m)=\mathcal{R}_n\int^\infty_{z_m}\psi(z_f)P(z_m|z_f)dz_f,
\end{equation}
where $\psi(z)$ is the Madau-Dickinson star formation rate:
\begin{equation}
    \psi(z)=\frac{(1+z)^\alpha}{1+(\frac{1+z}{C})^\beta},
\end{equation}
with $\alpha=2.7$, $\beta=5.6$, $C=2.9$ \citep{2014ARA&A..52..415M}, 
and $P(z_m|z_f,\tau)$ is the probability that the BBH merger at $z_m$ if the binary is formed at $z_f$, which we refer to as the distribution of delay time with the form \citep{2019ApJ...886L...1V}:
\begin{equation}
    P(z_m|z_f,\tau)=\frac{1}{\tau}\exp{\left[-\frac{t_f(z_f)-t_m(z_m)}{\tau}\right]}\frac{dt}{dz}.
\end{equation}
In the above equation, $t_f$ and $t_m$ are the look back time corresponding to $z_f$ and $z_m$ respectively. 

\begin{figure}
    \centering
    \includegraphics[width=0.45\textwidth]{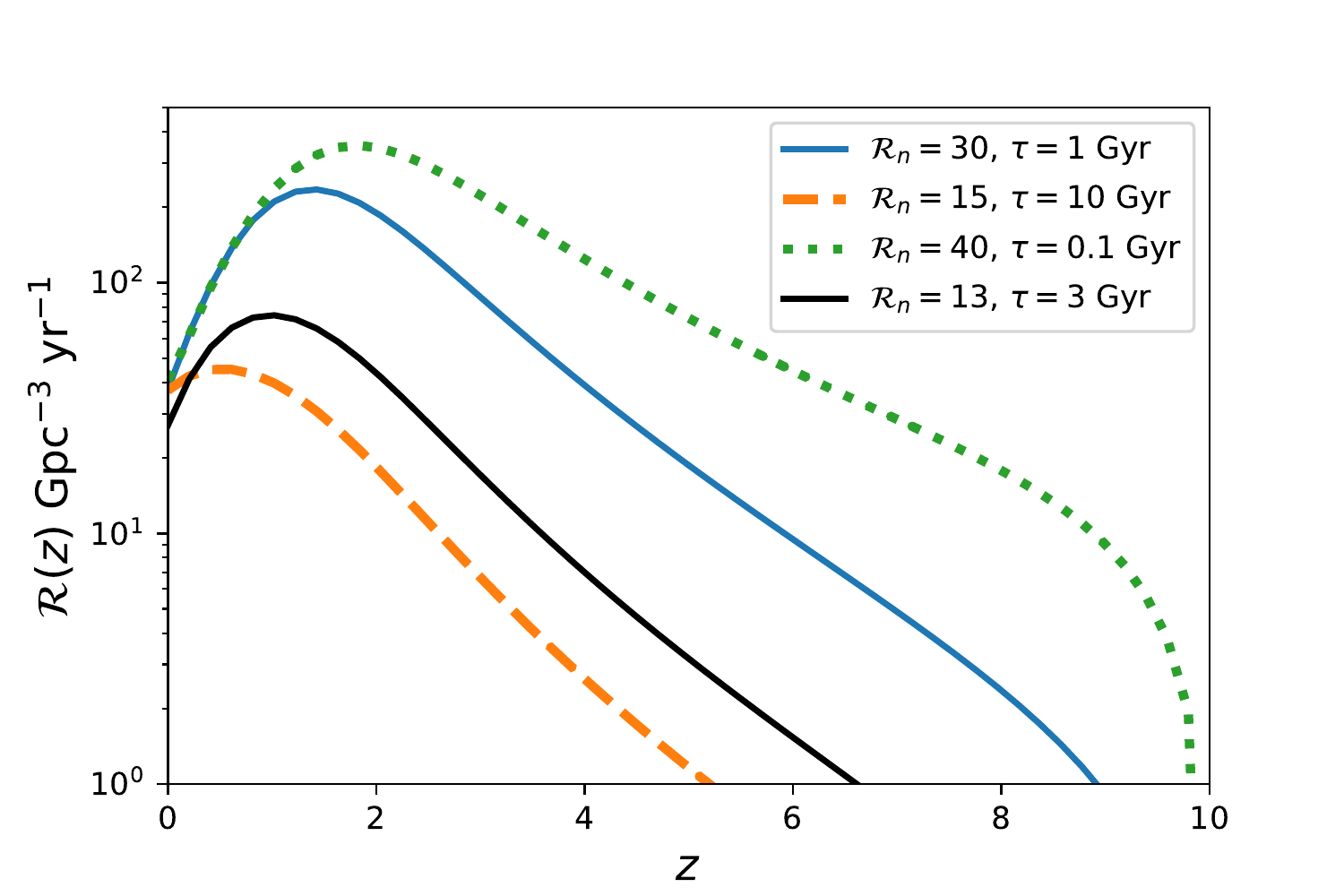}
    \caption{$\mathcal{R}(z)$, [eq.~(\ref{eq:Rz})]  with different $\mathcal{R}_0$ and $\tau$. }
    \label{fig:ndot}
\end{figure}

We give plots of $\mathcal{R}(z)$ with different $\mathcal{R}_n$ and $\tau$ in figure \ref{fig:ndot}. The default parameters are set to $\mathcal{R}_n=13\,$Gpc$^{-3}$\,yr$^{-1}$ and $\tau=3$\,Gyrs, which are compatible with the local merger rate of BBH found in O3a  \citep{2020arXiv201014533T}. 

Although some information is known about the masses of the observed BBH \citep{2020arXiv201014533T}, we assume a generic mass function $p(m_1)$:
\begin{eqnarray}\label{eq:pm1}
    p(m_1)
    \propto &
    \begin{cases}
    \exp\left(-\frac{c}{m_1-\mu}\right)(m_1-\mu)^{-\gamma}, & m_1\le m_{\text{cut}} \\
    0, & m_1>m_{\text{cut}}\\
    \end{cases}
    \label{eqn:BBH-Pop2-mass}
\end{eqnarray}
The distribution of $p(m_1)$ is defined for $m_1>\mu$, which has a power law tail of index $-\gamma$ and a cut-off above $m_{\rm cut}$. When $\gamma=3/2$, the distribution becomes a shifted L\'evy distribution. $p(m_1)$ peaks at $m_1=c/\gamma+\mu$. We set $\mu=3$, $\gamma=2.5$, $c=6$, $m_{cut}=95\,M_\odot$ as default, which result in simulated catalogue that fits with the observed one (see section 5.1). 
The normalization of $p(m_1)$ is 
    $$c^{1-\gamma}\Gamma(\gamma-1, \frac{c}{m_{\text{cut}}-\mu}),$$
where $\Gamma(a,b)$ is the upper incomplete gamma function;

\begin{figure}
    \centering
    \includegraphics[width=0.5\textwidth]{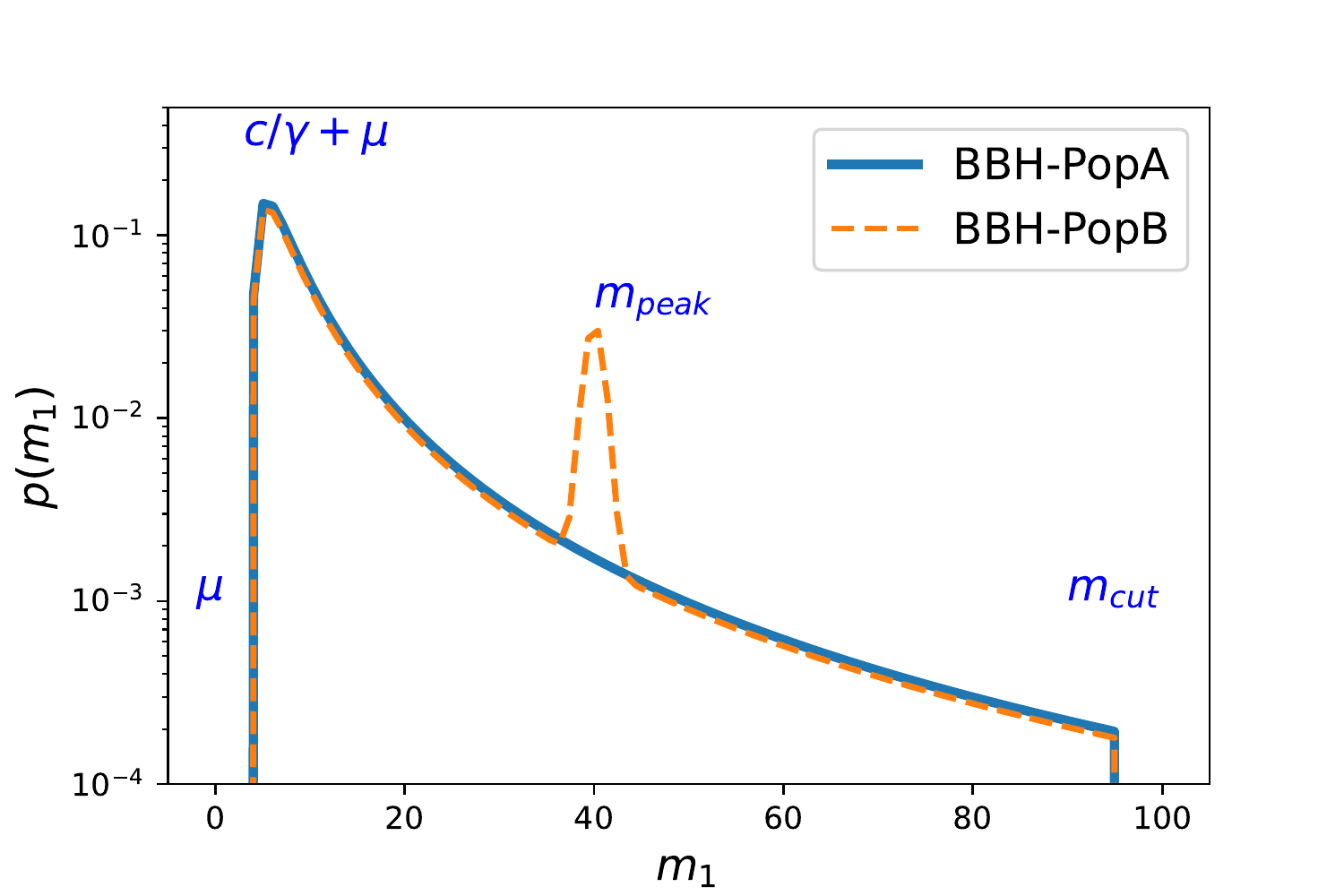}
    \caption{The primary BH mass distribution in the default models of BBH-PopA/B, see eqns~(\ref{eq:pm1}), (\ref{eq:ppeak}). Here we mark $\mu$, $\mu+c/\gamma$ and $m_{\rm peak}$ in the plot, to give an intuition of these quantities.}
    \label{fig:BBH_mass_distri}
\end{figure}

In order to provide more flexibility, we also provide an alternative $p(m_1)$, which has an extra Gaussian peak component $p_{\rm{peak}}(m_1)$ on top of that in equation (\ref{eqn:BBH-Pop2-mass}):
\begin{equation}\label{eq:ppeak}
    p_{\rm{peak}}(m_1)=A_{\rm{peak}}\exp\left[-\left(\frac{m_1-m_{\rm{peak}}}{\sigma_{\rm{peak}}}\right)^2\right],
\end{equation}
the normalization of the alternative $p(m_1)$ is:  
$$c^{1-\gamma}\Gamma(\gamma-1, \frac{c}{m_{\text{cut}}-\mu})+\sqrt{2\pi}\sigma_{\rm{peak}}A_{\rm{peak}}.$$ We denote the population model without/with the peak component in the mass function as BBH-PopA/B. 
Our default parameters for the peak component are $A_{\rm{peak}}=0.002$, $m_{\rm{peak}}=40\,M_\odot$, $\sigma_{\rm{peak}}=1\,M_\odot$, which are compatible with that implied from GWTC-3. In figure \ref{fig:BBH_mass_distri}, we plot the mass distributions for BBH-PopA/B.

For $\pi(q)$  we use a uniform distribution between [$q_{\rm cut}$, 1] and assume $\chi$ follows a truncated Gaussian distribution centered at zero with standard deviation $\sigma_{\chi}$, which is limited between -1 and 1 as demanded by general relativity. The actual mass ratio distribution is still poorly constrained from observation. The discovery of the GW190412 with asymmetric masses implies that the mass ratio distribution can be quite board \citep{2020PhRvD.102d3015A}. The Gaussian spin model is consistent with the finding in \cite{2020arXiv201014533T}. The default parameters we use are $q_{\rm cut}=0.4$ and $\sigma_{\chi}=0.1$. Those parameters in the population model can all be reset by users in both the web interface or in the python package. 


\subsubsection{Double neutron star mergers}\label{sec:DNS_model}

For the population model of DNS mergers, the merger rate density is similarly expressed as:
\begin{equation}
    \dot{n}(z, m_a, m_b,\chi)=\mathcal{R}(z)p(m_a)p(m_b)P(\chi),
\end{equation}
where $\mathcal{R}(z)$ is taking the same form as in BBH population model, but with a different default setting: $\mathcal{R}_n=300\,$Gpc$^{-3}$\,yr$^{-1}$ and $\tau=3$\,Gyrs, which are compatible with the local merger rate of DNS found with GWTC-2; 

$p(m_{a,b})$ is the mass function of the neutron stars. Note that we use $m_{a,b}$ instead of $m_{1,2}$. The latter are the primary and secondary stars based on their masses, while the former we do not distinguish between the primary and the secondary. We assume both $m_a$ and $m_b$ are following the same mass function. We use a truncated Gaussian with upper and lower cuts as the mass function. The default parameters are the mean $\overline{m}=1.4\,M_\odot$, the mass dispersion $\sigma_m=0.5\,M_\odot$, upper cut $m_{\rm cut, low}=1.1\,M_\odot$, $m_{\rm cut, high}=2.5\,M_\odot$; we apply also a truncated Gaussian spin model, with a smaller dispersion $\sigma_\chi=0.05$. These simplified choices are roughly agree with observations \citep{2011MNRAS.414.1427V,2012ApJ...757...55O,2013ApJ...778...66K,2015PhR...548....1M,2019ApJ...876...18F,2019MNRAS.488.5020Z}, while they can be substitute with more realistic models. 


\subsubsection{Neutron star/black hole mergers}\label{sec:NSBH_model}

For the population model of BHNS, we again assume the merger rate density to be:
\begin{equation}
    \dot{n}(z, m_1, m_2, \chi)=\mathcal{R}(z)f(m_\bullet)p(m_{\rm n})P(\chi),
\end{equation}
where $\mathcal{R}(z)$ is taking the same form as for BBH and DNS, with a different default setting: $\mathcal{R}_n=45\,$Gpc$^{-3}$\,yr$^{-1}$ and $\tau=3$\,Gyrs; which is broadly consistent with the  merger rate, which is implied by the number of BHNS detection in LVK O3a+b \citep{2021ApJ...915L...5A}. 
$f(m_\bullet)$ is the mass function of the BH, which has the same function forms same as BBH. We denote the population model without/with the peak component in $f(m_\bullet)$ as BHNS-PopA/B; The default parameters for $f(m_\bullet)$ are $\mu=3$, $\gamma=2.5$, $c=6$, $m_{cut}=95\,M_\odot$, $A_{\rm{peak}}=0.002$, $m_{\rm{peak}}=40\,M_\odot$, $\sigma_{\rm{peak}}=1\,M_\odot$.
$p(m_n)$ is the mass function of the NS, which is the same as in the DNS case. 

\subsection{Sources for space-borne detectors}

\subsubsection{Suppermassive Black Hole Binaries}\label{sec:MBHB_model}
\begin{figure}[h]
    \centering
    \includegraphics[width=0.5\textwidth]{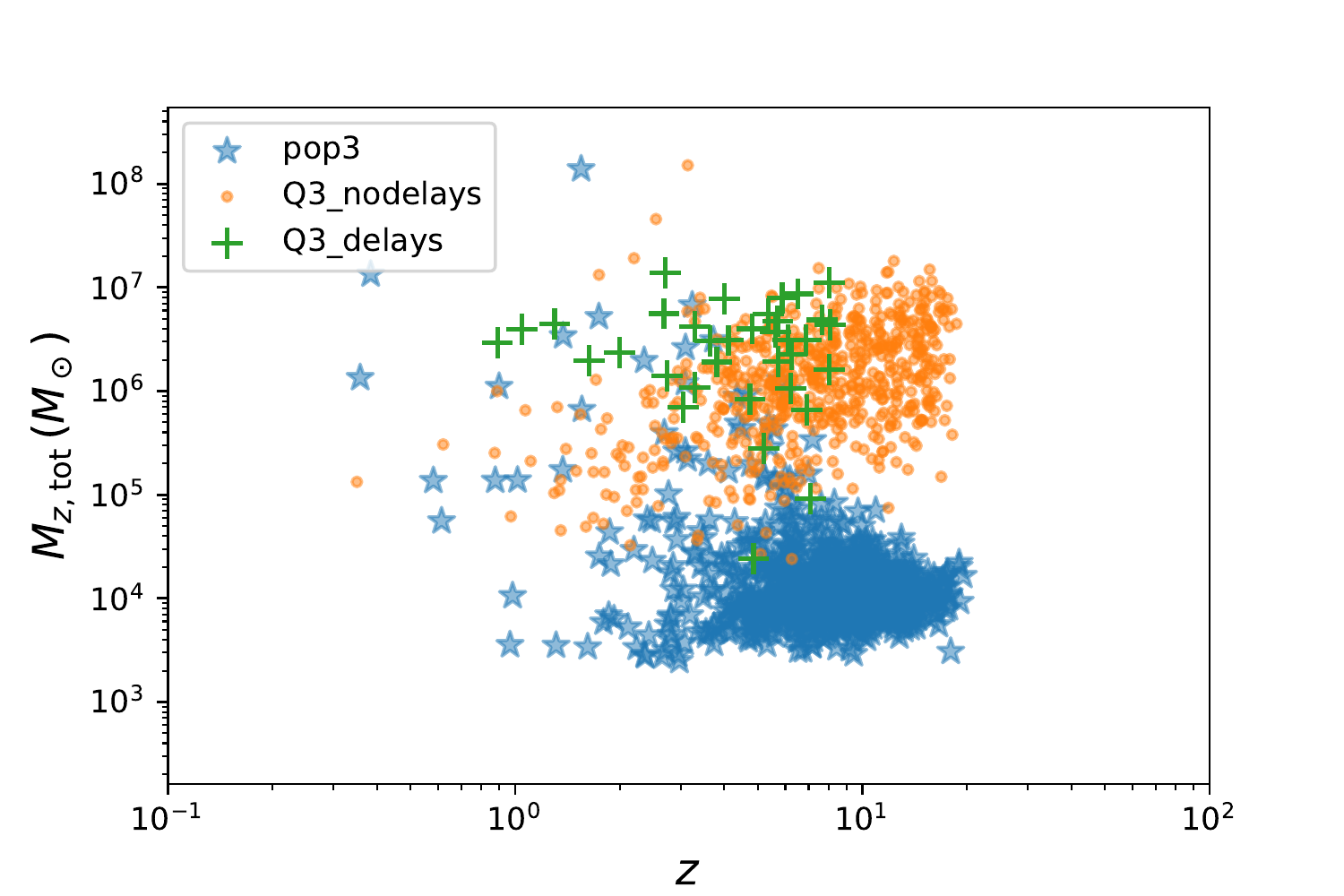}
    \caption{$M_{z,\rm tot}$ vs. $z$ of SMBBH mergers that occur in the Universe over a timescale of five years for three population models from Klein16. The data corresponds to one realisation.}
    \label{fig:mbhb_cat}
\end{figure}

In the last two decades, it has been established that in the center of most galaxies there is a supermassive black hole (SMBH, with mass from $10^4\,M_\odot$ to $10^{10}\,M_\odot$). Since the mergers of galaxies are thought to be ubiquitous under the hierarchical clustering process, there are expected to be close binaries of supermassive binary black holes (SMBBH) in the merger galaxies, which emit GW during their inspiral and merger phase \citep[e.g.][]{2014SSRv..183..189C}. Such SMBBH insprials are the main targets of LISA \citep{2017arXiv170200786A}, TianQin \citep{2016CQGra..33c5010L} and PTA \citep{2004ApJ...606..799J,2005ApJ...625L.123J}, since the frequency of their GW falls in the $10^{-8}-1$\,Hz range.     
We use the SMBBH merger catalogues from \citet{2016PhRvD..93b4003K} (Klein16 hereafter), which are based on \cite{2012MNRAS.423.2533B}. There are three population models being considered, namely \texttt{pop3}, \texttt{Q3\_nodelays} and \texttt{Q3\_delays}. They mainly differ in the origin of their SMBH seeds, and whether the delays between SMBH mergers and galaxy mergers are included (see Klein16 for a detailed description; see also a review on supermassive BH formation and evolution by \citealt{2020ARA&A..58...27I}). For \texttt{pop3}, the SMBH seeds are from PopIII stars (light seeds), and the delay between SMBH and galaxy mergers is accounted; for \texttt{Q3\_nodelays} and \texttt{Q3\_delays}, the SMBH seeds are assumed to be from the collapse of protogalactic disks (heavy seeds). The former accounts for the delay between SMBH and galaxy mergers, while the latter does not.  
For each population model, there are ten catalogues, each corresponding to a realisation of all sources in the Universe within five years. The number of total events for each population is $\sim890$ for \texttt{pop3}, $\sim630$ for \texttt{Q3\_nodelays} and $\sim40$ for \texttt{Q3\_delays}. They are compatible with the reported averaged merger rates in Klein16 (\texttt{pop3}: 175.36 yr$^{-1}$, \texttt{Q3\_nodelays}: 121.8 yr$^{-1}$ and \texttt{Q3\_delays}: 8.18 yr$^{-1}$). In figure \ref{fig:mbhb_cat} we plot $M_{z,\rm tot}$ (red-shifted total mass) and $z$ of SMBBH mergers that occur in the Universe over a timescale of five years for three population models as a direct demonstration of the population models. The distribution agrees with those shown in Fig 3 of Klein16.

\subsubsection{Double White Dwarfs}\label{sec:DWD_model}
Another important population of LISA sources are Ultra Compact Galactic Binaries. Among those Galactic binaries, close double white dwarfs (DWDs) are the dominant, and are long expected to be promising targets for LISA and other space-borne GW detectors \citep[e.g.][]{2001A&A...375..890N,2010A&A...521A..85Y,2012ApJ...758..131N,2012A&A...546A..70T,2017MNRAS.470.1894K,2019MNRAS.490.5888L,2020arXiv200507889H}. 

\begin{figure}[h]
    \centering
    \includegraphics[width=0.5\textwidth]{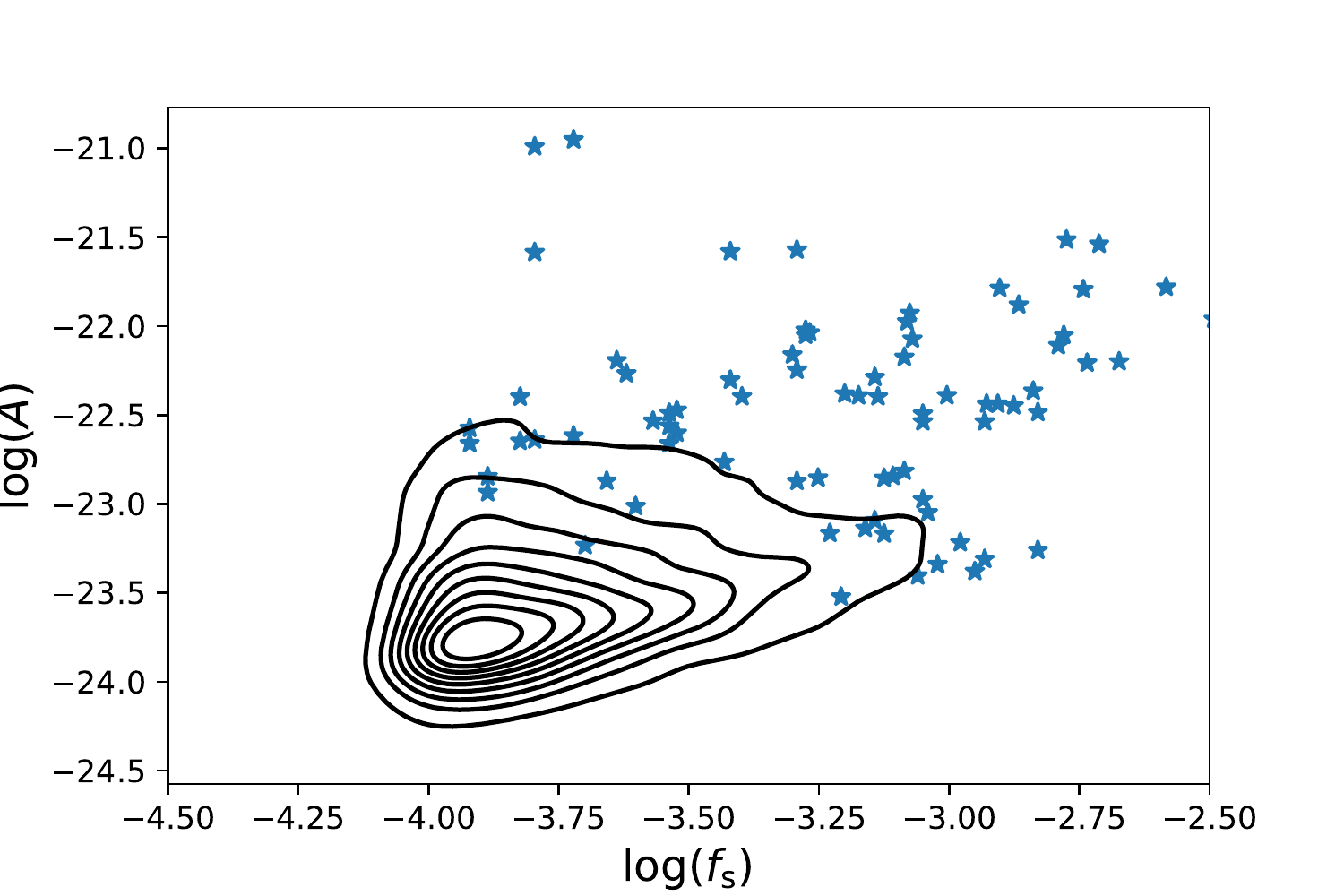}
    \caption{The distribution density of GW properties of DWD binaries in the simulated catalogue (black contours), as function of GW frequency $f_{\rm s}$ and and GW amplitude $A$. The blue stars are known DWDs as verification binaries. }
    \label{fig:GWDcat}
\end{figure}

We use a synthetic catalogue of close DWD in the whole Galaxy \citep{2001A&A...375..890N}. In figure \ref{fig:GWDcat} we plot the joint distribution density of the frequencies $f_{\rm s}=2/P$ (where $P$ is the orbital period of the binaries) and the intrinsic amplitudes $A=2\left(G\mathcal{M}\right)^{5/3}\left(\pi f\right)^{2/3}/(c^4d)$ of GW emitted from binaries in the catalogue. The contours mark levels which correspond to uniform iso-proportions of the density. The total number of sources in the catalogue is $\sim2.6\times10^7$. 

Beside the synthetic catalogue, we also include a catalogue of 81 known DWDs
\citep{2020arXiv200507889H} a.k.a. verification binaries (VBs, see also \citealt{2018MNRAS.480..302K}). Those VBs are plotted along in figure \ref{fig:GWDcat} with star markers. Note that the distribution of the verification binaries do not follow that of the simulated DWD in the whole Galaxy, due to very strong observational biases which favour close by sources to be detected as VBs. 

\subsubsection{Extreme Mass-Ratio Inspirals}\label{sec:EMRI_model}
In the nuclei region of galaxies, surrounding the supermassive black holes (SMBH), there are abundant stellar populations. Among them, compact objects, including stellar mass BHs, NS and white dwarfs, can inspiral into the central SMBH, radiating a large amount of energy in GW. These systems are referred to as extreme mass ratio insprials (EMRIs). EMRIs are very interesting targets for space-borne detectors such as LISA \citep[see][]{2007CQGra..24R.113A}.        

\begin{figure*}
    \centering
    \includegraphics[width=\textwidth]{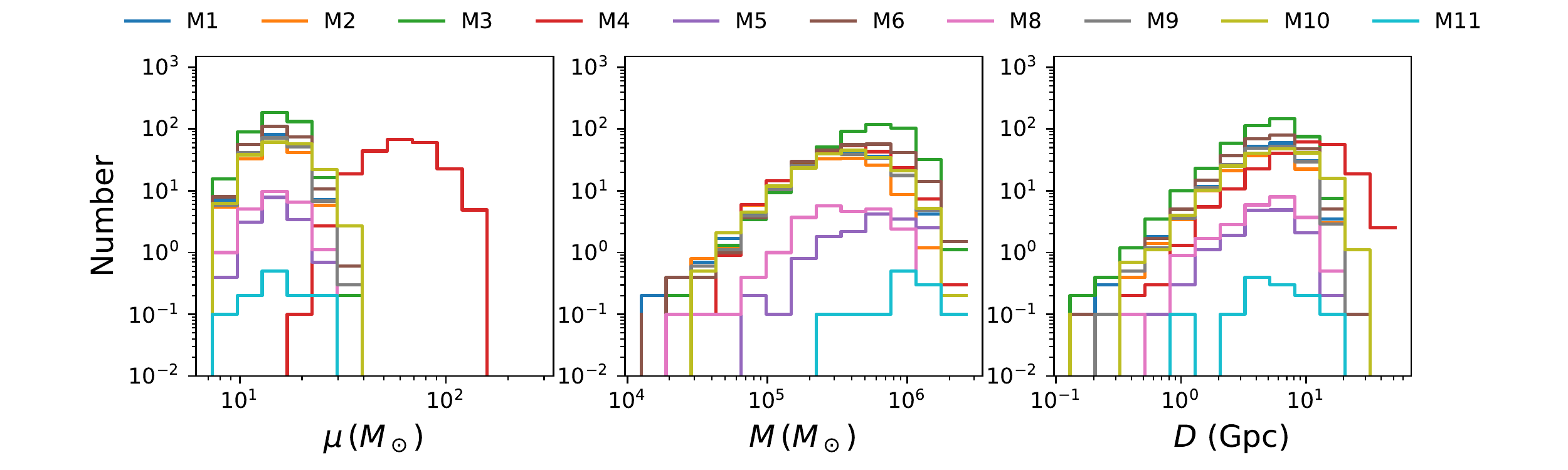}
    \caption{The averaged histograms of $\mu$, $M$ and $D$ of EMRIs happen in one year, assuming different population models from Babak17}
    \label{fig:emris_cat}
\end{figure*}

In the \GWT, we use the simulated catalogues from \citet{2017PhRvD..95j3012B} (Babak17 hereafter) and use their population models M1-M11 (skipping M7, explained later). These populations differ in several ingredients: the mass function and spin distribution of SMBHs, the $M-\sigma$ relation, the ratio of plunges to EMRIs and the characteristic mass of the compact objects (see Babak17) for a detailed description of models).
For each population model, there are ten realizations of the catalogues, which contain detectable EMRIs within one year with the assumption of standard LISA noise properties. The distributions of $\mu$ (mass of the stellar BH), $M$ (mass of the massive BH) and $D$ (luminosity distance) in the catalogues are plotted in figure \ref{fig:emris_cat}. The histogram for each population is averaged among the ten realizations. Since we are using an SNR limited sample, instead of a complete one of the whole Universe (which is $\sim10$ times larger), the \GWT will give an underestimated detectable number and an incomplete catalogue of detections, especially when using a lower SNR cutoff or a more sensitive LISA configuration. For now, we exclude M7 and M12 from the Toolbox, because in their population models the direct plunges are ignored, therefore the total number of EMRIs are about one order of magnitude larger than others, which will make the computation take too long.  

\subsection{Sources for Pulsar Timing Arrays}
\subsubsection{Individual Massive Black Hole}\label{sec:Indiv_BH_model}

Long before (millions or tens of millions of years, depends on their chirp mass; see \citealt{2010PhRvD..81j4008S,2013PhRvD..87f4036P,2019A&ARv..27....5B}) the GW from supermassive black hole binaries entering the band of space-borne GW detectors, it lies in the PTA frequency range. If there would be such MBH binaries sufficiently close to the Earth, PTAs could detect their signals. Since no sources are known yet, we incorporate this source class in the \GWT in the form of a free form in which the user can fill in the frequency and GW amplitude, and the \GWT will determine if this binary, as a monochromatic GW source, can be detected by the selected PTA.

\subsubsection{Stochastic background}\label{sec:SGWB_model}
The second class of target for PTA are Stochastic GW background (SGWB). It can originate from the incoherent overlapping of many unresolvable SMBBH \citep{2001astro.ph..8028P,2008MNRAS.390..192S,2014ApJ...789..156M}, the relic of primordial GW \citep{1976JETPL..23..293G,1977NYASA.302..439G,1982PhLB..108..389L,1980PhLB...91...99S}, or the collision of cosmic-strings \citep{2001PhRvD..64f4008D,2005PhRvD..71f3510D,2006PhRvD..73j5001S,2007PhRvL..98k1101S,2010PhRvD..81j4028O}. Each of these gives rise to a power-law GW signal 
\begin{equation}
    h_c^2(f)=Cf^\gamma,
\end{equation}
where the index $\gamma$ corresponds to the origin of SGWB. For incoherent overlapping of SMBBH, $\gamma=-2/3$; for relic GW, $\gamma=-1$ and for cosmic-strings, $\gamma=-7/6$. We also enable users to customize $\gamma$. 
\section{Implementation 2: Gravitational Wave detectors}

\subsection{Ground-based interferometers}
\subsubsection{Noise model of interferometers}
\begin{figure}
    \centering
    \includegraphics[width=0.5\textwidth]{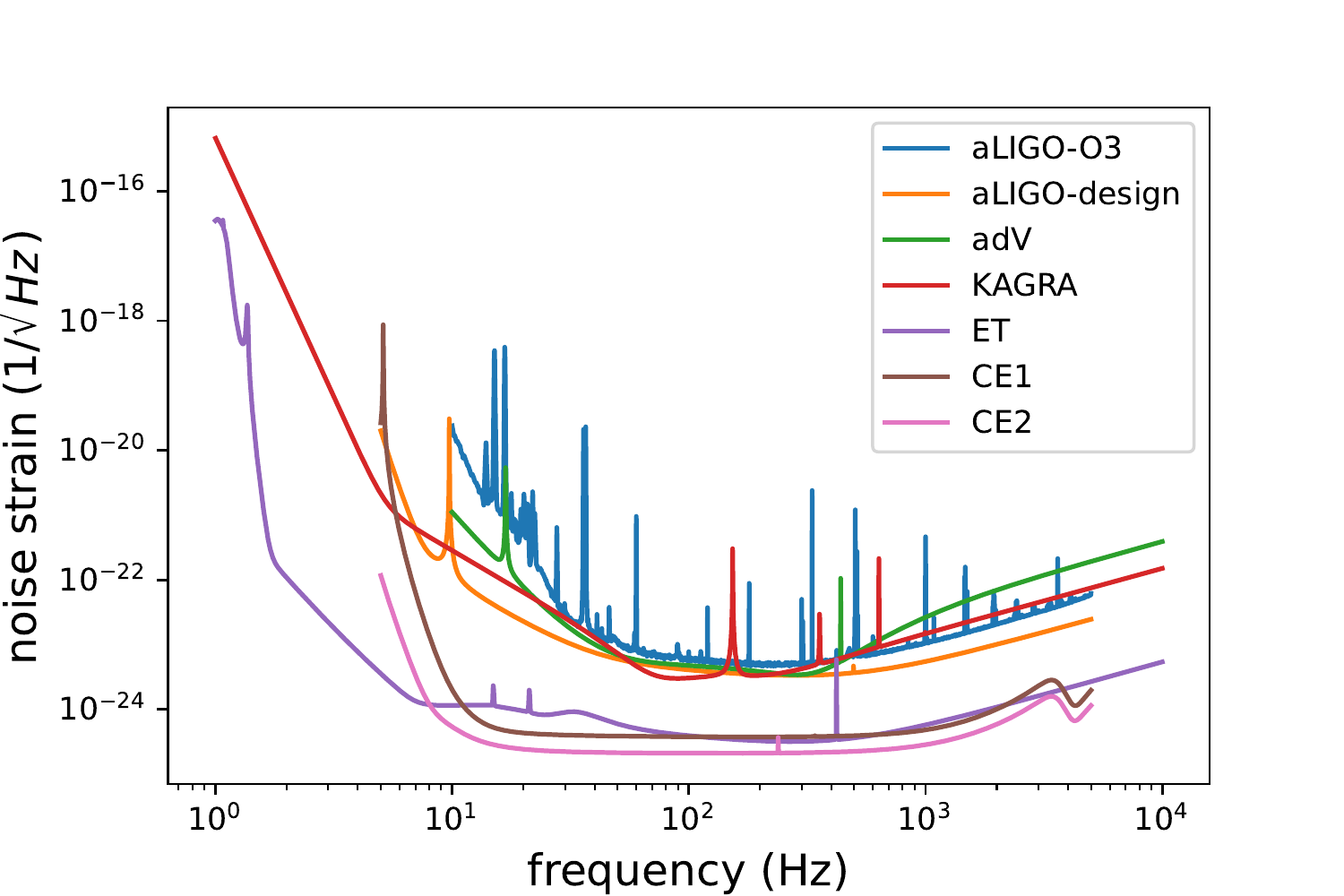}
    \includegraphics[width=0.5\textwidth]{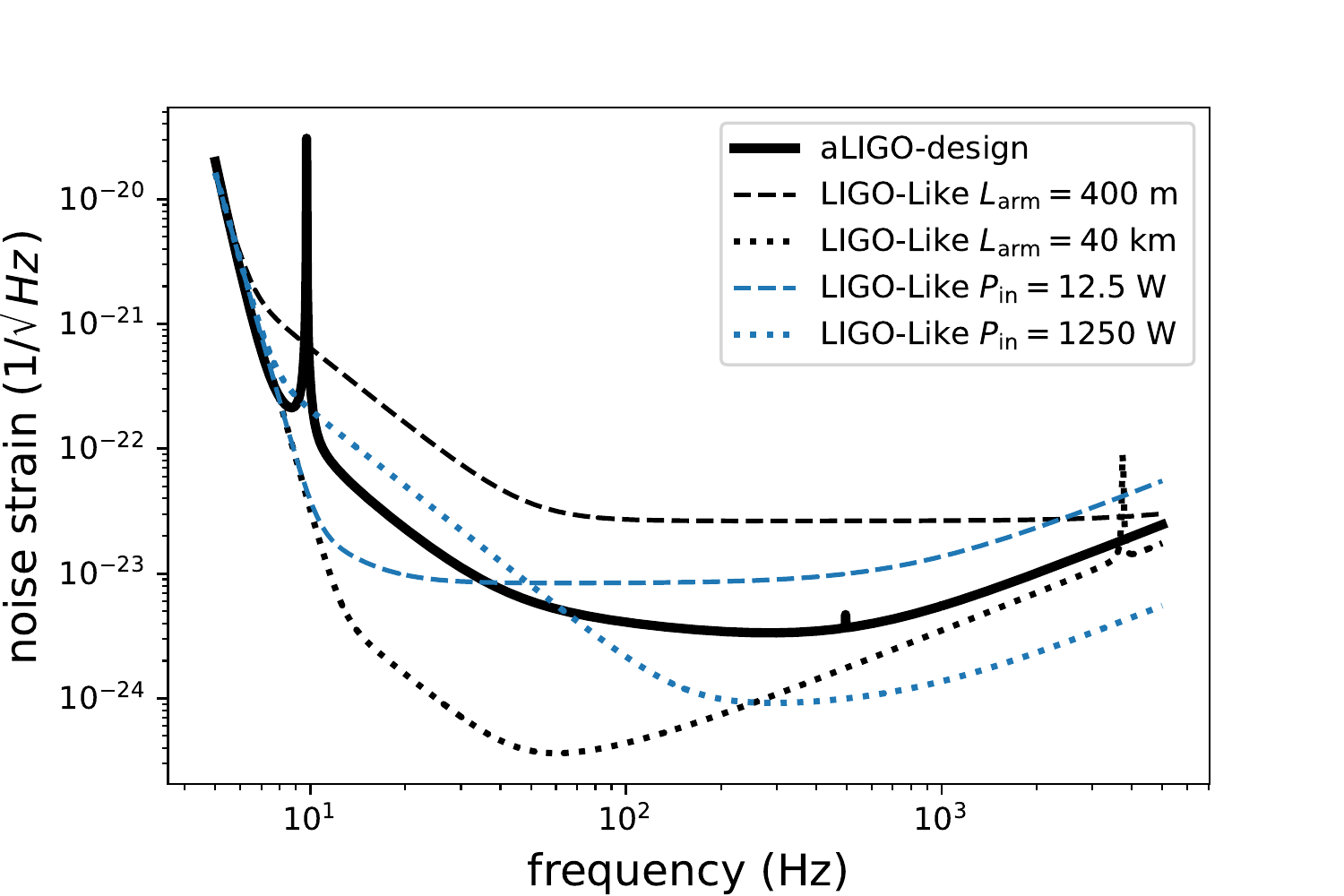}
    \caption{\textbf{Upper Panel:} Noise Curves of the default detectors: aLIGO-O3 corresponds to the noise performance in O3 period, while aLIGO-design is the designed sensitivity of the aLIGO; CE1 and CE2 correspond to the expected first and second stages of CE respectively; ET-D-sum curve is applied for ET. \textbf{Lower Panel:} aLIGO in design vs. Customised settings (arm length=4 km, laser power=125 W)}
    \label{fig:earth_det}
\end{figure}

For ground-based interferometers, the \GWT integrates the design performance of advanced LIGO (aLIGO), Advanced Virgo (AdV), KAGRA, CE and ET instruments. The noise models for the above-mentioned interferometers are taken from the following resources:
\begin{itemize}
    \item aLIGO: \\\url{https://dcc.ligo.org/LIGO-T1800044/public}, see also \cite{2015CQGra..32g4001L};
    \item adV and KAGRA: \\\url{https://dcc.ligo.org/LIGO-T1500293/public};
    \item CE: \\\url{https://dcc.ligo.org/LIGO-P1600143/public};
    \item ET: \\\url{http://www.et-gw.eu/index.php/etsensitivities}, see also  \cite{2008arXiv0810.0604H,2010CQGra..27a5003H,2011CQGra..28i4013H};
\end{itemize}
In the upper panel of figure \ref{fig:earth_det}, we plot the noise curves that are used for the default detectors. aLIGO-O3 corresponds to the noise performance in O3 period, while aLIGO-design is the designed sensitivity of the aLIGO; CE1 and CE2 correspond to the expected first and second stages of CE respectively; for ET we use the ET-D-sum curve.

In addition, the \GWT employs the package \texttt{FINESSE} to calculate $S_{\rm{n}}$ of a LIGO-like or a ET-like interferometer with customised settings \citep{2020SoftX..1200613B}. Users can define the following parameters of the detector: 
\begin{itemize}
    \item Arm Length
    \item Laser power
    \item Arm mirror mass
    \item Arm mirror transmission coefficient
    \item Signal recycling mirror transmission coefficient
    \item Power recycling mirror transmission coefficient
    \item Signal recycling phase factor
    \item Power recycling cavity length
    \item Signal recycling cavity length
\end{itemize}
The most important parameters that affect the sensitivity are the arm length and laser power. In the lower panel of figure \ref{fig:earth_det}, we show the effects of varying the arm length and the laser power starting from the design aLIGO sensitivity. One of the other influential parameters is the mass of the arm mirror. Heavier arm mirrors will decrease the noise in the low frequency ends slope and leave the high frequency end unaffected. 


\subsubsection{SNR of GW from compact binary merger}
\label{sec:SNRandFIM}
The core of the method with which the \GWT determines detectability of sources is to compare the SNR threshold $\rho_\star$ with that of a source, which can be calculated with \citep{maggiore2008gravitational}:
\begin{equation}
        \rho^2=4\int^{f_{\rm{high}}}_{f_{\rm{low}}}\frac{|h^2(f)|}{S_{\rm{n}}(f)}df,
    \label{eqn:snrLVK}
\end{equation}
where $h(f)$ is the frequency domain response of the interferometer to the GW signal, and $S_{\rm n}$ is the noises power density. For a binary system, the detector response can be expressed as\footnote{in general, the detector response to the ($\ell$, $m$) multipole mode is $$h^{\ell m}=F_+h^{\ell m}_++F_\times h^{\ell m}_\times.$$ When we limit the waveform to the dominant (quadrupole) mode, the response takes the form in equation \ref{eqn:snrLVK}, with the terms $(1+\cos^2\iota)/2$, $\cos\iota$ from the $h^{22}_+$ and $h^{22}_\times$ respectively. For expressions of higher order mode $h^{\ell m}_{+,\times}$, see \cite{2017PhRvD..96l4010M,2021PhRvD.103b4042M}. We discuss the influence of including the higher order modes in Section \ref{sec:discussion}.}:
\begin{equation}
    h(f)=\mathcal{C}\sqrt{\left(\frac{1+\cos^2\iota}{2}\right)^2F_+^2+\cos\iota^2F_\times^2}A(f)e^{-i(\Psi(f)+\phi_{\rm{p}})}.
    \label{eqn:response}
\end{equation}
In the above equation, the constant
\begin{equation}
    \mathcal{C}=\frac{1}{2}\sqrt{\frac{5}{6}}\frac{\left(G\mathcal{M}\right)^{5/6}}{c^{3/2}\pi^{2/3}D_{\rm{L}}},
\end{equation}
where $\mathcal{M}$ is the red-shifted chirp mass of the binary, $D_{\rm{L}}$ is the luminosity distance, $\iota$ is the inclination angle between the orbital angular momentum and the line of sight; $A(f)$ is the frequency dependence of the GW amplitude. As a proof of principle of the \GWT, we apply the waveform approximant, which has hybrid degrees of simplification. For the amplitude $A(f)$, we use a broken power-law:
\begin{equation}
       A(f)=
\begin{cases}
f^{-7/6}, &f\le f_{\rm trans}, \\
f^{-2/3}, &w_mf_{\rm trans}<f\le f_{\rm cut}; \\
0, &f\ge f_{\rm cut},\\
 \end{cases}
\end{equation}
where $w_m$ is the scaling factor that to make $A(f)$ continuous. The above formula represents the inspiral, merger and ring-down phases of a circular, non-spinning point mass binary. It corresponds to the $A(f)$ in the approximant \texttt{IMRPhenomD} \citep{2011PhRvL.106x1101A}, when the effective spin $\chi=0$. The transition frequency $f_{\rm trans}$ and the cut-off frequency $f_{\rm cut}$ corresponds to $f_1$ and $f_2$ of \citet{2011PhRvL.106x1101A}. $A(f)$ and $m_1,m_2$ and $D$ via $\mathcal{C}$ essentially determine the SNR of sources, because the effective spin has a much weaker effect on the SNR. Therefore the dependence on $\chi$ can be separated from that on $m_1,m_2$ and $D$. It makes the catalogue sampling process easier and more effective (the Markov Chain mixes faster with less dimension, see below).

$\Psi(f)$ is the phase of the waveform. It is essential for evaluating the uncertainties on the intrinsic parameters. Therefore we use $\Psi(f)$ from the frequency domain waveform \texttt{IMRPhenomD} \citep{2011PhRvL.106x1101A} and restore its dependency on $\chi$. It corresponds to a hybrid waveform of a circular binary with parallel spin, that matches post Newtonian and numerical relativity waveforms.

$F_{+,\times}$ are the antenna patterns of the interferometer, which are function of position angles of the source $(\theta,\varphi)$ and the polarization angle of the GW $\psi$, For LIGO/Virgo/KAGRA-like interferometers, which have two perpendicular arms, the antenna patterns are: 
\begin{align}
    F_{+,90^\circ}&=\frac{1}{2}(1+\cos^2\theta)\cos2\varphi\cos2\psi+\cos\theta\sin2\varphi\sin2\psi \nonumber\\
    F_{\times,90^\circ}&=\frac{1}{2}(1+\cos^2\theta)\cos2\varphi\sin2\psi+\cos\theta\sin2\varphi\cos2\psi,
\end{align}
and for ET-like interferometers with $60^\circ$ angles between the arms \citep{2012PhRvD..86l2001R}:, 
\begin{align}
    F_{+,60^\circ}&=-\frac{\sqrt{3}}{2}F_{+,90^\circ}\nonumber\\
    F_{\times,60^\circ}&=\frac{\sqrt{3}}{2}F_{\times,90^\circ}.
    \label{eqn:et-at}
\end{align}
ET will have three nested interferometers, $60^\circ$ rotated with respect to each other. The antenna pattern for each interferometers are $F_{i,+,\times}(\theta,\varphi,\psi)=F_{0,+,\times}(\theta,\varphi+2/3i\pi,\psi)$, where $i=0,1,2$ is the index of the interferometers, and $F_{0,+,\times}$ are those in equation (\ref{eqn:et-at}). The joint response can be calculated with equation (\ref{eqn:response}), where the antenna pattern squared should be substituted with:
\begin{equation}
    F^2_{+,\times}=\sum^2_{i=0}F^2_{i,+,\times}.
\end{equation}

In our treatment, the modulation on the antenna patterns due to the Earth rotation is neglected, for the duration of the GW signal is typically much less than the period of Earth rotation. The situation would change for 3G detectors, which has much broader frequency window and the ability to better resolve the waveform. 

\subsubsection{Determining the sample of detected sources}
Given the differential cosmic merger rate density for compact binary mergers $\dot{n}$, the theoretical number distribution for each source class in the catalogue is:
\begin{equation}
    N_{\rm{D}}(\mathbf{\Theta},\theta,\varphi,\psi,\iota)=\frac{\Delta T}{1+z}\frac{dV_{\rm{c}}}{4\pi dz}\dot{n}(\mathbf{\Theta},\theta,\varphi,\iota,\psi)\mathcal{H}(\rho^2-\rho^2_\star).\label{eqn:eight}
\end{equation}
where $\Delta T$ is the time span of observation, $dV_{\rm{c}}/dz$ is the differential cosmic comoving volume (volume per redshift), $\mathcal{H}$ is the Heaviside step function and $\rho_\star$ is the SNR threshold, $\mathbf{\Theta}$ denotes the intrinsic parameters and the luminosity distance of the source. Marginalising over the directional parameters and assuming that $\dot{n}$ is isotropic yields
\begin{equation}
    N_{\rm{D}}(\mathbf{\Theta})=\frac{T}{1+z}\frac{dV_{\rm{c}}}{dz}\dot{n}(\mathbf{\Theta})\mathcal{D}(\mathbf{\Theta}),\label{eqn:detectables}
\end{equation}
where 
\begin{equation}
    \mathcal{D}(\mathbf{\Theta})=\oiint d\Omega d\Omega^\prime\mathcal{H}(\rho^2-\rho^2_\star)/(4\pi)^2,
\end{equation}
is the detectability of the source, which is determined by the detector properties and the waveform of the source. Since we use the same waveform for BBH, DNS and BH-NS, the difference among these three populations are only in the cosmic merger rate $\dot{n}$ discussed above in Sections~\ref{sec:BBH_model}-\ref{sec:NSBH_model}. 

The total number of expected events catalogue is:
\begin{equation}
    N_{\rm{tot}}=\int d\mathbf{\Theta}N_D(\mathbf{\Theta}),
\end{equation}
and the number of detections thus is Poisson realisation of the expectation value $N_{\rm{D}}(\mathbf{\Theta})$. The synthetic catalogue is then obtained by a Markov Chain-Monte Carlo sampling from $N_{\rm{D}}(\mathbf{\Theta})$. We apply the elliptical slice sampling algorithm \citep{ESS}, which converges faster to the target distribution comparing with the traditional Metropolis-Hasting algorithm \citep{MH} and requires less tuning on the initial parameters. 

We also give the estimated uncertainties on the parameters using the Fisher Information Matrix (FIM):
the covariance matrix is related to the Fisher matrix with:
\begin{equation}
    \left<\delta\mathbf{\Theta}_i\delta\mathbf{\Theta}_j\right>=\mathcal{F}^{-1}_{ij},
    \label{eqn:covariance}
\end{equation}
where the Fisher matrix is defined as:
\begin{equation}
    \mathcal{F}_{ij}=(\partial h/\partial\mathbf{\Theta}_i|\partial h/\partial\mathbf{\Theta}_j).
\end{equation}
The partial derivatives in the above equation are calculated numerically:
\begin{equation}
\frac{\partial h}{\partial\mathbf{\Theta}_i}=\frac{h(\mathbf{\Theta}_i+\Delta\mathbf{\Theta}_i)-h(\mathbf{\Theta}_i-\Delta\mathbf{\Theta}_i)}{2\Delta\mathbf{\Theta}_i}.
\label{eq:derivatives}
\end{equation}
In the \GWT, we use $\Delta\Theta_i=10^{-8}\Theta_i$, as it is small enough to give stable results. We only calculate the FIM for intrinsic parameters ($m_{1,z}$, $m_{2,z}$ and $\chi$) and an overall scaling factor that represents the effect of all extrinsic parameters. As a result, we cannot give an estimate of the uncertainties of extrinsic parameters, i.e., the distance, sky locations, inclination and polarization angle, for these parameters contribute to the overall scaling factor in a highly correlated way. In reality, the sky locations are determined largely by triangulation with a network of detectors, and the precision of other extrinsic parameters are also depended on triangulation. In the current version of the \GWT we do not include a detector network. The calculation of FIM is thus simplified. However, in order to obtained an estimation on the uncertainties on the source frame masses, one still needs to propagate the uncertainties on the redshift. The uncertainties on the redshift itself is also an interesting quantity that users want to obtain from the synthetic catalogue of \GWT. We give evaluation on $\delta z$ from empirical relations. For 2G detectors, we use $\delta z=0.5z$ roughly represents the trend of $\delta z-z$ in GWTC-3; for 3G detectors, we use $\delta z=0.017z+0.012$, which is a fit from the simulation results of \cite{2017PhRvD..95f4052V}. $\delta z$ is propagated to the uncertainties on the source frame masses with the following equation: 
\begin{equation}
    \delta m_i=m_i\sqrt{\big(\frac{\delta m_{i,z}}{m_{i,z}}\big)^2+\big(\frac{\delta z}{1+z}\big)^2}.
    \label{eq:error_propagation}
\end{equation}

It is worth mentioning that although FIM is a quick, simple and widely applied method of uncertainties, it something gives overestimated uncertainties comparing with those from full Bayesian inference \citep[e.g.][]{2015PhRvD..91d2003V}, especially for events with low SNR \citep[e.g.][]{2008PhRvD..77d2001V,2013PhRvD..88h4013R}. 

The returned catalogue is composed by the mean values of the parameters and their estimated uncertainties. A more realistic simulation of the observed catalogue can be obtained by shifting the mean values according to the corresponding uncertainties, which is straightforward and easy. Since the uncertainties given here are conservative, the \GWT return the un-shifted catalogue, and let the user to decide whether or how to further shift the catalogue.   

\subsection{Space-borne interferometers}
The Space-borne interferometers module of the \texttt{Toolbox} enables users to simulate observations with LISA-like space-borne GW observatories \citep[see][]{2015CQGra..32i5004B}. We work with the codes of the LISA Data Challenge (LDC, \url{https://lisa-ldc.lal.in2p3.fr}, a successor program of the earlier Mock LISA Data Challenge \citep{MLDC}), and make it possible for users to customize the arm length, the laser power and the telescope diameter of LISA. In the ground-based interferometers section the theoretical probability distribution of parameters of the detectable sources are first calculated, afterwards samples are drawn from such distribution as synthetic catalogues of observations. The procedure for LISA-like detectors is different: we go through pre-generated synthetic catalogues of different source populations in the whole Universe/Galaxy and calculate the SNR of each source to be detected by LISA. The SNR is still calculated with:
\begin{equation}
    \rho^2=4\int\frac{|h^2(f)|}{S_{\rm{n}}(f)}df,
    \label{eqn:lisasnr}
\end{equation}
where $h(f)$ is the LISA response to a waveform of a source, and $S_n(f)$ is the noise power spectrum density (PSD). The time-delay interference (TDI) channels are combinations of data streams such that the noises arising from the fluctuation of the laser frequency can be exactly cancelled while the signal in GW can be preserved \citep{1999PhRvD..59j2003T,1999ApJ...527..814A,2000PhRvD..62d2002E}. In the \GWT, we consider the LISA responses and the noise spectral density in the first generation TDI-$X$ channel \citep{1999ApJ...527..814A}. The noise spectrum will be introduced in the next section. Three classes of sources are included in the \GWT for LISA, namely: mergers of SMBBH, resolved DWDs in the Galaxy and EMRIs. Waveforms and the corresponding LISA responses will be introduced in the following subsections. Examples of synthetic observations are given and compared with the literature in Section~\ref{Results}. Uncertainties of the parameters are again estimated with the FIM method.   

\subsubsection{Noise TDI}
The PSD of the noise TDI-$X$ response is formulated as \citep{1999ApJ...527..814A}:
\begin{equation}
    S_X(f)=[4\sin^2(2x)+32\sin^2x]S^{\rm{accel}}_y+16\sin^2xS^{\rm{optical}}_y,\label{eqn:noisepsd}
\end{equation}
where $x=2\pi fL/c$, and $L$ is the arm length of LISA and $c$ is the speed of light, $S^{\rm{accel}}_y$ and $S^{\rm{optical}}_y$ are the fractional frequency fluctuations due to the acceleration noise of the spacecrafts and the optical meteorology system noise respectively. For the acceleration noise, we use \citep{2017arXiv170200786A}:
\begin{equation}
    S^{\rm{accel}}_a=9\times10^{-30}\frac{[\text{m}\,\text{s}^{-2}]^2}{[\rm{Hz}]}\left(1+\left(\frac{[0.4\text{mHz}]}{f}\right)^2\right)\left(1+\left(\frac{f}{[8\,\text{mHz}]}\right)^4\right).
\end{equation}
Note that the above noise is in the form of acceleration. To convert it into a fractional frequency fluctuation, one needs to divide it by a factor $4\pi^2f^2c^2$ \citep{1999ApJ...527..814A}, resulting in
\begin{align}
    &S^{\rm{accel}}_y\\
    &=\frac{3.9\times10^{-44}}{[\text{Hz}]}\left(1+\left(\frac{[0.4\text{mHz}]}{f}\right)^2\right)\left(\left(\frac{[8\,\text{mHz}]}{f}\right)^2+\left(\frac{f}{[8\,\text{mHz}]}\right)^2\right).
\end{align}

The noise of the optical metrology system can be split
\begin{equation}
    S^{\rm{optical}}_y=S^{\rm{ops}}+S^{\rm{opo}},
\end{equation}
where $S^{\rm{ops}}$ is the laser shot noise, which scales with the arm length $L$, the laser power $P$ and the diameter of the telescope $D$ as \citep{2017arXiv170200786A}:
\begin{equation}
    S^{\rm{ops}}=5.3\times10^{-38}\times\left(\frac{f}{[\rm{Hz}]}\right)^2\frac{[2W]}{P}\left(\frac{L}{2.5[\rm{Gm}]}\right)^2\left(\frac{[0.3\rm{m}]}{D}\right)^2\,\rm{Hz}^{-1};
\end{equation}
and 
\begin{equation}
   S^{\rm{opo}}=2.81\times10^{-38}(f/[\rm{Hz}])^2\,\rm{Hz}^{-1}
\end{equation}
denotes the contribution from other noise in the optical meteorology system. 

We also include the TDI-$X$ noise PSD originating from the foreground GW emission from Galactic DWDs (GWD). In practice this confusion noise will be modulated with the orbital phase of the spacecraft. For simplicity, we adopt an analytic approximation for the averaged equal-arm Michelson PSD of GWD as \citep{2019CQGra..36j5011R}:
\begin{equation}
    S_{\rm GWD}(f, T_{\rm{obs}})=Af^{-7/3}e^{-f^\alpha+\beta f\sin(\kappa f)}\left(1+\tanh(\gamma(f_k-f))\right),
\end{equation}
Note that the noise depends on the observation duration, because for longer observations, more and more individual DWDs can be resolved and removed from the confusion noise foreground. This time dependent is represented by using different parameters with different $T_{\rm{obs}}$ as shown in table~\ref{tab:GWD}.
\begin{table}
\begin{center}
        \begin{tabular}{ | c | c | c | c | c | }
                \hline
                & $\le$6 mo & $\le$1 yr & $\le$2 yr & $\le4$ yr \\ [0.5ex]
                \hline \hline
                $\alpha$&0.133&0.171&0.165&0.138\\
                $\beta$&243&292&299&-221 \\
                $\kappa$&482&1020&611&521\\
                $\gamma$&917&1680&1340&1680\\
                $f_{k}$&0.00258&0.00215&0.00173&0.00113\\
                \hline\hline
        \end{tabular}
\end{center}
\caption{The parameters of the confusion noise of the unresolved GWD background. }\label{tab:GWD}
\end{table}
The amplitude $A$ is fixed at $9\times10^{-45}$ for $T_{\rm{obs}}\le4$ years, and is set to zero for larger $T_{\rm{obs}}$ since the foreground then disappears. 

\begin{figure}
    \centering
    \includegraphics[width=0.5\textwidth]{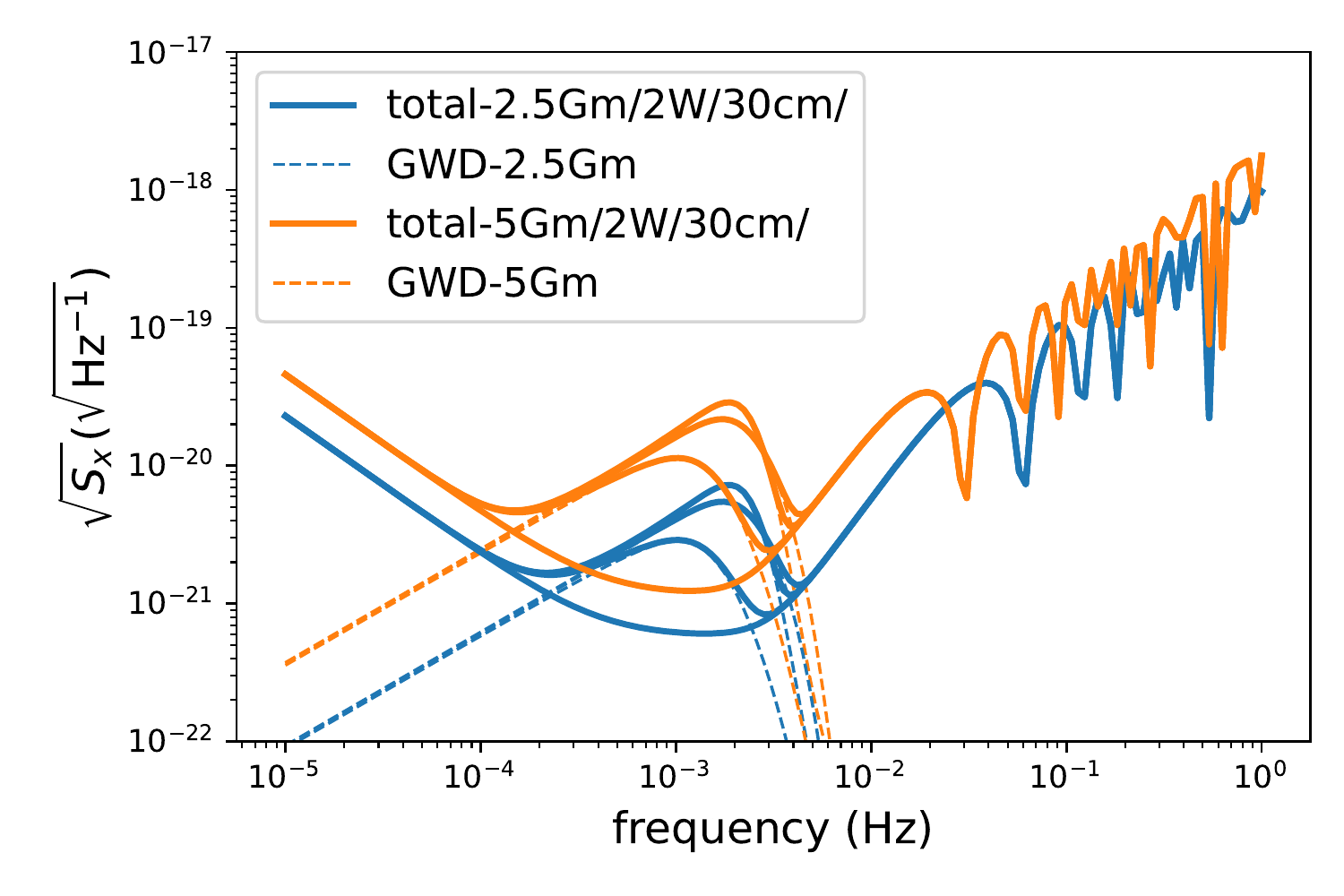}
    \includegraphics[width=0.5\textwidth]{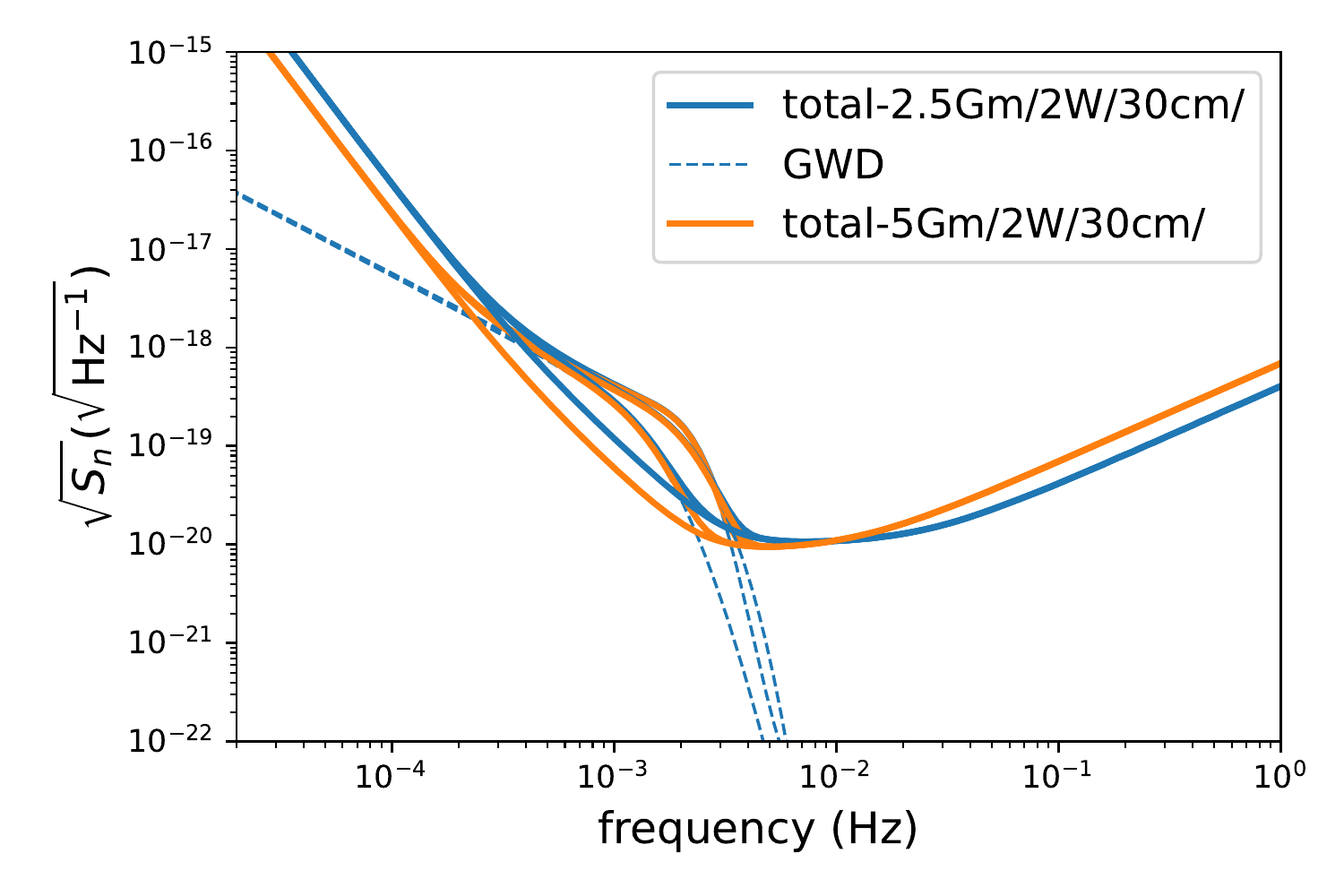}
    \caption{\textbf{Upper Panel:} Square root of the PSD noise TDI-$X$ corresponding to different LISA configurations. The solid curves are the total PSD, while the dashed curves are the contribution from the confusion GWDs. The bundle of curves in the same colour correspond to $T_{\rm{obs}}=1,2,4,5$ years from top to bottom; \textbf{Lower Panel:} Sensitivity curves corresponding to different arm length and $T_{\rm{obs}}$. The solid curves are the total curve, while the dashed curves are the contribution from the confusion GWDs. The bundle of curves in the same colour correspond to $T_{\rm{obs}}=1,2,4,5$ years from top to bottom.}
    \label{fig:lisa_sens}
\end{figure}

The equal-arm Michelson response (fractional displacement) PSD term $S_{\rm GWD}$ can be converted to the fractional frequency by timing a factor $x^2$ and then added to $S^{\rm{ops}}$ calculated with equation (\ref{eqn:noisepsd}). The upper panel of figure \ref{fig:lisa_sens} shows the square root of the noise PSD with various LISA parameters. 
Note that our $S_X$ should not be confused with the PSD in the Michelson response. The latter is more commonly applied and sometimes referred as the sensitivity curve. The \GWT also provide the latter with the following analytic model \citep{2019CQGra..36j5011R}:
\begin{equation}
    S_{\rm{n}}=\frac{10}{3L^2}\left(S^{\rm{op}}_{\rm{dis}}+2(1+\cos^2(x))\frac{S^{\rm{acc}}_{\rm{a}}}{(2\pi f)^4}\right)\left(1+\frac{3}{5}x^2\right)+S_{\rm{GWD}},
\end{equation}
where $S^{\rm{op}}_{\rm{dis}}$ is the noise in the optical system in term of the displacement, which can be converted into the previous Doppler $S^{\rm{optical}}_y$ by multiplication by a factor $2\pi f/c$. We plot the sensitivity curves corresponding to different arm length and $T_{\rm{obs}}$ in figure \ref{fig:lisa_sens}. In the appendix, we give a summary plot of the conversion among different detector responses.

\subsubsection{TDI response to the SMBBH waveform}
\begin{figure}
    \centering
    \includegraphics[width=0.5\textwidth]{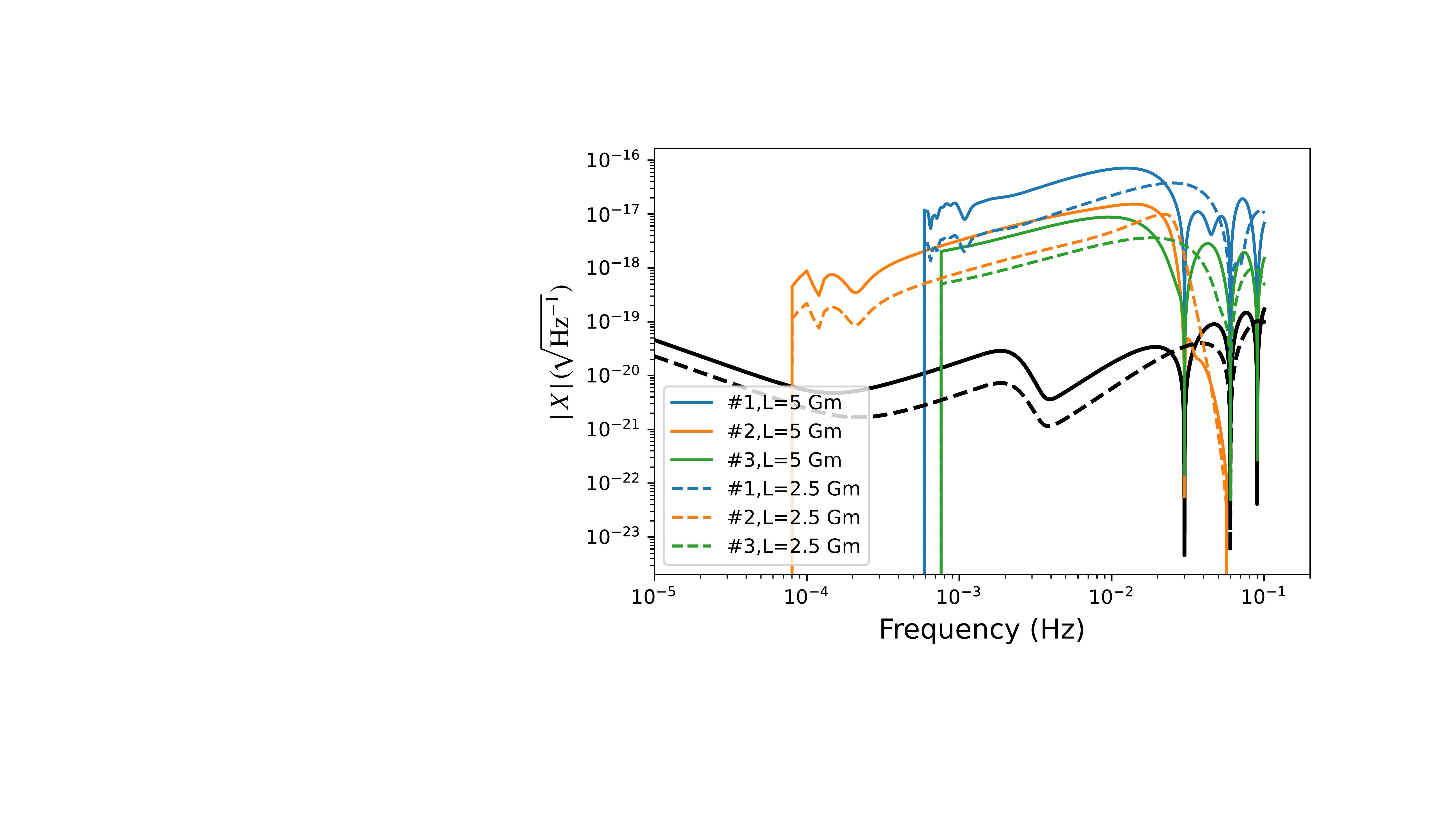}
    \caption{Modulus of frequency domain TDI-$X$ responses to GW from different SMBBHs (whose parameters are listed in Table \ref{tab:para_mbhb}). The solid curves correspond to LISA with 5 Gm laser arms, and dashed curves correspond to a 2.5 Gm arms configuration.}
    \label{fig:mbhbh_x}
\end{figure}

The TDI-$X$ response of LISA due to a GW from a SMBBH merger is calculated using the LDC code \citep{MLDC}, where the \texttt{IMRPhenomD} waveform is adopted \citep{2011PhRvL.106x1101A}. Figure \ref{fig:mbhbh_x} shows the modulus of the TDI-$X$ responses in the frequency domain, for three different sources. The parameters of the example sources are listed in Table \ref{tab:para_mbhb}. The low-frequency limit corresponds to the time to coalescence at the beginning of the observation, and the dips at the high frequency end are due to the term $\sin(x)$ when converting to TDI. The sample cadence is fixed to 5\,s, which corresponds to a high frequency cut-off at $0.1$ Hz. For systems with heavy BHs, like \#2 in the example (masses $> 10^5 M_\odot$), the frequency at coalescence is lower than the cut-off frequency, therefore the current cadence will not lose any power from the signal; On the other hand, for systems with light BHs, like \#1 in the example (masses $<50,000 M_\odot$), the high frequency part ($>0.1$ Hz) of the waveform will be lost. However, the decrease of SNR is less than 1\% comparing to that using a cadence of 1 s. Therefore it is acceptable to fix the cadence to 5 s for all sources, in order to perform the simulation fast.  

\begin{table*}[]
\small
\begin{tabular}{l|lllllllllllllllll}
\# & $\beta$ (rad) & $\lambda$ (rad) & $\theta_{\chi_1}$ (rad) & $\theta_{\chi_2}$ (rad)& $\varphi_{\chi_1}$ (rad)& $\varphi_{\chi_2}$ (rad)& $\chi_1$ & $\chi_2$ & $m_1\,(M_\odot)$ & $m_2\,(M_\odot)$ & $\theta_L$ (rad)& $\varphi_L$ (rad)& $z$   & $t_{\rm c}$ (yr) \\
\hline
1  & -1.3   & 0.44     & 0.8             & 2.6   & 4.5  & 5.98     & 0.8    &  0.2 &  37695 & 4582    & 2.18 & 1.3 & 0.069 & 0.76 \\
2  & -0.44  & 4.7     & 0.08   &    0.037          & 4.26              & 5.48   & 0.04    & 0.2    & 420555      & 298237  & 1.14      & 3.26     & 5.1 & 0.0038 \\
3  & -0.01   & 2.7    & 0.12       & 0.12        & 0.72            & 6.08           & 0.6   & 0.13 & 76476      & 28854 & 1.673     & 1.47       & 2.8 & 0.5
\end{tabular}
\caption{Parameters of example sources corresponding the figure \ref{fig:mbhbh_x}. The meaning of the parameters are: $\beta$-Ecliptic Latitude; $\lambda$-Ecliptic Longitude; $\theta_{{\chi_1}/{\chi_1}}$-Polar angle of spin 1/2; $\chi_{1,2}$-Spin 1/2; $m_{1,2}$-(Intrinsic) masses of primary/secondary BH; $\theta_L$-Initial polar Angle of the orbital angular momentum; $\varphi_L$ Initial azimuthal Angle of the orbital angular momentum; $z$ red-shift; $t_{\rm{c}}$ time to coalescence at the beginning of observation;} \label{tab:para_mbhb}
\end{table*}

\subsubsection{TDI Waveform of DWD}
We first derive the frequency domain TDI-$X$ waveform in a monochromatic plane wave approximation. The equal-arm Michelson response of a plane GW in the long-wavelength region can be approximated as a sine wave with a orbit averaged amplitude $\langle\mathcal{A}\rangle$ and frequency $f_{\rm s}$. The relation between $\langle\mathcal{A}\rangle$ and the intrinsic amplitude $A$ of the source binary can be found in equations (A12,A13) of \cite{2017MNRAS.470.1894K}.

The Fourier transform of such a signal with the duration $T_{\rm obs}$ is:
\begin{equation}
    \Tilde{h}_{\rm Mich}(f)=\frac{1}{2}\langle\mathcal{A}\rangle T_{\rm obs}{\rm sinc}((f-f_{\rm s})T_{\rm obs}).
\end{equation}
Note that here we use the convention that ${\rm sinc}(x)=\sin(\pi x)/(\pi x)$, such that the integration of $|\Tilde{h}(f)|^2$ equals $T_{\rm obs}A^2$.

\begin{figure}
    \centering
    \includegraphics[width=.5\textwidth]{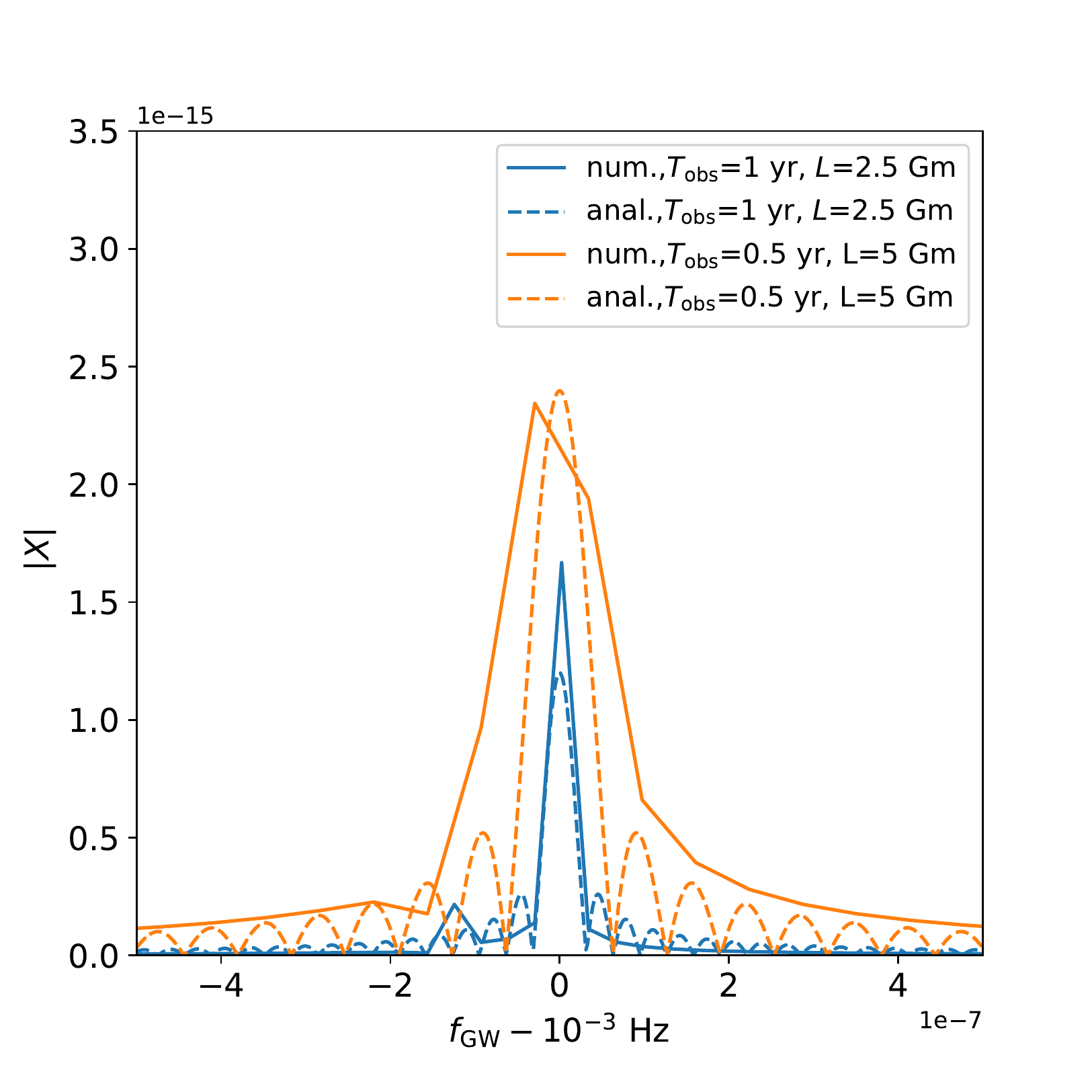}
    \caption{Frequency domain LISA responses to GW from a DWD: The blue and oranges lines correspond to two different LISA configurations; solid and dashed lines correspond to responses calculated with numerical and analytical methods respectively.}
    \label{fig:GBwave}
\end{figure}

To convert the equal-arm Michelson into TDI-$X$, we multiply by a factor $4x\sin x$, where $x=f(2\pi L/c)$.
\begin{equation}
    X(f)=2x\sin x\langle\mathcal{A}\rangle T_{\rm obs}{\rm sinc}\left[(f-f_{\rm s})T_{\rm obs}\right]
    \label{eqn:GBwave_ana}
\end{equation}
In figure~\ref{fig:GBwave} we show the waveforms of a DWD with $A=10^{-20}$, $f_{\rm s}=10^{-3}$\,Hz, calculated analytically from equation (\ref{eqn:GBwave_ana}) and compare them with those calculated numerically with the LDC code (which is based on \citealt{2007PhRvD..76h3006C}). Although simplified, the overall agreement between the two is good.

From equations (\ref{eqn:lisasnr},\ref{eqn:GBwave_ana}), we obtain an approximated squared SNR expression:
\begin{equation}
    \rho^2_{\rm approx.}=\frac{16x^2\sin^2x<\mathcal{A}^2>T_{\rm obs}}{S_X(f_{\rm s})}. 
    \label{eqn:rho_ana}
\end{equation}
When we replace the TDI-$X$ noise PSD with the Michelson noise PSD, and drop the $16x^2\sin^2x$ term, the above equation becomes equation (10) of \cite{2017MNRAS.470.1894K}. 

\subsubsection{TDI Waveform for EMRIs}
\begin{figure}
    \centering
    \includegraphics[width=.45\textwidth]{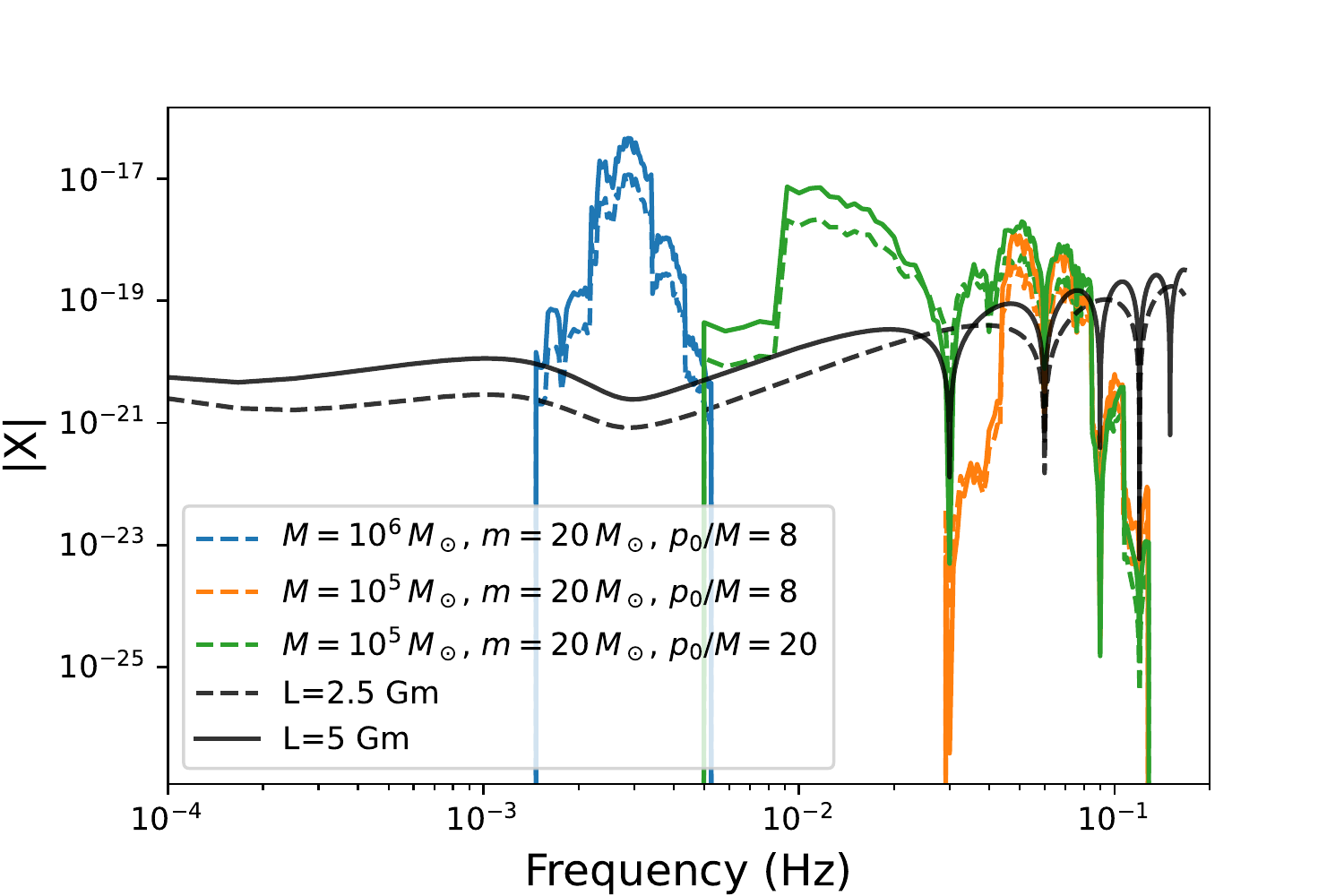}
    \caption{Frequency domain TDI-$X$ responses to EMRI AK waveform, which correspond to three EMRIs systems and two $L$ designs of LISA}
    \label{fig:EMRIs_waveforms}
\end{figure}

The analytic kludge (AK) waveforms \citep{2004PhRvD..69h2005B} for EMRIs are applied and the corresponding TDI-$X$ responses are calculated with the code package \texttt{EMRI\_Kludge\_Suite}\footnote{https://github.com/alvincjk/EMRI\_Kludge\_Suite} \citep{2015CQGra..32w2002C,2017PhRvD..96d4005C}. As examples, in figure \ref{fig:EMRIs_waveforms} we plot the frequency domain TDI-$X$ responses to the AK waveforms, which correspond to three EMRIs systems and two $L$ designs of LISA. The first system (blue curves) has a supermassive BH with mass $M=10^6\,M_\odot$ and stellar mass BH with mass $m=20\,M_\odot$. Here the masses are all measured in the observer's frame, i.e., red-shifted. The frequency domain response corresponds to a time-domain waveform simulated from the semi-latus rectum $p=8GM/c^2$ to the final plunge. The time resolution is $dt=25$\,s, which is set to ensure that the highest frequency cut-off set by $1/(2dt)$ is larger than the Kepler frequency around the innermost stable circular orbit (ISCO) of the supermassive BH. The lower frequency cut corresponds to initial orbital inspiral, and the higher frequency cut corresponds to the orbital frequency at the plunge, which approximates to the Kepler frequency at ISCO; The second system (orange curves) has a supermassive BH with mass $M=10^5\,M_\odot$ and stellar mass BH with mass $m=20\,M_\odot$. The initial semi-latus rectum is also $p=8GM/c^2$. Since the Kepler frequency at ISCO is inversely proportional to the mass of the supermassive BH, we use $dt=2.5$\,s. The third system (green curves) is identical with the second one. The difference is the initial semi-latus rectum of the third one is $p=20GM/c^2$. To track the evolution from this larger initial semi-latus rectum to the final plunge, the simulation includes $\sim10$ times longer time steps. As a result, more low-frequency components are included in the third waveforms than the second. Other physical parameters are identical for the three system and are (values used in parenthesis) \citep{2004PhRvD..69h2005B}:
\begin{itemize}
    \item $s$: dimension-less spin of the massive BH ($s=0.5$);
    \item $e$: the initial eccentricity ($e=0$);
    \item $\iota$: the initial inclination ($\iota=0.524$ rad);
    \item $\gamma$: the pericenter angle in AK Waveform ($\gamma=0$);
    \item $\psi$: the initial phase ($\psi=0.785$ rad);
    \item $\theta_{\rm{S}}$: the sky position polar angle of source in an ecliptic-based coordinate system, equals to $\pi/2$ minus the ecliptic latitude ($\theta_{\rm{S}}=0.785$ rad);
    \item $\phi_{\rm{S}}$: the sky position azimuth angle of source in an ecliptic-based coordinate system, equals to the ecliptic longitude ($\phi_{\rm{S}}=0.785$ rad);
    \item $\theta_{\rm{K}}$: the polar angle of the massive BH spin ($\theta_{\rm{K}}=1.05$ rad);
    \item $\phi_{\rm{K}}$: the azimuth angle of the massive BH spin ($\phi_{\rm{K}}=1.05$ rad);
    \item $\alpha$: the azimuthal direction of the orbital angular momentum ($\alpha=0$);
    \item $D$: luminosity distance ($D=1$\, Gpc).
\end{itemize}

As mentioned above, if the frequency domain waveform were to be simulated in real time for every candidate event, it is difficult to reconcile both the speed of simulation and to include the full GW signal from the beginning of the observation to the plunge. As a solution, we generate the frequency domain TDI waveform corresponding to each candidate event in the catalogue in advance, and store their modulus in files. The pre-generated TDI waveform corresponds to signal from the beginning of the observation to the final plunge. The initial semi-latus is calculated with a Newtonian formula equation 4.136 of \cite{maggiore2008gravitational} according to its masses, eccentricity and time to plunge at the beginning of observation. The pre-calculated TDI corresponds to a LISA arm length 2.5 Gm. The conversion to a different LISA arm length can be done by rescaling with $x_{1,i}\sin x_{1,i}/(x_{0,i}\sin x_{0,i})$, where:
\begin{equation}
    x_{0,i}=2\pi f_i L_{\rm{default}}/c,
\end{equation}
and 
\begin{equation}
    x_{1,i}=2\pi f_i L_{\rm{new}}/c.
\end{equation}
%

\subsection{Pulsar Timing Arrays}
Pulsars are rotating neutron stars. Some of the known pulsars which are very stable, i.e., their spin period only changes a tiny fraction in a very long epoch. Therefore, the arrival time of each pulse from such a pulsar can be modeled with high precision. The passing of a series of GW will cause additional changes to the expected time of arrival of the pulses (TOAs), and thus provide a way to detect GW, at frequencies between $10^{-8}$ and $10^{-5}$ Hz, where the lower frequency limit corresponds to the observation span of years, and the high frequency limit corresponds to the average cadence of a couple of days \citep[see][]{1978SvA....22...36S,1979ApJ...234.1100D,1983ApJ...265L..39H,2005ApJ...625L.123J}. A pulsar timing array is a group of pulsars, which are stable and have been monitored with radio telescopes for a long time. The existing PTA consortia are EPTA \citep{2013CQGra..30v4009K}, PPTA \citep{2013CQGra..30v4007H}, NANOGrav \citep{2013CQGra..30v4008M} and IPTA \citep{2010CQGra..27h4013H}. The standard procedure of pulsar timing is to first fit a timing model to the TOAs of individual pulsars, that includes inaccuracies in the pulsars' astrometric parameters, a model for spin evolution, refractive effects of interstellar medium and solar wind, the orbital and spin motion of the Earth, delays due to general relativity {\it etc.} \citep[see][]{2006MNRAS.369..655H}. The difference between the timing model and the observed TOAs, the timing residuals, are used to extract information about possible GW signals with frequentist \citep{2005ApJ...625L.123J,2012PhRvD..85d4034B,2012ApJ...756..175E} or Bayesian methods \citep{2009MNRAS.395.1005V,2013CQGra..30v4004E}. Here we want to use a simplified way to represent the properties of PTAs, without the need to make use of the full time series of the timing residuals, and obtain results which agree to an order of magnitude with the published results. We base our method on measuring the excess power from GW over analytic timing noise power spectra. Such a practice was also used by some early work \citep{2004ApJ...606..799J,2010MNRAS.407..669Y,2014MNRAS.445.1245Y}. 

\subsubsection{Representing the timing noises}
Suppose that we have already removed every known effect that contributed to differences in the TOAs, the timing residuals that remain are purely intrinsic to the pulsars due to their spin irregularity. Previous studies found that such timing noise can be decomposed into a red noise component and a white noise component \citep{2010MNRAS.402.1027H}. The red noise component can be represented with a power-law spectrum, with increasing power towards the lower frequencies, while the white noise component has a frequency independent power level. In the \GWT, we use the following equation to represent the noise spectrum density of the timing residuals of an individual pulsar:
\begin{equation}
    S_{total}(f)=\sigma^2_{\rm w}/(f_{\rm high}-f_{\rm low})+S_{\rm n,red}(f),
\end{equation}
where $\sigma_{\rm w}$ is the level of the white noise, $f_{\rm high}=N/(2T)$ is the high frequency cut-off defined by the observation cadence and $f_{\rm low}=1/T$ is the low frequency cut-off defined by the inverse of the duration, $S_{\rm n,red}(f)$ is the red noise component, which has a power-law form:
\begin{equation}
    S_{\rm n,red}(f)=\frac{A^2_{\rm{red}}}{12\pi^2}\left(\frac{f}{\text{yr}^{-1}}\right)^{-\alpha}. 
\end{equation}
Therefore, we define the noise spectrum of a pulsar with five parameters, namely: $N$ the number of observations, $T$ the duration of observation, $A_{\rm red}$ the normalization of the red noise, $\alpha$ the power index of the red noise, and $\sigma_{\rm w}$ the level of the white noise. The last three are intrinsic properties of the pulsar. These parameters for the pulsars in the above mentioned PTAs are fitted and published \citep{2016MNRAS.458.3341D, 2019MNRAS.483.4100P, 2020arXiv200506495A}. The \GWT includes 42 pulsars in EPTA, 26 pulsars in PPTA, 47 pulsars in NANOGrav and 87 pulsars in IPTA. Besides the pulsars in the current PTAs, the \GWT also includes simulated future observations, with customised observation cadence and duration, and an increasing number of newly discovered pulsars during the observation period. The parameters (sky coordinates RA, DEC and noise parameters $A_{\rm{red}}$, $\alpha$, $\sigma_{\rm{w}}$) of the simulated pulsars are assigned in the following way: randomly select two pulsars from the current PTA with replacement, and draw a uniformly random number between the parameters of the selected pair of pulsars, and assign the random variable as the corresponding parameter of the new pulsar. In this method, the noise properties and sky distribution of the new pulsars reflect those of the known pulsars.  

\begin{figure}
    \centering
    \includegraphics[width=0.45\textwidth]{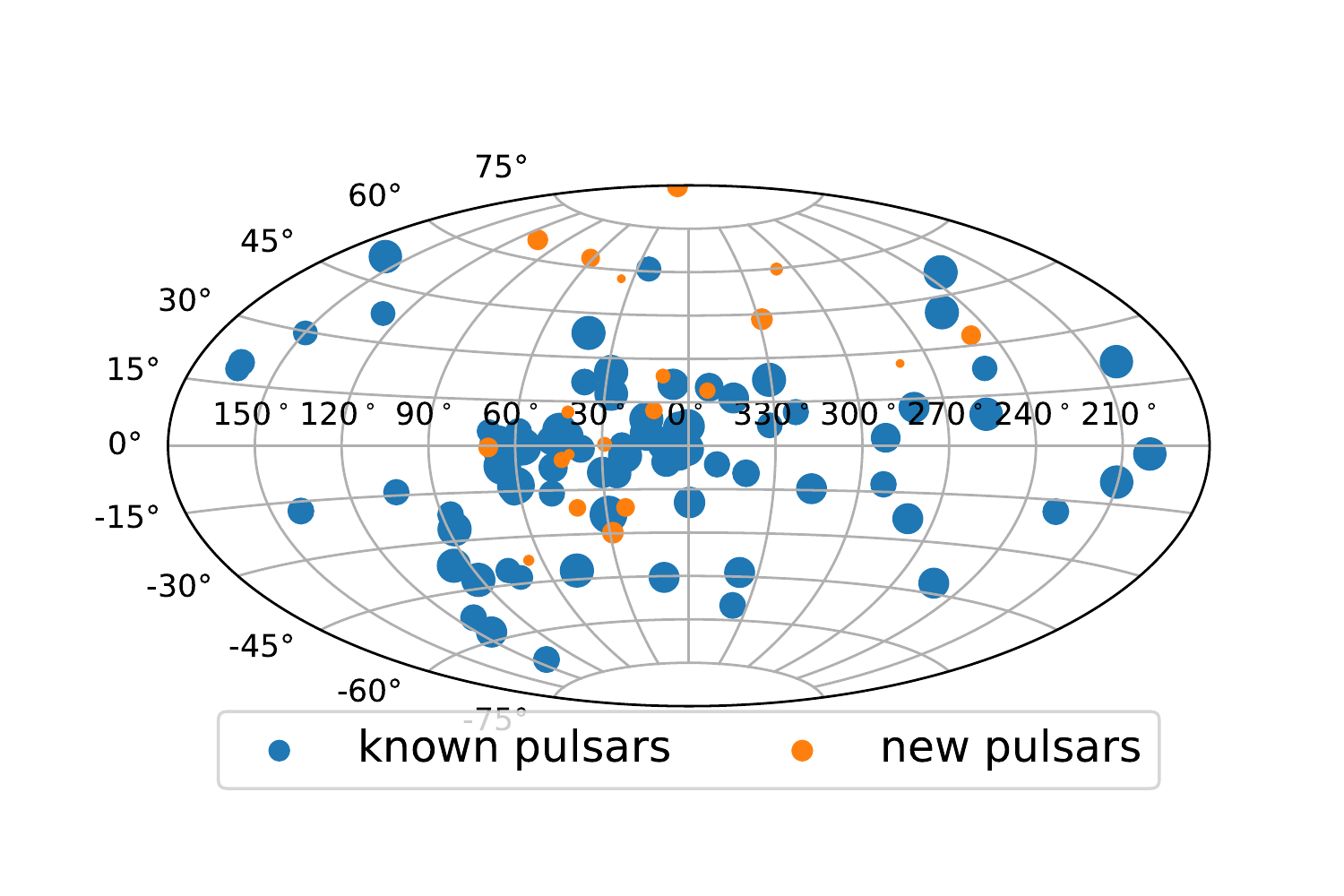}
    \caption{Galactic distribution of the PTAs. The blue dots indicate the pulsars from current IPTA, and orange dots indicate simulated new pulsars. The size of the markers is proportional to the number of TOAs of the corresponding pulsar.}
    \label{fig:pta_sky}
\end{figure}

\begin{figure}
    \centering
    \includegraphics[width=0.4\textwidth]{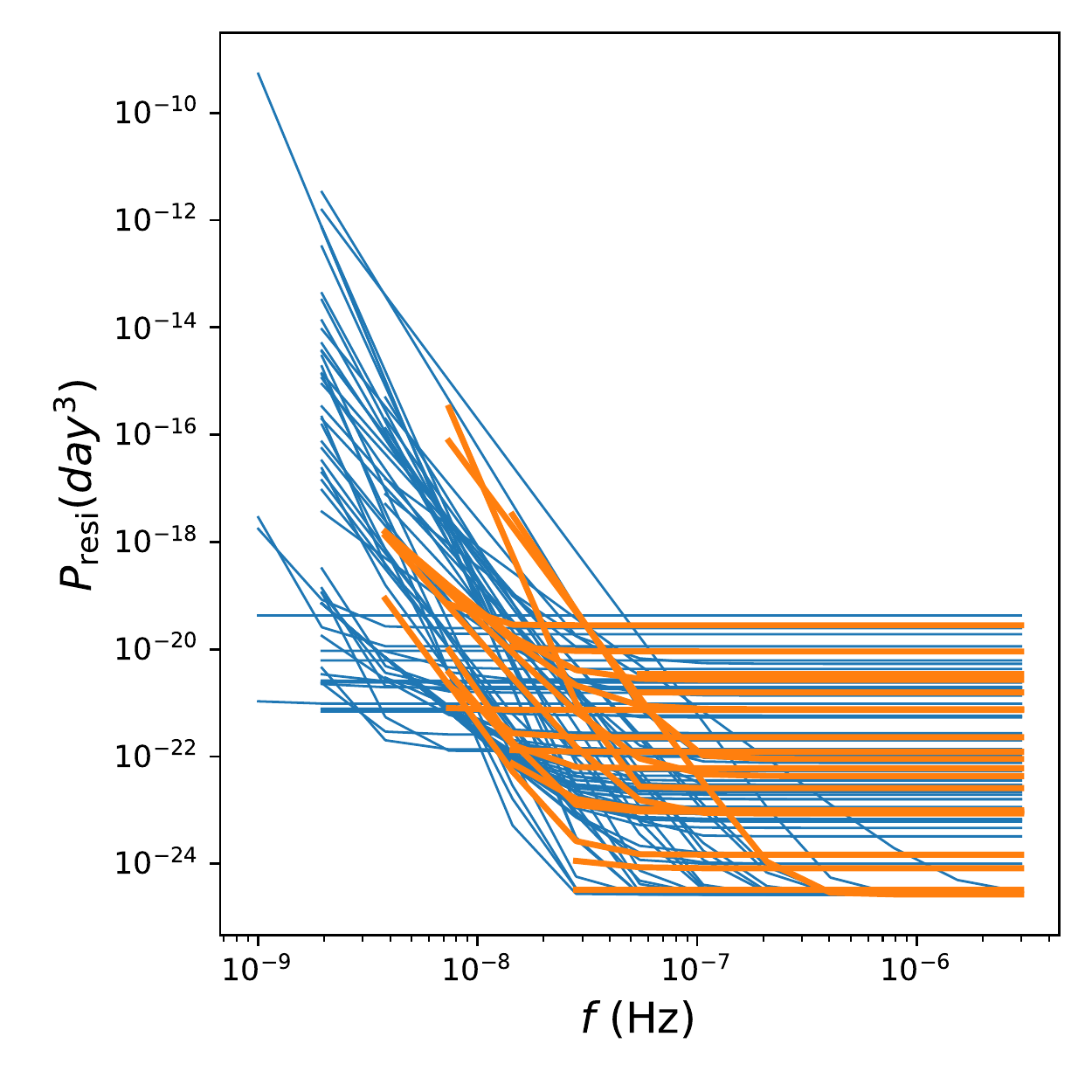}
    \caption{The noise spectra of the pulsars. Blue curves correspond to known pulsars in existing PTAs, and orange curves are simulated new pulsars.}
    \label{fig:presi}
\end{figure}

In figure {\ref{fig:presi}}, we plot the noise spectra density of pulsars in the PTAs used by the \GWT. Blue curves correspond to known pulsars in existing PTAs, and orange curves are simulated new pulsars. In figure {\ref{fig:pta_sky}}, we plot the sky coordinates of pulsars in the PTA. The blue dots indicate the pulsars from IPTA, and orange dots indicate simulated new pulsars. The size of the markers is proportional to the number of TOAs of the corresponding pulsar. 
\begin{figure}
    \centering
    \begin{minipage}{0.45\textwidth}
    \includegraphics[width=\textwidth]{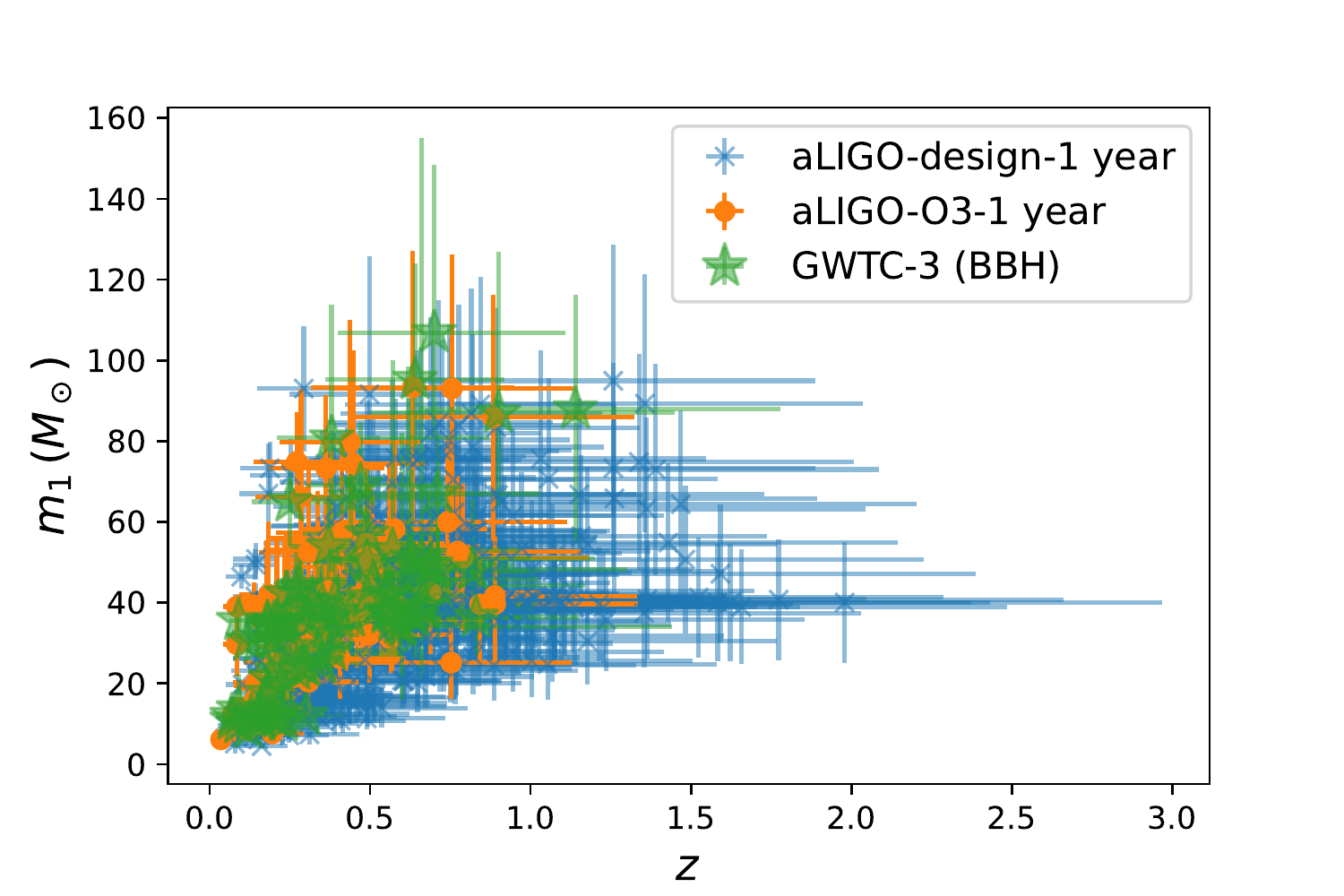}
    \end{minipage}
    \begin{minipage}{0.45\textwidth}
    \includegraphics[width=\textwidth]{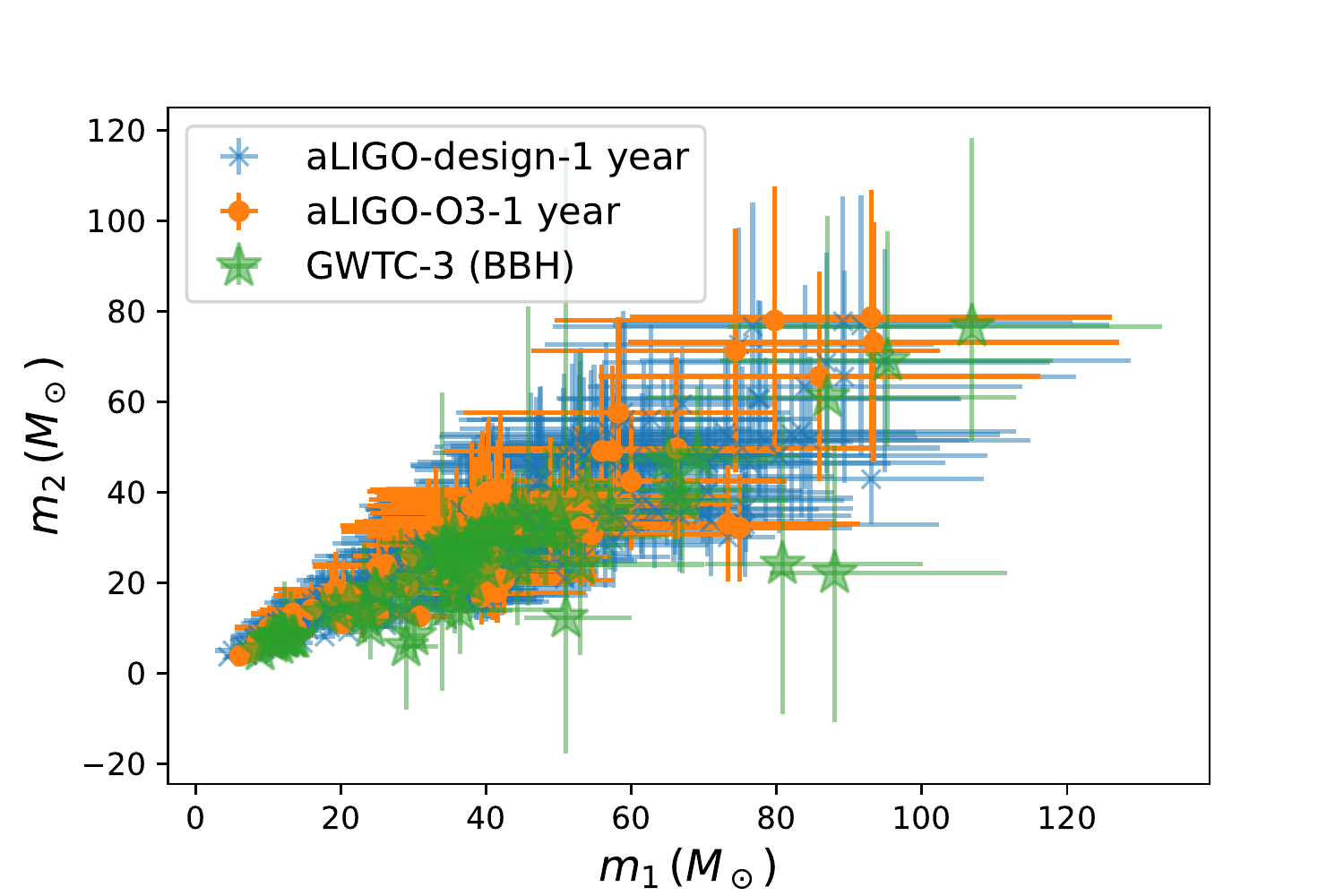}
    \end{minipage}
\caption{\small{\textbf{aLIGO-design-1 year:} The simulated catalogue from one year observation by aLIGO with designed noise performance on BBH, marked with blue crosses; \textbf{aLIGO-O3-1 year:} The simulated catalogue from one year observation by aLIGO with O3 noise performance on BBH, marked with orange dots; \textbf{GWTC-3 (BBH)}: BBH events in GWTC-3, marked with green star symbols.}}\label{fig:LIGO-GWTC}
\end{figure}

\subsection{PTA detections}
The GW from an individual supermassive BH can be approximated with a monochromatic wave. The timing residuals induced in the $i$-th pulsar is \citep{2014MNRAS.445.1245Y}:
\begin{equation}
    \mathcal{A}_i=\frac{h_{\rm{s}}}{\omega}\left(1+\cos\theta\right)\sqrt{\cos^22\psi\left(\frac{1+\cos^2\iota}{2}\right)^2+\sin^22\psi\cos^2\iota}
\end{equation}
where $\theta$, $\iota$, $\Psi$ are the angle between the direction towards the pulsar and the GW source, the inclination of the source binary plan and the polarization angle respectively. The SNR of the GW in the $i$-th pulsar is:
\begin{equation}
    \rho^2_i=\mathcal{A}_{i}^2/S_{n,i}(f)\times T_i, 
\end{equation}
where $S_{n,i}(f)$ is the noise spectrum density of the pulsar, and $T_i$ is the observation duration. The total SNR squared of a PTA is:
\begin{equation}
    \rho^2=\sum\rho_i^2.
\end{equation}

The effects of a stochastic GW background (SGWB) to the timing residual are an additional red noise:
\begin{equation}
    h_c^2(f)=Cf^\gamma,
\end{equation}
where $h_{\rm c}$ is the characteristic GW strain at the frequency 1 yr$^{-1}$, and the index $\gamma$ depends on the origin of the SGWB. 
For incoherent overlapping of MBH, $\gamma=-2/3$; for relic GW, $\gamma=-1$ and for cosmic strings, $\gamma=-7/6$. Besides the additional red noise, it is also expected that the timing residuals due to the SGWB are correlated between pairs of pulsars. The correlation as a function of the angular separation between the pair is:
\begin{equation}
        \Gamma_0=3\Big\{\frac{1}{3}+\frac{1-\cos\xi}{2}[\ln(\frac{1-\cos\xi}{2}-\frac{1}{6}]\Big\},
\end{equation}
which is referred as the Hellings and Downs Curve \citep{1983ApJ...265L..39H}. The SNR squared in a pair of pulsars is \citep{2009PhRvD..79h4030A}:
\begin{equation}
     \rho_{1,2}=\frac{H_0^2}{4\pi^2}\sqrt{2T\int^\infty_0df\frac{\Omega^2_{\rm{gw}}(f)\Gamma^2_0}{f^6P_1(f)P_2(f)}},
     \label{eqn:SNRpair}
\end{equation}
where $H_0$ is the Hubble constant, $\Omega_{\rm gw}(f)$ is the energy density of the SGWB relative to the critical density of the Universe. The relation between $h_c(f)$ and $\Omega_{\rm gw}(f)$ is:
\begin{equation}
    h^2_c(f)=\frac{3H_0^2}{2\pi^2}\frac{1}{f^2}\Omega_{\rm{gw}}(f).
\end{equation}
In equation (\ref{eqn:SNRpair}) $P_{1,2}(f)=S_{\rm n 1,2}(f)f^2$. 
The total SNR squared is the summation of SNR squared over all pairs in the PTA. 
\newpage

\section{Results and examples}\label{Results}

In order to test and validate the \GWT, we discuss the outcome of the simulations for the different source populations and different detectors, and compare these with the literature where possible. 

\subsection{Examples for stellar mass BBH detected with Earth-based detectors}\label{sec:BBH_example}
For BBHs, we generate catalogues for a one year observation of aLIGO, both with the noise spectrum of O3 (aLIGO-O3) and that of the design (aLIGO-design). The underlying population model is BBH-PopB with the default parameters (delay time of 3\,Gyr and a mass function with an extra Gaussian peak at 40\msun, see section~\ref{sec:BBH_model}). The simulated number of detection of aLIGO-O3 is 99 (with a Poisson expectation 87.4). It is compatible with the rate of detection in O3a ($\sim36$ BBH in six month of observation with duty cycle $\sim80\%$, see \cite{2020arXiv201014527A}); the simulated number of detection for the aLIGO-design is 379 (expectation 376.0). In the panels of figure \ref{fig:LIGO-GWTC}, we plot the simulated catalogues in the $z-m_1$ and $m_1-m_2$ planes. To compare, we also plot the BBH events in GWTC-3\citep{GWTC-3}, which is comprised by GWTC-1 \citep{2019PhRvX...9c1040A} and events detected in O3a+b. The simulated sets agree well with the observed ones, except that the uncertainties of parameters are in general larger in our simulation comparing with those in the real observation in GWTC-3 (up to $\sim50\%$), especially in the low SNR region. The origin of this overestimation is double folded: 1, the intrinsic difference between FIM and Bayesian methods, as mentioned above; 2, the localisation information from triangulation is not included in our method as were in real catalogue. Therefore, the estimated uncertainties can serve as a conservative upper limits. 

\begin{figure}
    \centering
    \includegraphics[width=0.45\textwidth]{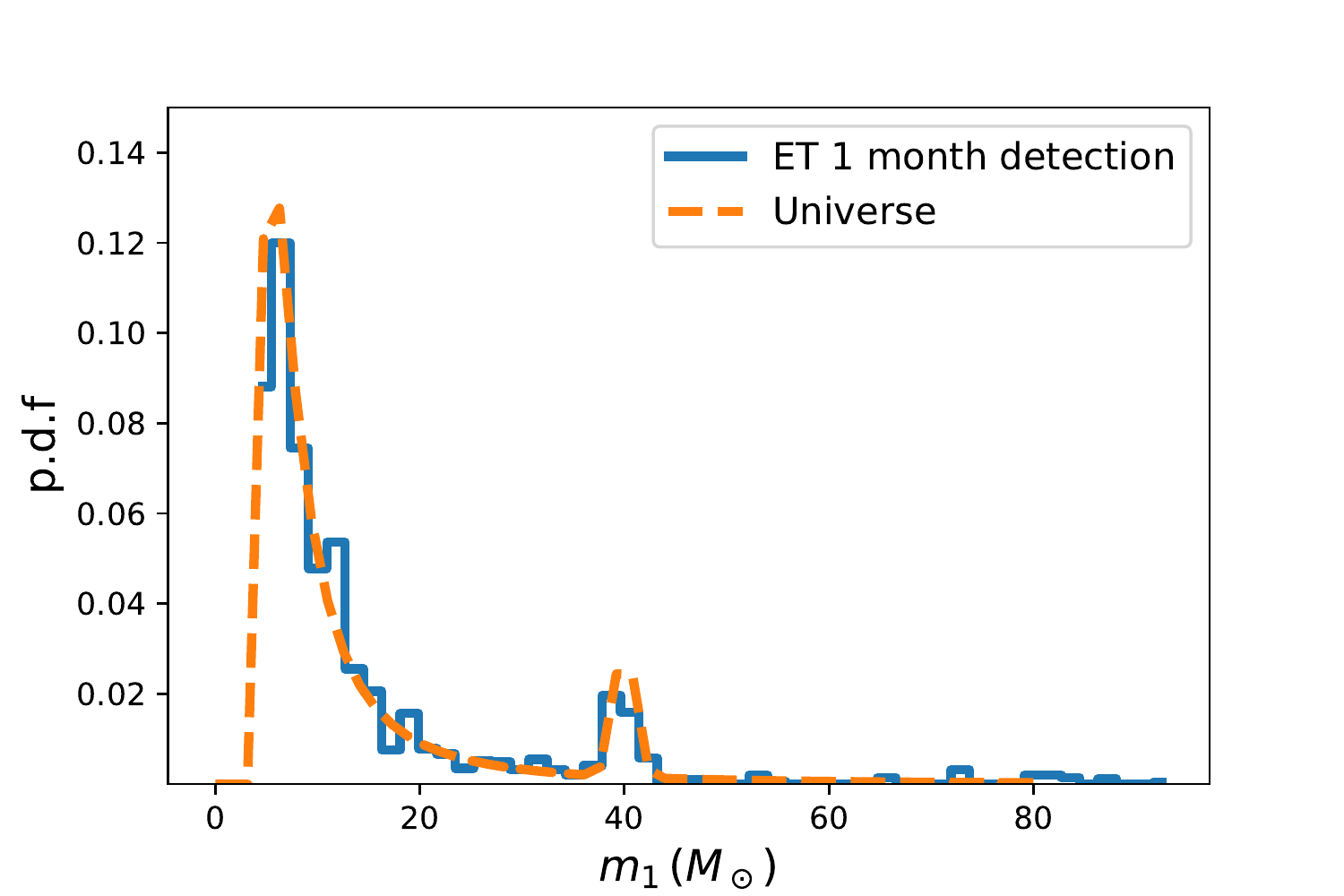}
    \includegraphics[width=0.45\textwidth]{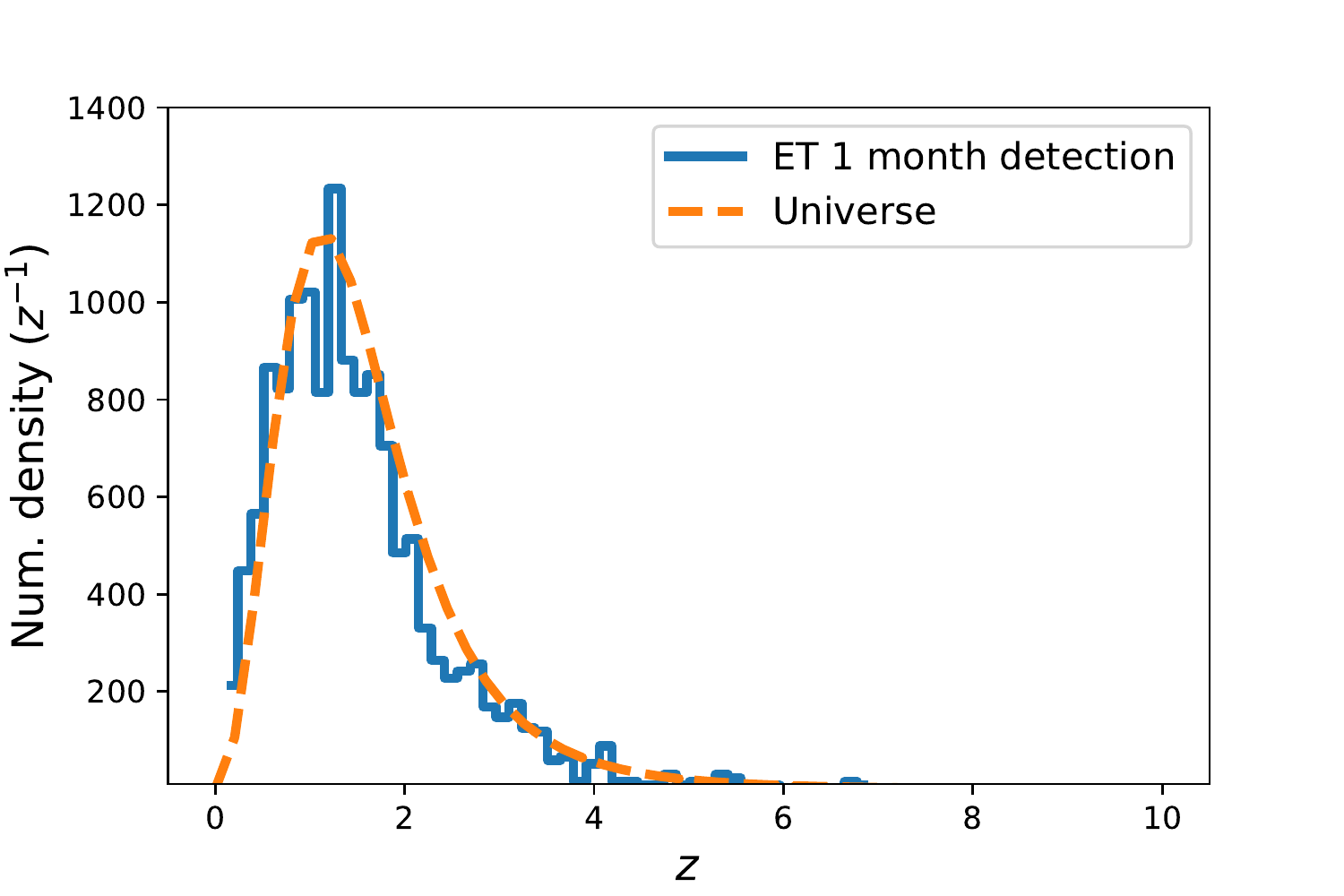}
    \caption{$m_1$ distribution (upper panel) and the number density as function of redshift (lower panel) in the catalogue for one month of observation of BBH mergers by ET (solid blue). The integral of the latter is the total number of detection. As a comparison, we plot that of the whole Universe within one month in dashed orange.}
    \label{fig:ET-BBH}
\end{figure}

We also simulate a catalogue of one month observation with ET. In this case, the number of detection is 1923 (expectation $1.9\times10^3$). In the panels of figure \ref{fig:ET-BBH}, we plot the distribution of the catalogue in $z$ and $m_1$. We also plot the distribution of BBH mergers in the whole Universe as defined by the population model for comparison. From this example, it is clear that ET will probe the distribution of sources throughout the Universe very well, as was shown before \citep[e.g.][]{2017PhRvD..95f4052V}. We will study this in more detail in a forthcoming paper (Yi et al. in preparation)
\begin{figure}
    \centering
    \begin{minipage}{0.45\textwidth}
    \includegraphics[width=\textwidth]{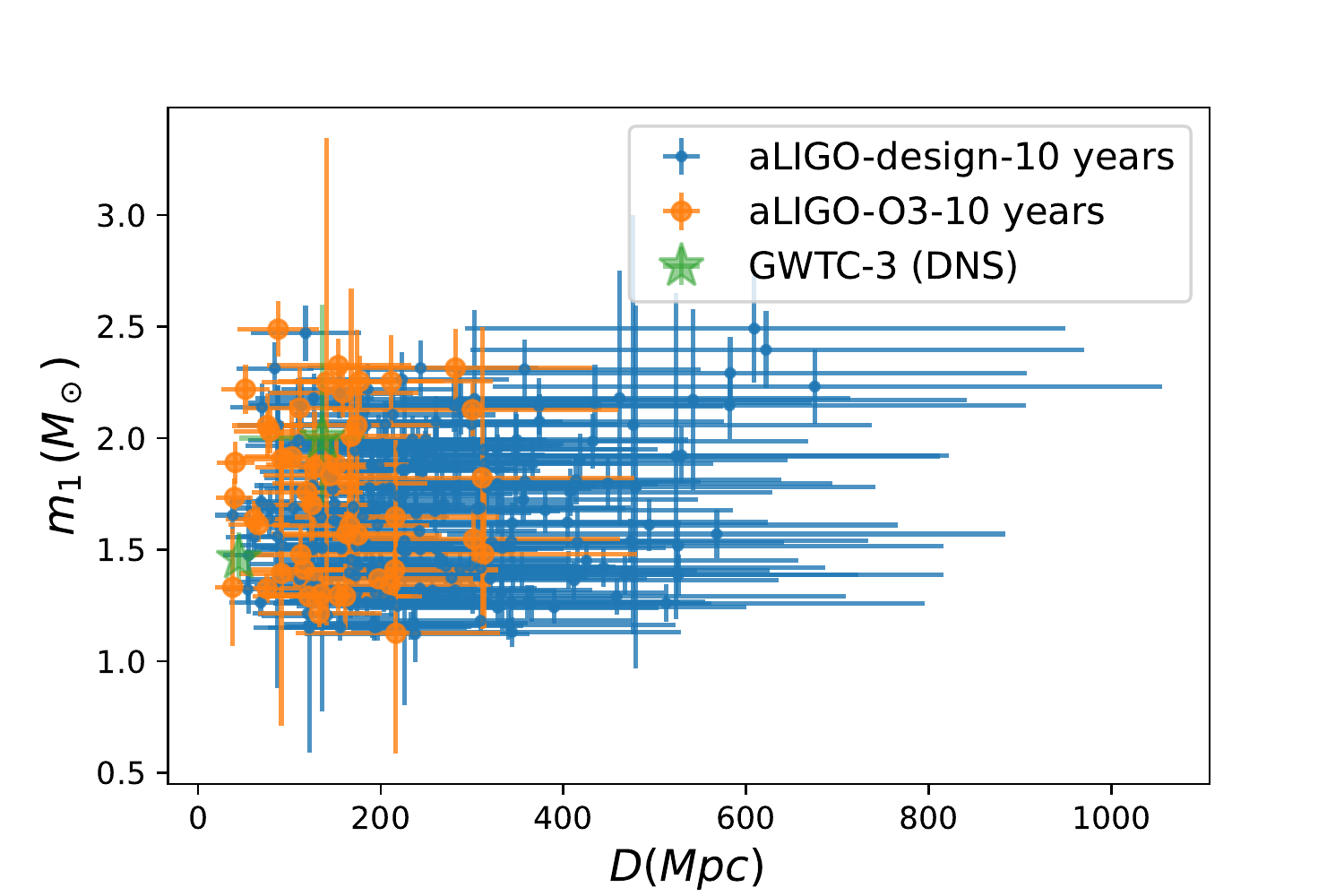}
    \end{minipage}
    \begin{minipage}{0.45\textwidth}
    \includegraphics[width=\textwidth]{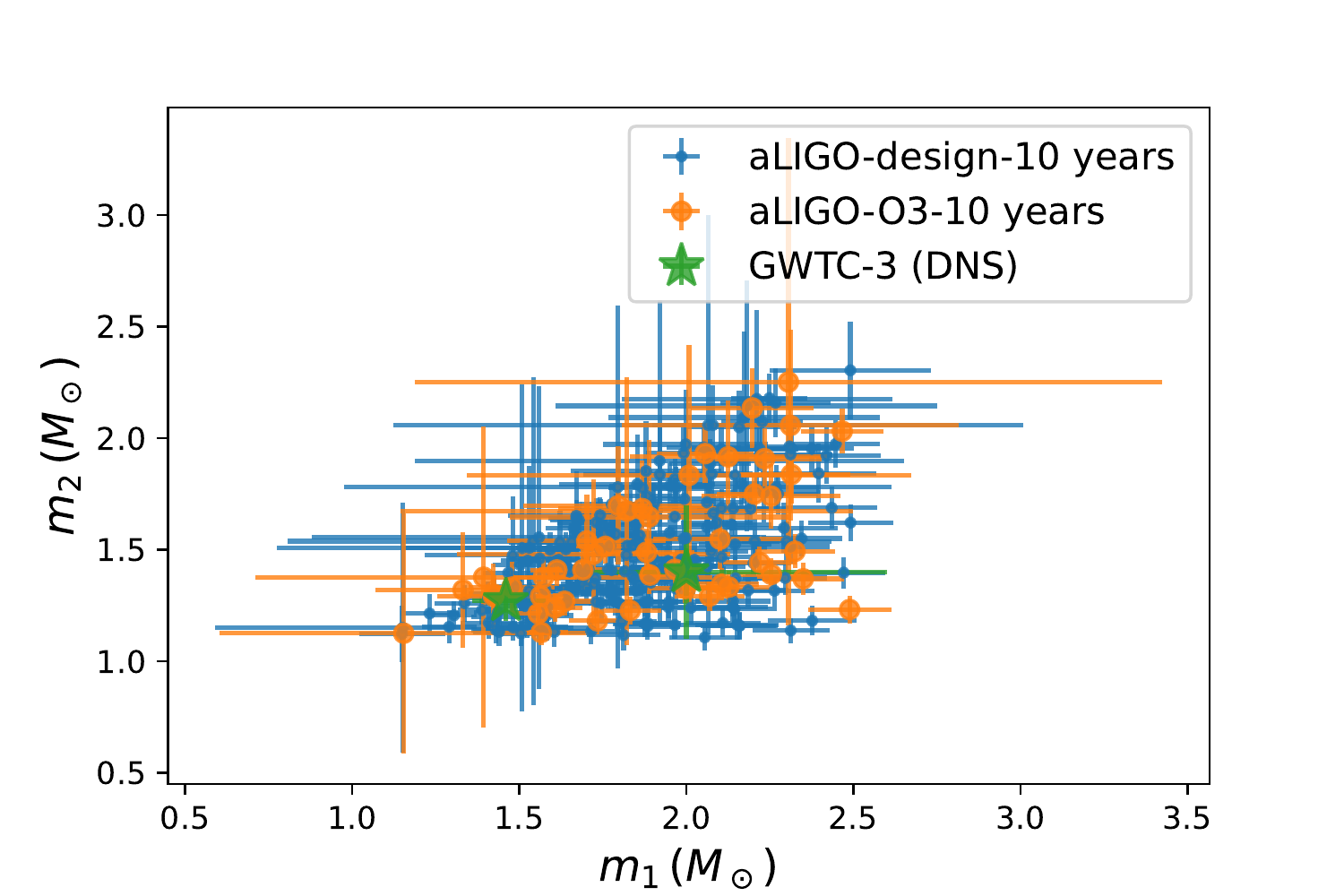}
    \end{minipage}
\caption{The same as figure \ref{fig:LIGO-GWTC}, but for ten years observation of DNS with eLIGO (design and O3 sensitivity).}\label{fig:LIGO-GWTC-NSNS}
\end{figure}


\begin{figure}
    \centering
    \includegraphics[width=0.45\textwidth]{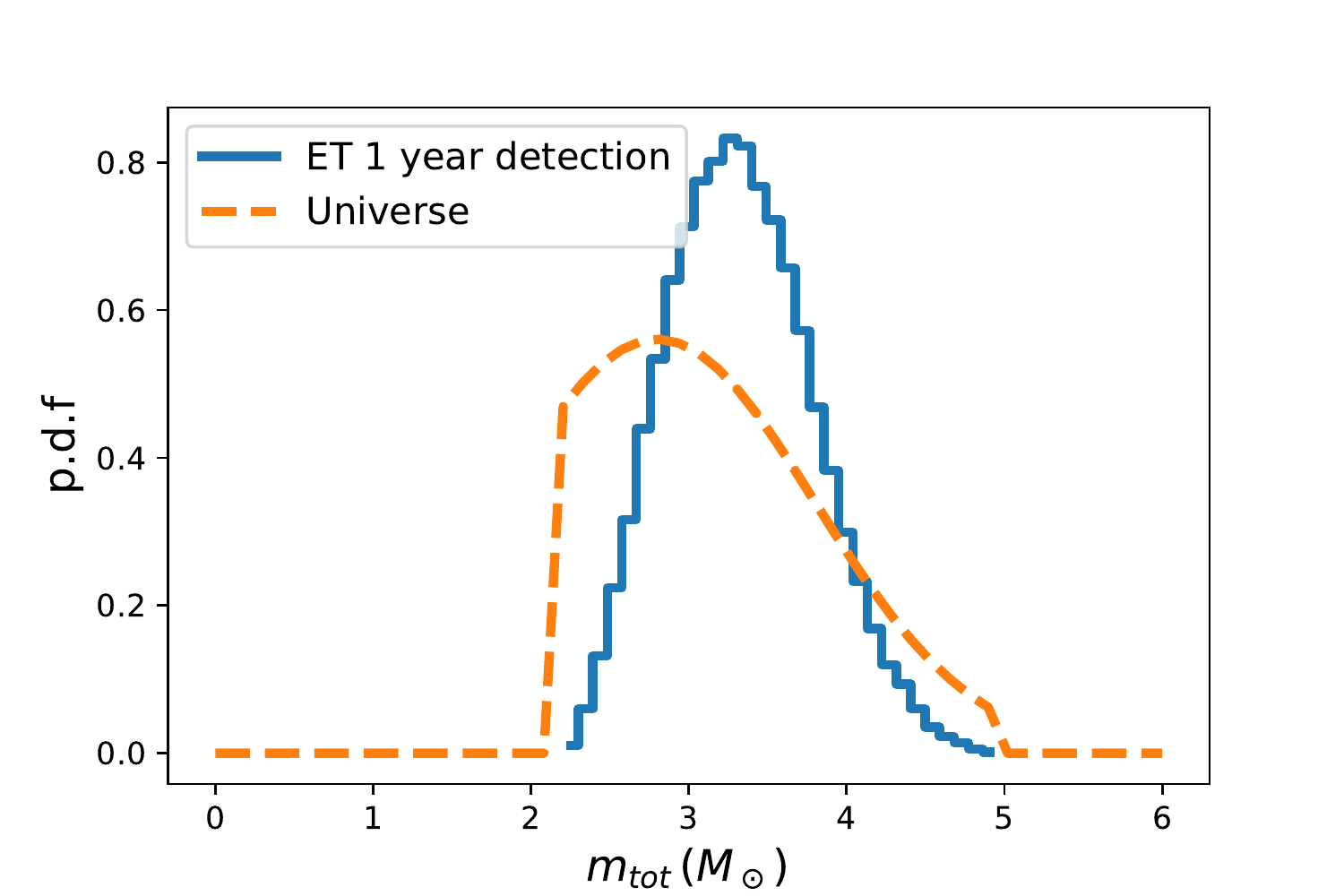}
    \includegraphics[width=0.45\textwidth]{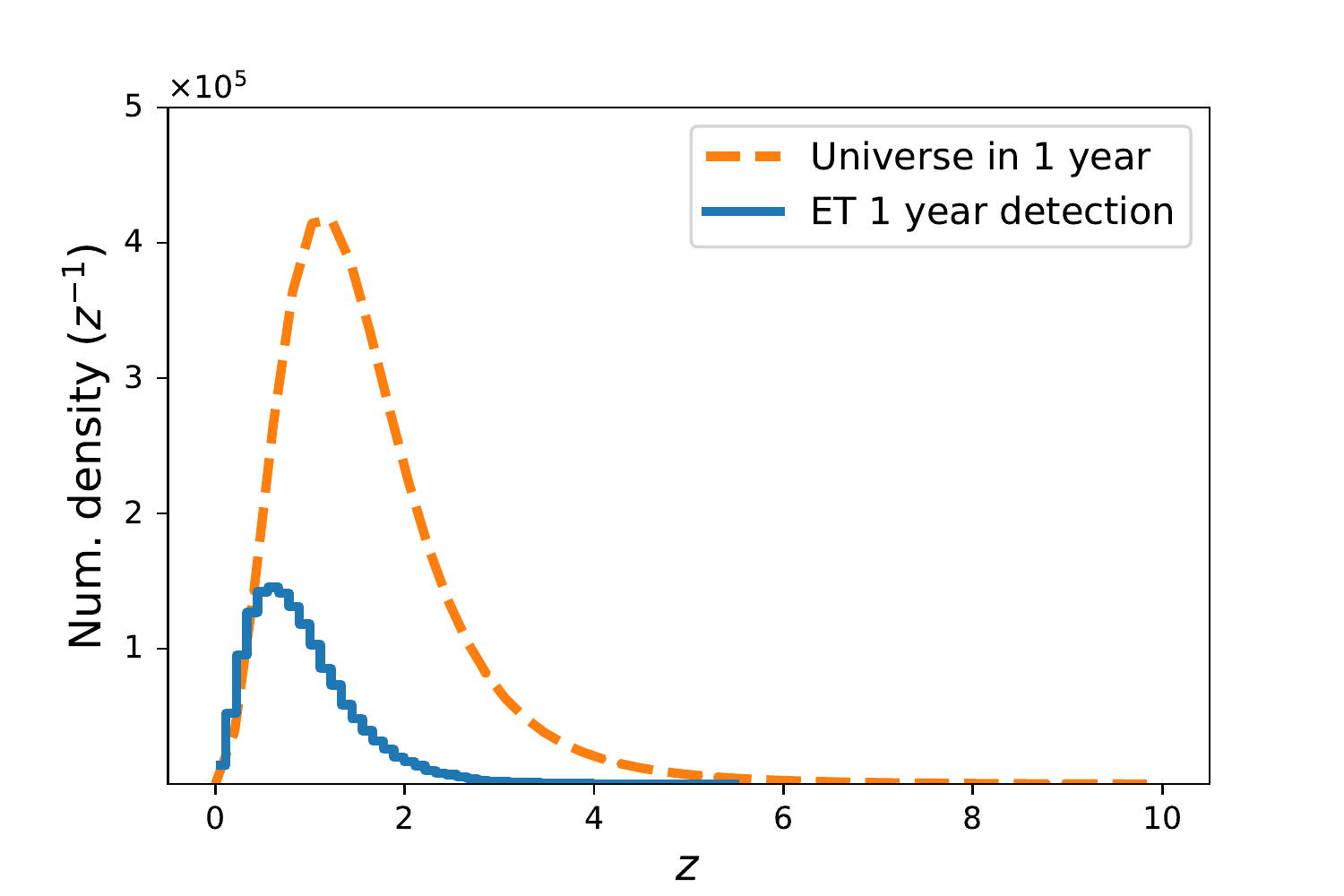}
    \caption{The total mass distribution (upper panel) and the number density as function of red-shift (right panel) in the catalogue of one year observation of DNS mergers by ET (solid blue). The integral of the latter is the total number of detection. The dashed orange curve is that in the whole universe within one year.}\label{fig:ET-NS}
\end{figure}
\begin{figure}
    \centering
    \begin{minipage}{0.45\textwidth}
    \includegraphics[width=\textwidth]{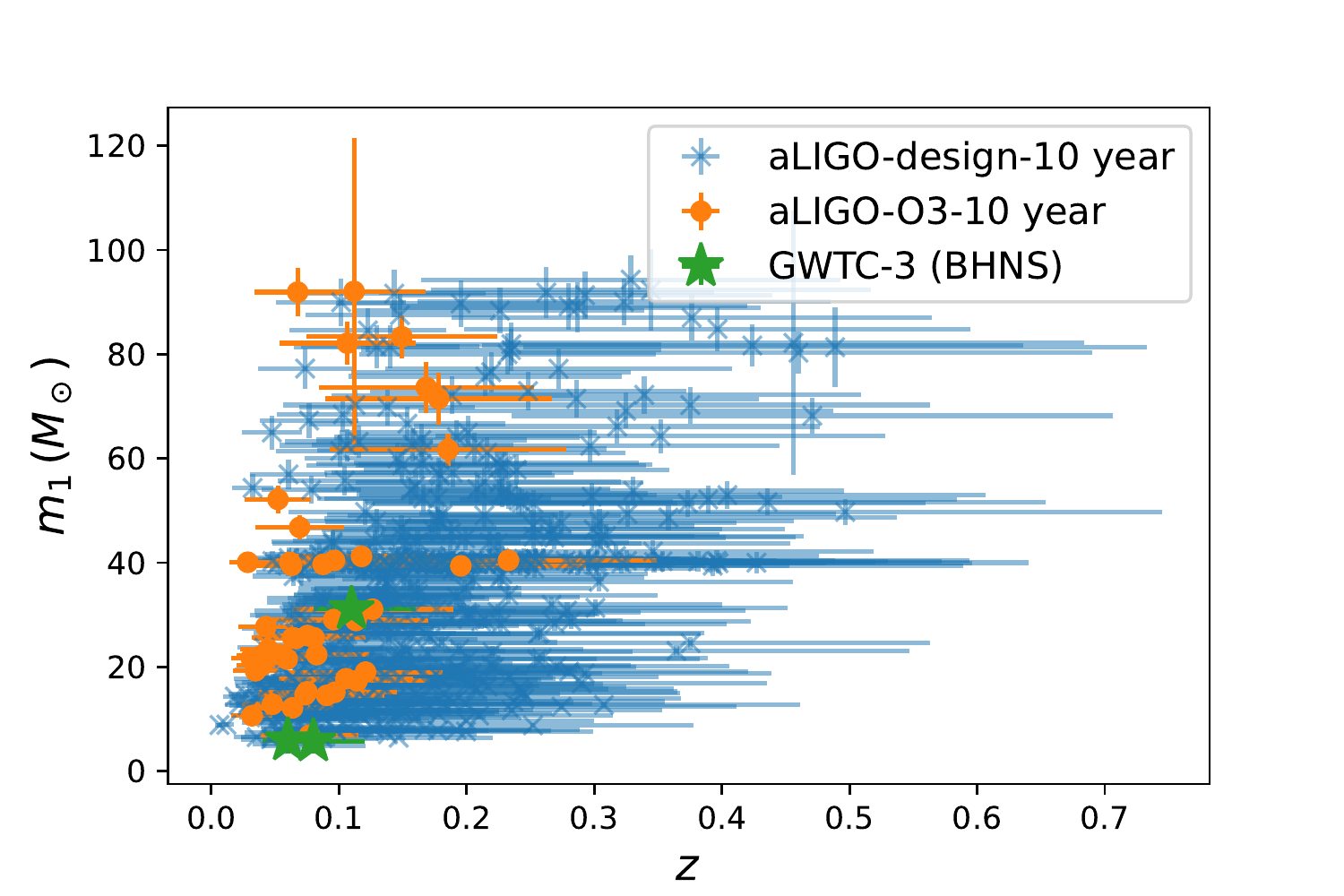}
    \end{minipage}
    \begin{minipage}{0.45\textwidth}
    \includegraphics[width=\textwidth]{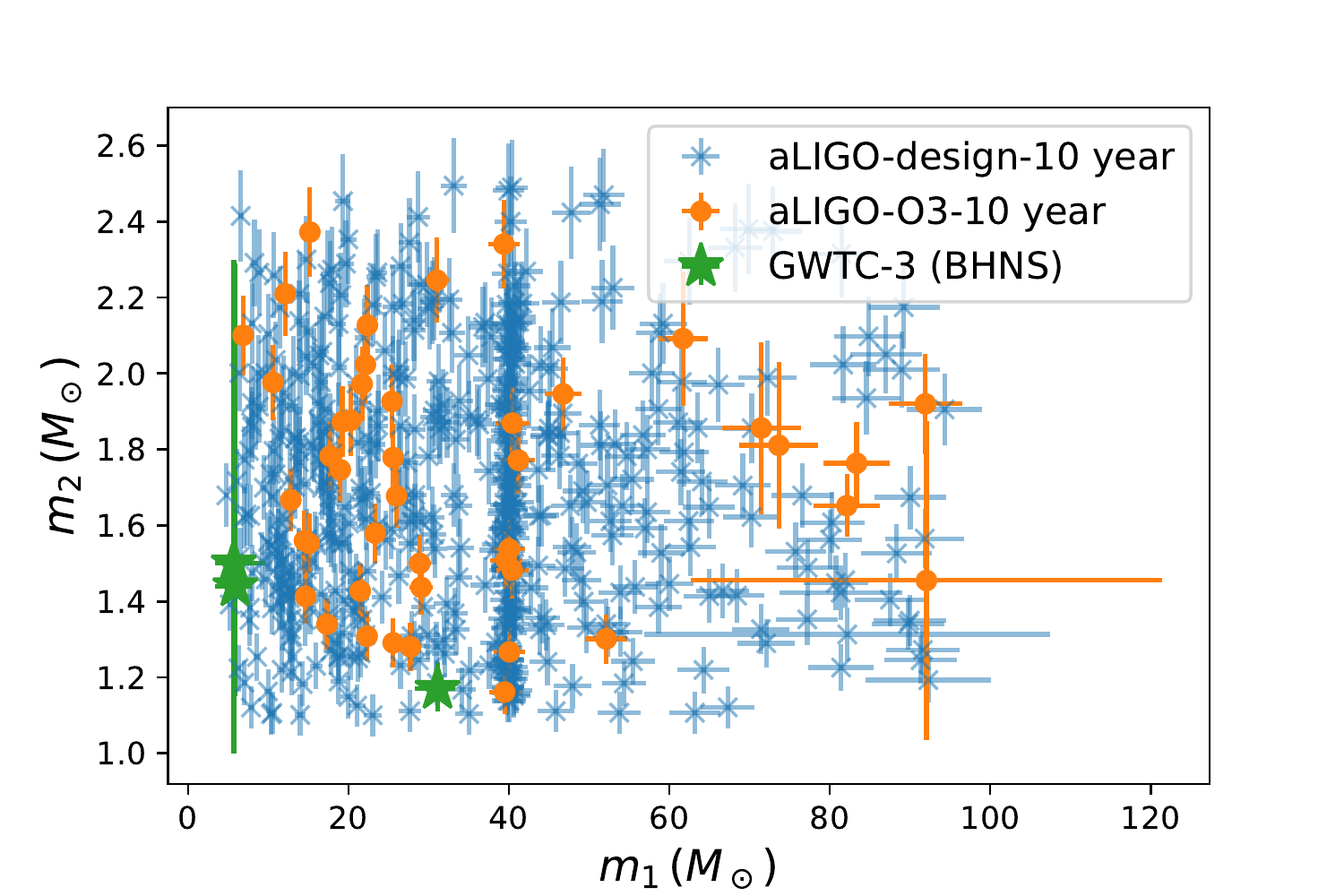}
    \end{minipage}
\caption{\textbf{aLIGO-design-10 year:} The simulated catalogue from ten years' observation by aLIGO with designed noise performance on BHNS, marked with blue crosses; \textbf{aLIGO-O3-10 year:} The simulated catalogue from ten year observation by aLIGO with O3 noise performance on BHNS, marked with orange dots.}\label{fig:LIGO-GWTC-BHNS}
\end{figure}

\subsection{Examples for double neutron stars detected with Earth-based detectors}
The number of DNS mergers detected so far is small (2-3). Therefore, we generate catalogues for ten years of detection of DNS mergers by aLIGO, both with the noise spectrum in O3 (aLIGO-O3) and that at design (aLIGO-design). The number of detection of aLIGO-O3 is 56 (expectation 48.7) in ten years, which is in accordance with the real detection rate in O3a (1-2 in six month); the number for aLIGO-design is 238 (expectation 236.3). In the panels of figure \ref{fig:LIGO-GWTC-NSNS}, we plot the simulated catalogues in the $z-m_1$ and $m_1-m_2$ planes. We also plot the DNS events in GWTC-3 (GW170817, GW190425) to compare. It is difficult to draw strong conclusions with so few detection, but broadly the results of the \GWT agree with the observations so far. 
To look further in the future, we also simulate the catalogue of one year observation by ET. The expected number of detection is 168455 (expectation $1.68\times10^5$). In figure \ref{fig:ET-NS}, we plot the distribution of the catalogue in redshifts and the total masses. We plot together the distribution of DNS mergers in the whole Universe as defined by the population model for comparison. As we can see from the upper panel of figure \ref{fig:ET-NS}, the detected mass distribution is shifted to the high mass side, due to higher detectability; and in the lower panel of figure \ref{fig:ET-NS}, we see the portion of detectable DNS merger decreases towards higher redshift, as expected.

\subsection{Example for neutron star/black hole mergers detected with Earth-based detectors}
In a similar way as above, we generate catalogues for ten years aLIGO observation for BHNS mergers, both with the noise spectrum in O3 (aLIGO-O3) and that at design (aLIGO-design). The underlying population model is BHNS-PopB with the default parameters. The event number in the simulated catalogue of aLIGO-O3 is 29 (a Poisson random with expectation value 32.0), which is compatible with the rate found in the O3 period (2-3 in one year); the number for aLIGO-design is 553 (expectation value 517.2). In the top and bottom panels of figure \ref{fig:LIGO-GWTC-BHNS}, we plot the simulated catalogues in $z-m_\bullet$ and $m_\bullet-m_{\rm{n}}$ planes. A more sensitive detector would help to properly characterize this population.

\begin{figure}
    \centering
    \includegraphics[width=0.45\textwidth]{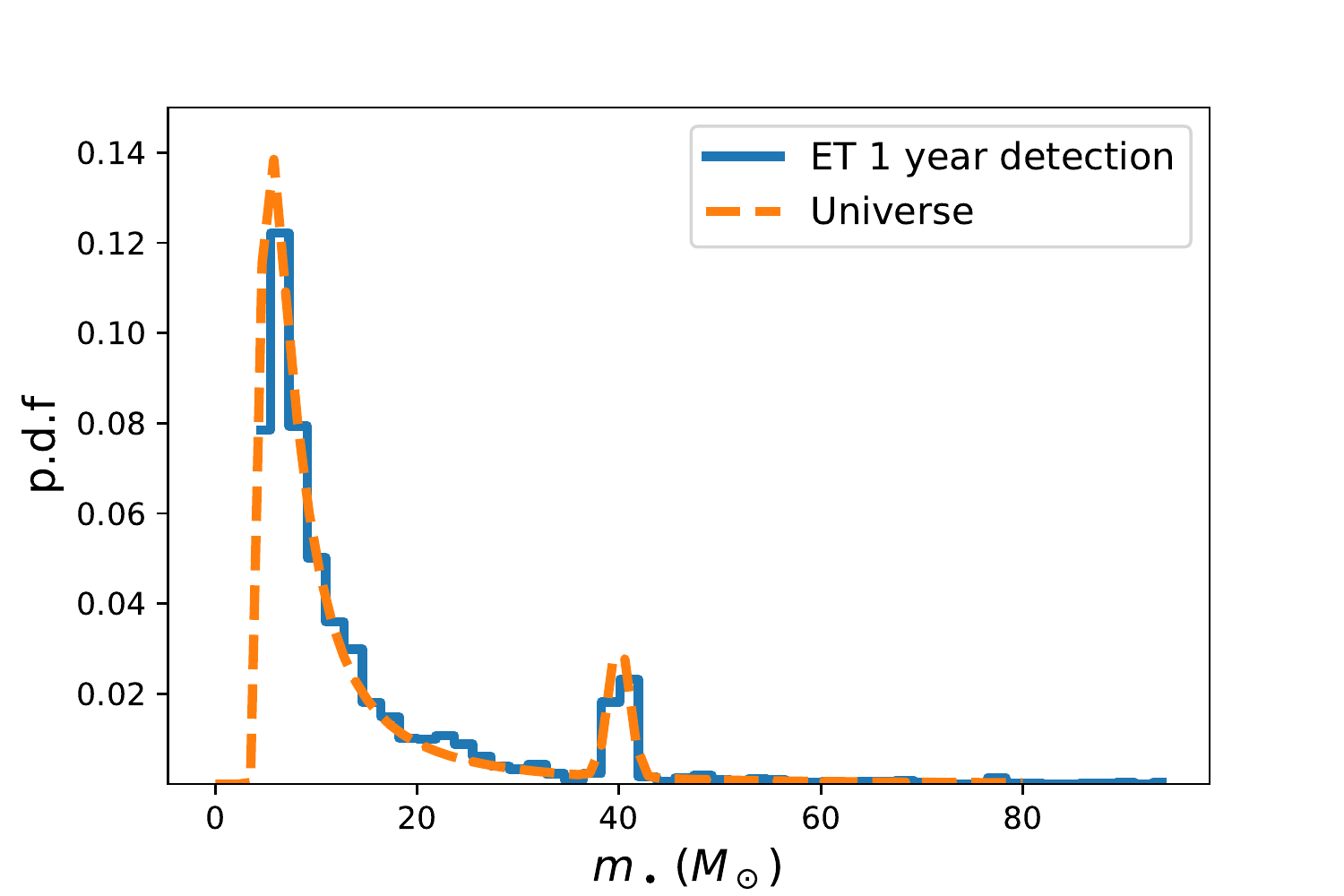}
    \includegraphics[width=0.45\textwidth]{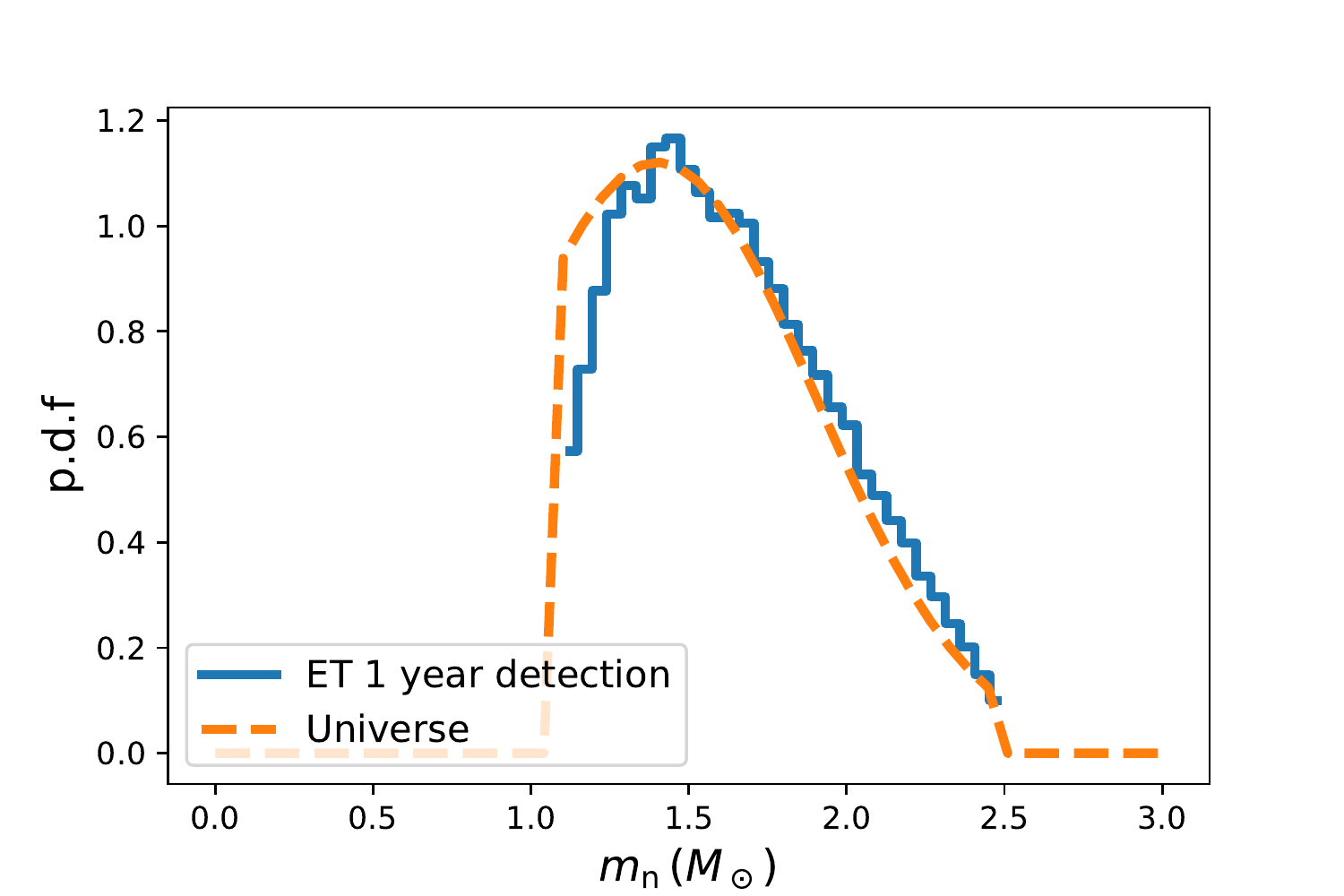}
    \includegraphics[width=0.45\textwidth]{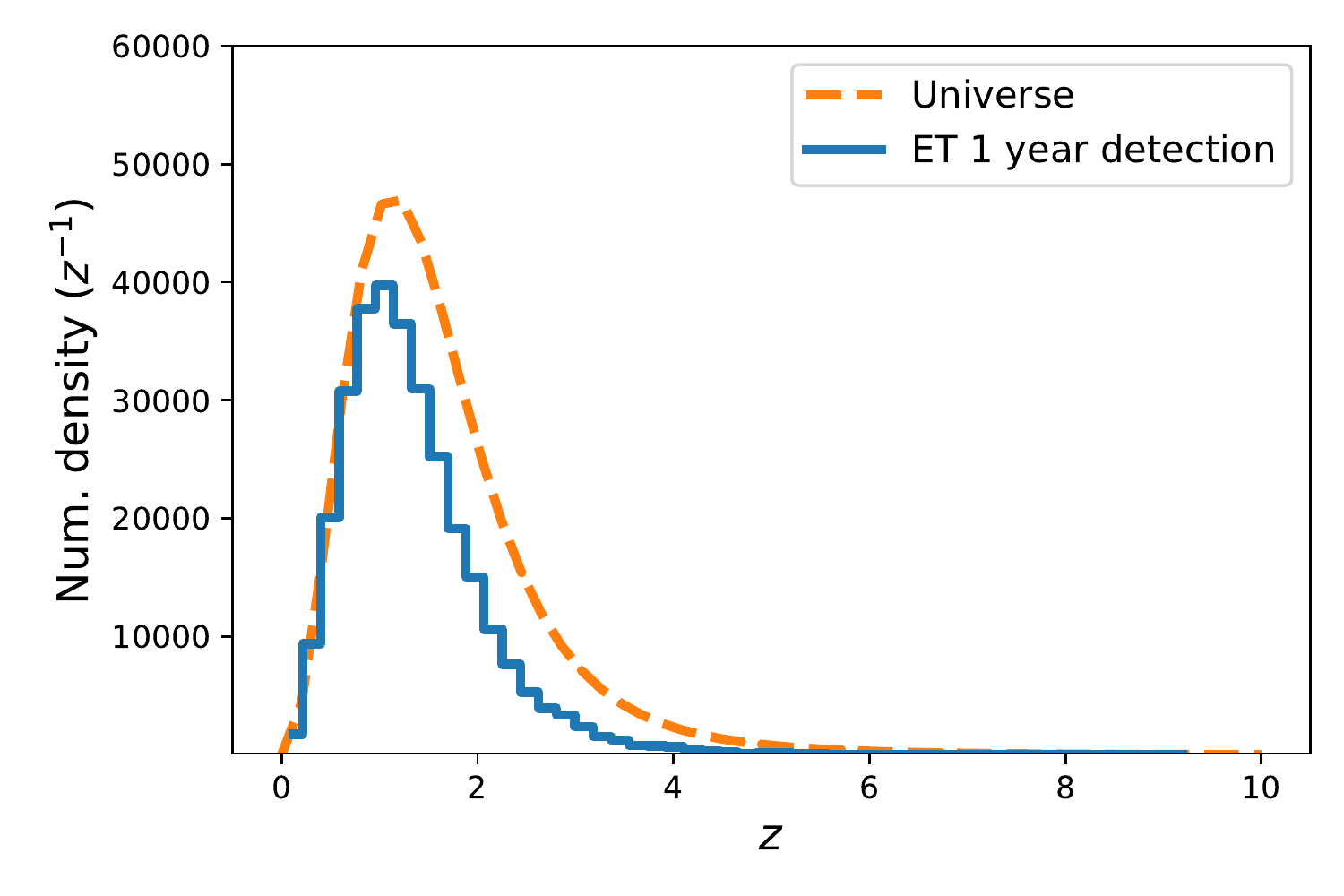}
    \caption{The Probability density distribution as function of $m_\bullet$ (upper panel), $m_{\rm{n}}$ (middle panel) and the number density as function of red-shift (lower panel) for 10 years observation of BHNS mergers by ET.  The integral of the latter is the total number of detection. The orange lines show the intrinsic distributions.}
    \label{fig:ET-BHNS}
\end{figure}

We also simulate a catalogue for one year observation by ET. The number of events is 6251 (expectation $6.32\times10^4$). In the panels of figure \ref{fig:ET-BHNS}, we plot the distribution of the catalogue in $z$, $m_\bullet$ and $m_{\rm{n}}$. We also plot the distribution of BHNS mergers in the whole Universe as defined by the population model for comparison. We see the the effect that the fraction of detectable BHNS mergers increases towards higher masses, and decreases towards higher redshift, although ET probes essentially the whole distribution.

\subsection{Examples for SMBBH mergers detected with LISA}

\begin{figure}
    \centering
    \includegraphics[width=0.45\textwidth]{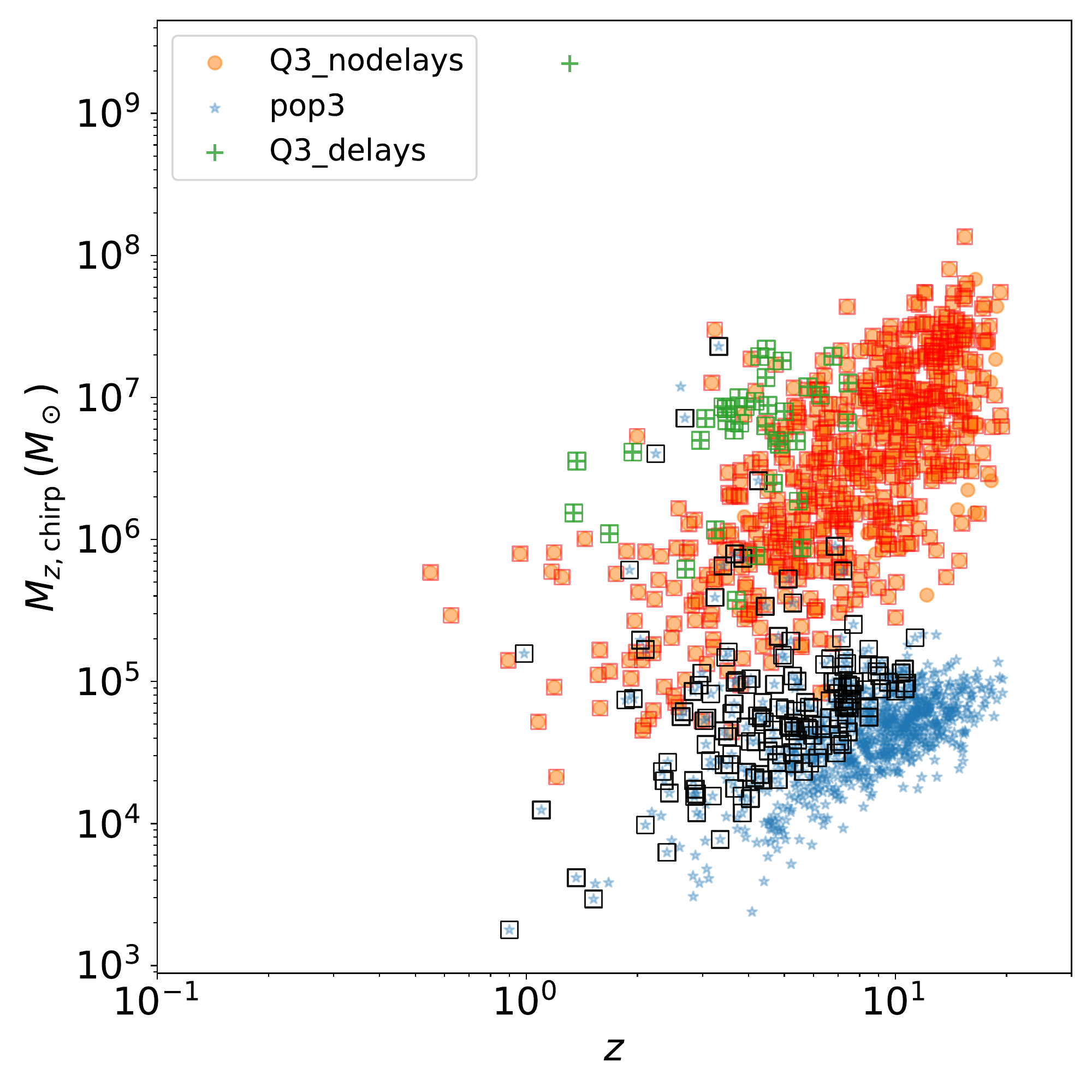}
    \caption{The red-shifted chirp masses and the redshift of the total events (markers) and detected ones (squares on top of markers) by the standard LISA configuration in five years. Orange, blue and green markers correspond to Q3\_nodelay, pop3 and Q3\_delays population models respectively. The underlying SNR cutoff is 8. }
    \label{fig:MBHBH_mtot_z}
\end{figure}

We now turn to the space-based detectors. The \GWT simulates the observed catalogue of SMBBH mergers by going through the catalogue in the Universe for a given observation duration, calculate the SNR for each SMBBH and select against the SNR cut-off, in this case SNR=8 as default. In figure \ref{fig:MBHBH_mtot_z}, we plot the red-shifted chirp masses and the redshift of the total events and detected ones (squares on top of markers) by the standard LISA configuration in five years, corresponding to different population models. As we can see from figure \ref{fig:MBHBH_mtot_z}, the detection horizon of our default LISA passes though the Pop3 population and below the Q3\_delays and Q3\_nodelays population. As a result, almost all events in Q3\_delays and Q3\_nodelays population are detectable, while the detectable fraction of Pop3 changes significantly with different LISA noise settings. 

\begin{figure}
    \centering
    \includegraphics[width=0.45\textwidth]{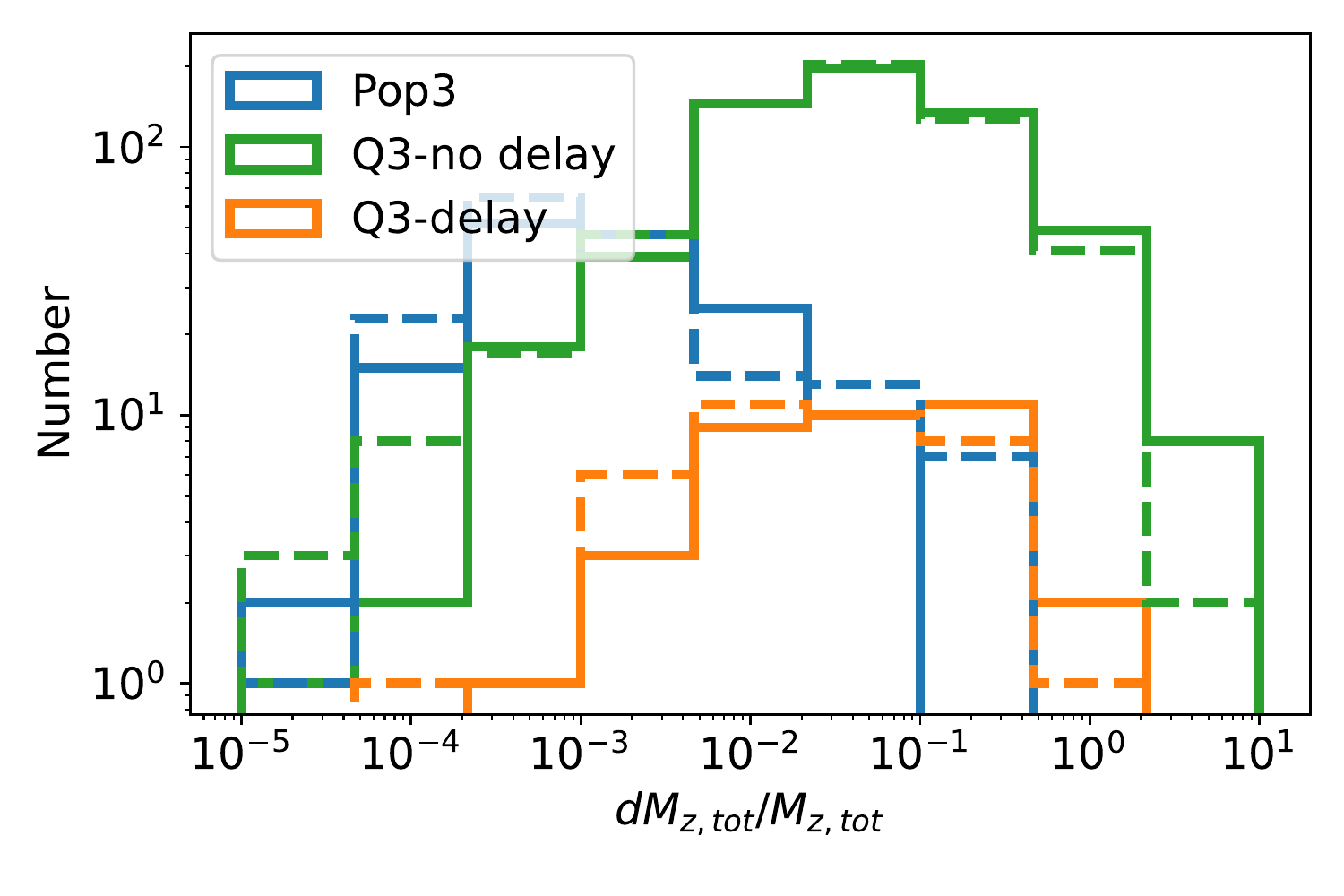}
    \includegraphics[width=0.45\textwidth]{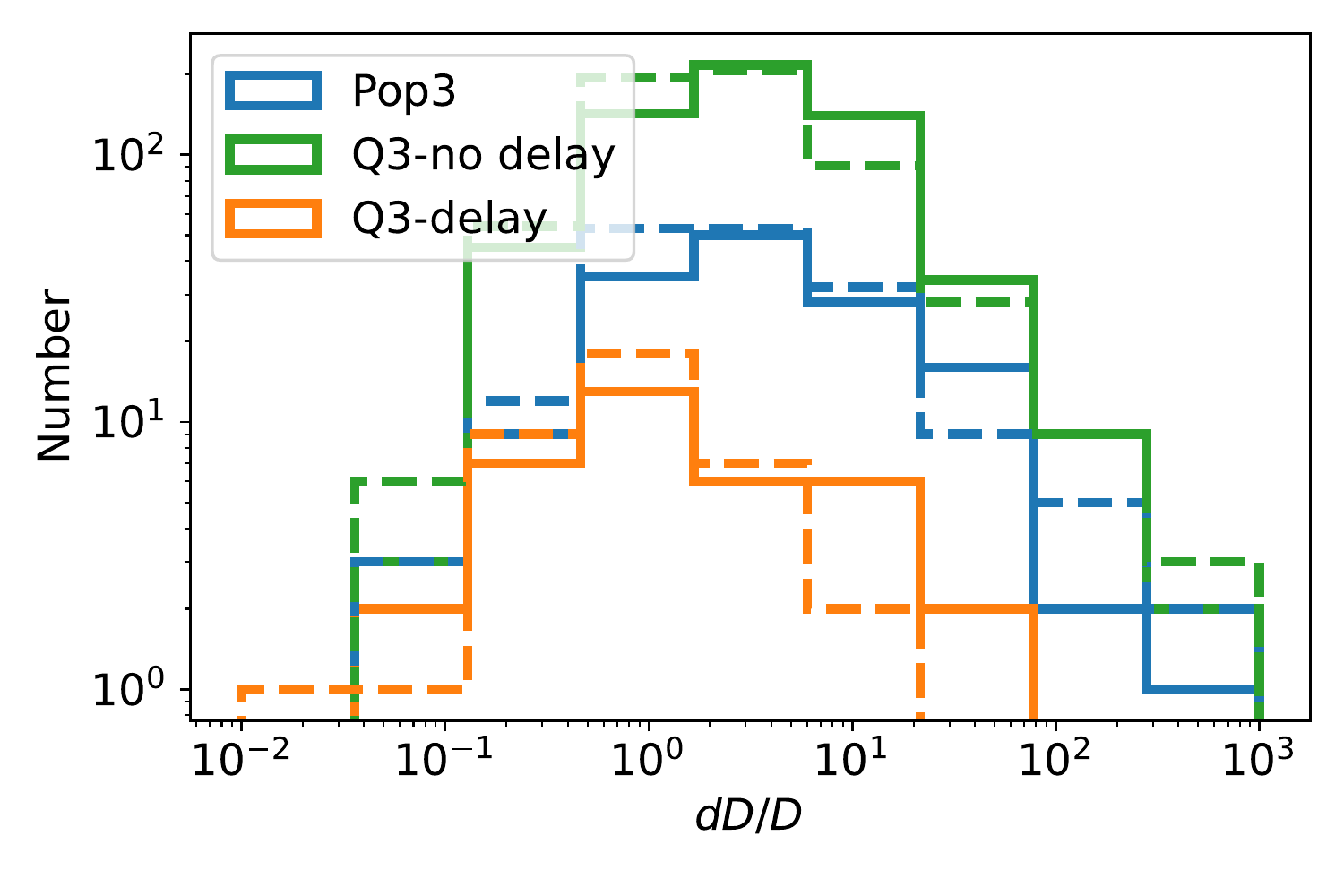}
    \includegraphics[width=0.45\textwidth]{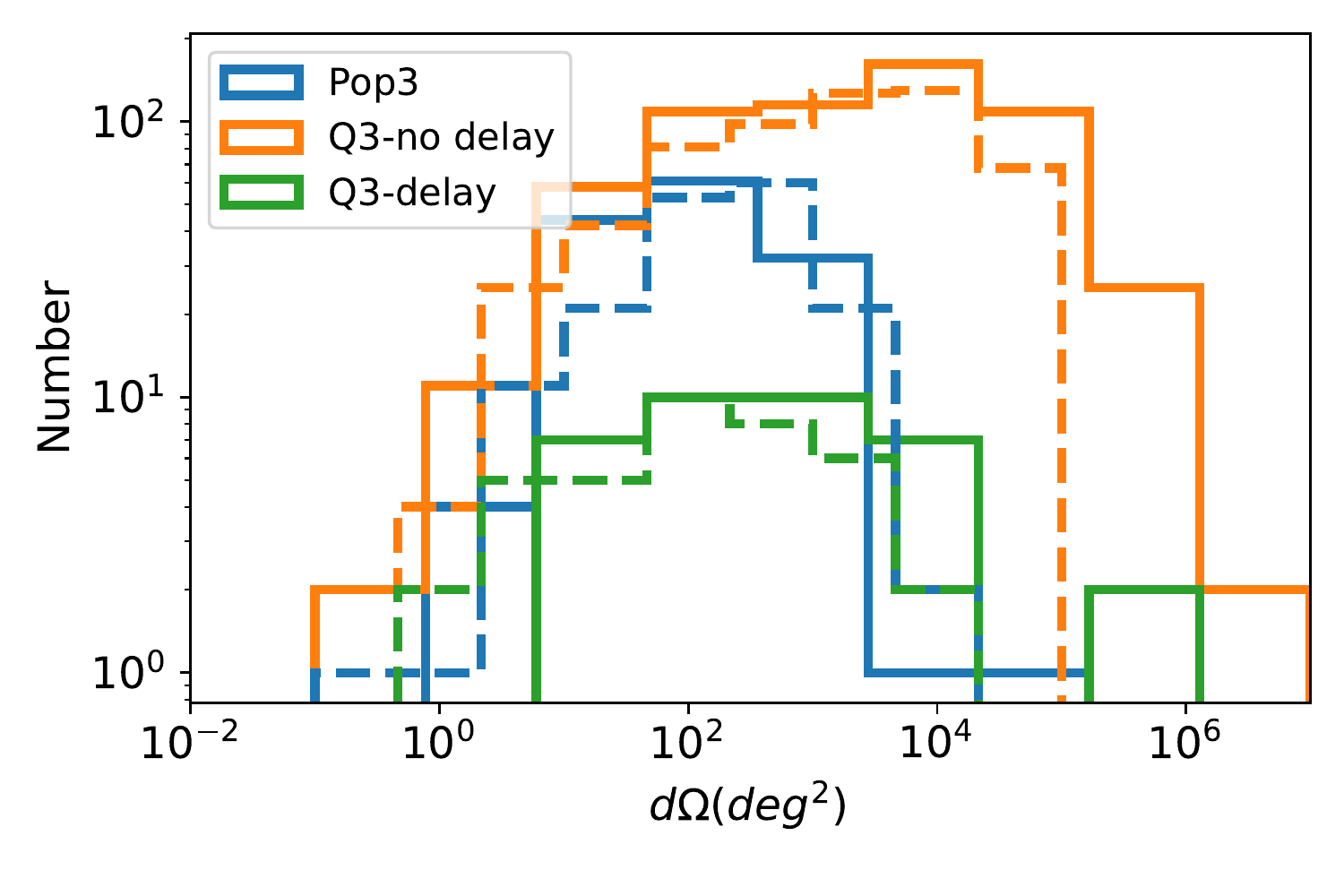}
    \caption{\textbf{Upper:} The distribution of relative uncertainties on the total masses in the catalogue of MBH mergers detected by LISA in five year. Solid lines are for default LISA and dashed lines are for 5Gm arm length; \textbf{Middle} Same as the upper panel, but on the relative uncertainties on the luminosity distances; \textbf{Lower:} Same as the upper two panels, but on uncertainties of the sky location (deg$^2$).}
    \label{fig:MBHB-uncertainties}
\end{figure} 
For these sources, we also calculate the uncertainties in the parameters. In panels of figure \ref{fig:MBHB-uncertainties}, we plot the distribution of the relative uncertainties of total masses, distances and sky localization $d\Omega$ in units of square degrees. The uncertainties are all estimated with a FIM method. The uncertainties span a wide range, but a significant fraction has quite well determined masses while only a small fraction has well determined distance and sky position. Our findings are in general in agree with previous results in Klein16.
\begin{figure}
    \centering
    \includegraphics[width=0.45\textwidth]{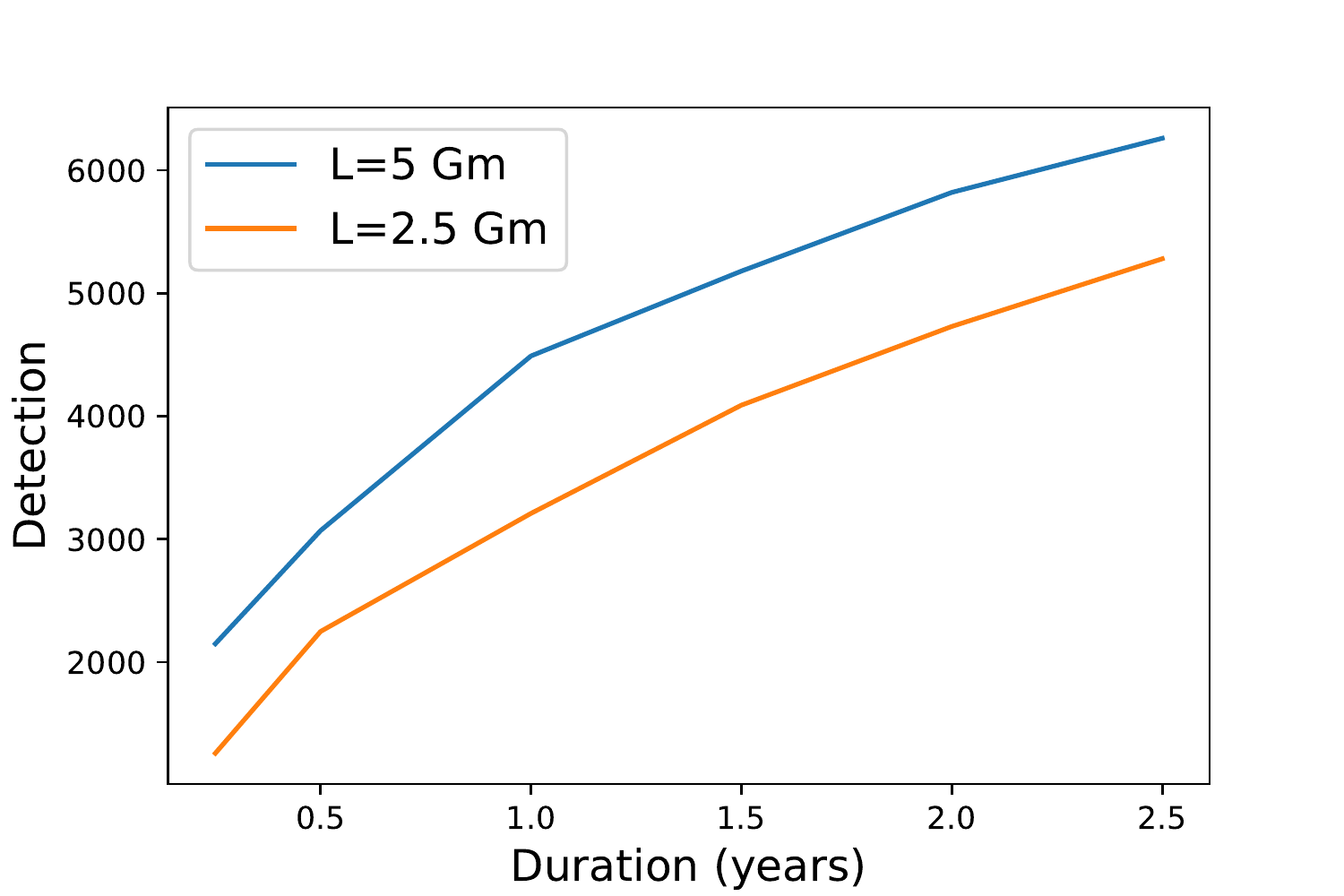}
    \caption{The detected number of DWDs with the default LISA and a larger LISA with 5 million km arms as function of the observation duration. We use a threshold SNR $\rho_\star=7$}
    \label{fig:GBnum}
\end{figure}

\begin{table*}
\begin{tabular}{ccccccc}
\toprule
                                       Name & $f$ (Hz) &  $\beta$ (deg) & $\lambda$ (deg) &                                          $A$ &           $d\Omega$ (deg$^2$) &      SNR \\
\midrule
J0806 &$6.2\times10^{-3}$$\pm$$3.1\times10^{-8}$ &   -4.704 &   120.442 &  $1.2\times10^{-22}$$\pm$$6.5\times10^{-23}$ &  $6.9\times10^{-3}$ &   91.851 \\
V407 Vul &$3.5\times10^{-3}$$\pm$$3.1\times10^{-8}$ &   46.783 &   294.995 &  $5.9\times10^{-23}$$\pm$$1.1\times10^{-22}$ &  $1.2\times10^{-1}$ &   65.930 \\
ES Cet &$3.2\times10^{-3}$$\pm$$3.1\times10^{-8}$ &  -20.334 &    24.612 &  $4.7\times10^{-22}$$\pm$$1.0\times10^{-22}$ &  $1.7\times10^{-1}$ &   46.347 \\
ZTF J153932.16+502738.8 &$4.8\times10^{-3}$$\pm$$3.1\times10^{-8}$ &   66.162 &   205.031 &  $3.1\times10^{-24}$$\pm$$1.8\times10^{-22}$ &  $1.3\times10^{-2}$ &  188.051 \\
SDSS J065133.34+284423.4 &$2.6\times10^{-3}$$\pm$$3.1\times10^{-8}$ &    5.805 &   101.340 &  $1.9\times10^{-23}$$\pm$$1.6\times10^{-22}$ &               2.306 &   15.613 \\
SDSS J093506.92+441107.0 &$1.6\times10^{-3}$$\pm$$3.1\times10^{-8}$ &   28.091 &   130.980 &  $7.6\times10^{-22}$$\pm$$3.0\times10^{-22}$ &              19.776 &   10.653 \\
\bottomrule
\end{tabular}
\caption{Detection of the verification binaries with the LISA detector.}\label{tab:VBs}
\end{table*}

\subsection{Examples for DWDs detected with LISA}
In order to simulate the DWDs observed by LISA, we go through a simulated catalogue of all Galactic DWDs and calculate their SNR with the analytic approximation in equation (\ref{eqn:rho_ana}). Then we select the sources with SNR larger than the SNR threshold of 10 as the detected sources. We do the same for the verification binaries (VBs). In Table \ref{tab:VBs}, we list the detected VBs and the estimated uncertainties of parameters. When calculating the uncertainties with FIM, the numerical waveform calculated with LDC are used. The results are compatible with earlier work, the SNR a bit lower than \cite{2018MNRAS.480..302K}, due the use of a slightly different LISA sensitivity. Since the size of the DWD catalogue is huge ($\sim2.6\times10^7$), we go through a smaller sub-catalogue that are randomly drawn from the whole synthetic DWD catalogue instead, and scale the number of detection from this small sub-sample to the whole GWD catalogue to obtain the total expected number of  detection. In figure \ref{fig:GBnum}, we plot the number of detection as function of the observation duration. Our results are about a factor of 0.5 lower than the earlier results e.g. in \cite{2001A&A...375..890N,2012ApJ...758..131N,2017MNRAS.470.1894K}. We attribute this to a slight difference in the LISA noise model or the DWD catalogue.  

\subsection{Examples for EMRIs detected with LISA}
\begin{figure}
    \centering
    \includegraphics[width=0.45\textwidth]{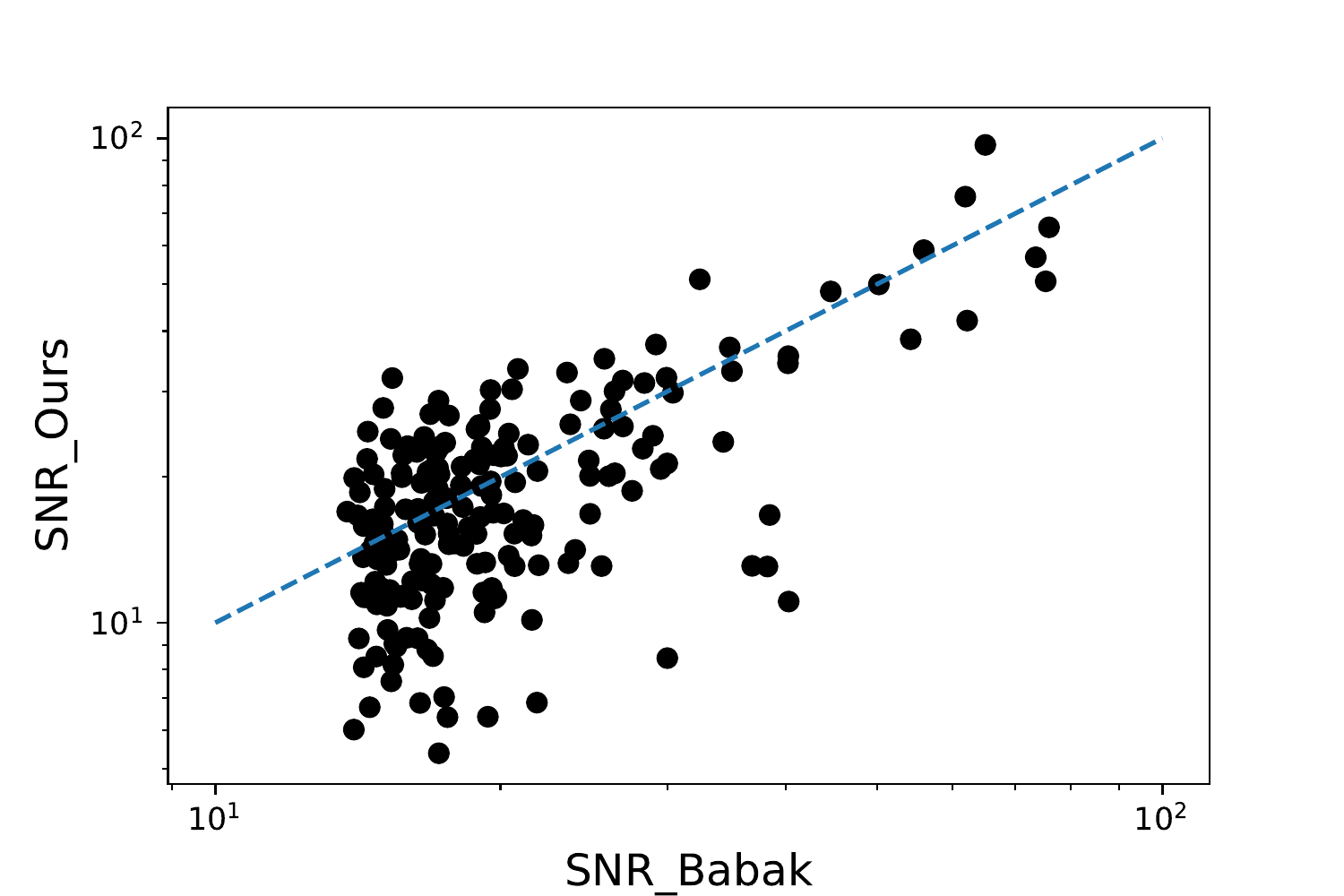}
    \caption{The pre-calculated SNR by Babak17 comparing with SNR of this work for the same catalogue, for one years' observation on M1 population.}
    \label{fig:ratio}
\end{figure}

\begin{figure}
    \centering
    \includegraphics[width=0.45\textwidth]{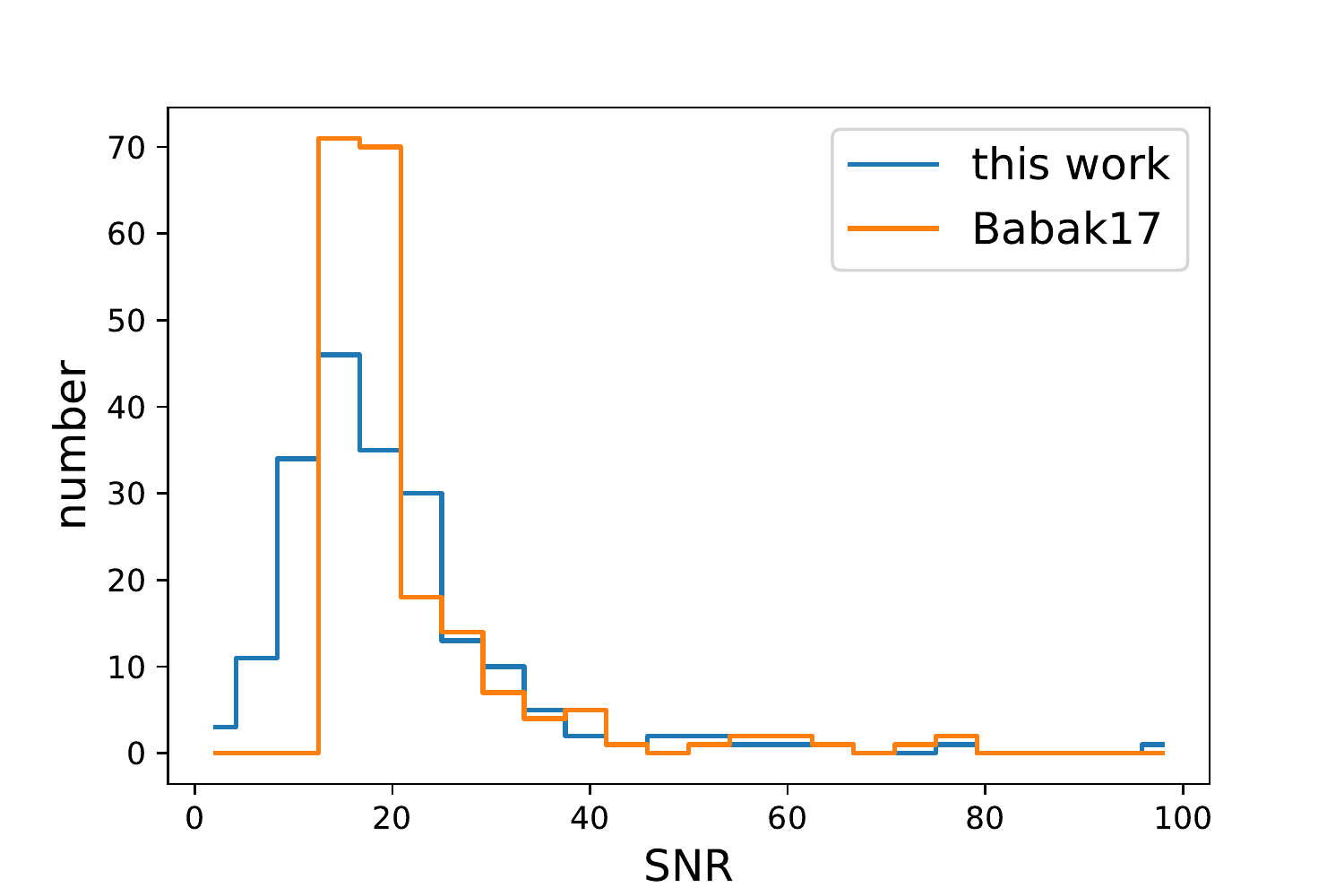}
    \caption{Histogram of the SNR of the bright EMRIs in catalogue of Babak17. The blue histogram indicate the SNR calculated in this work, and the orange histogram is that pre-calculated by Babak17 using a slightly different LISA noise and waveform. }
    \label{fig:snrstars}
\end{figure}

\begin{figure}
    \centering
    \includegraphics[width=0.4\textwidth]{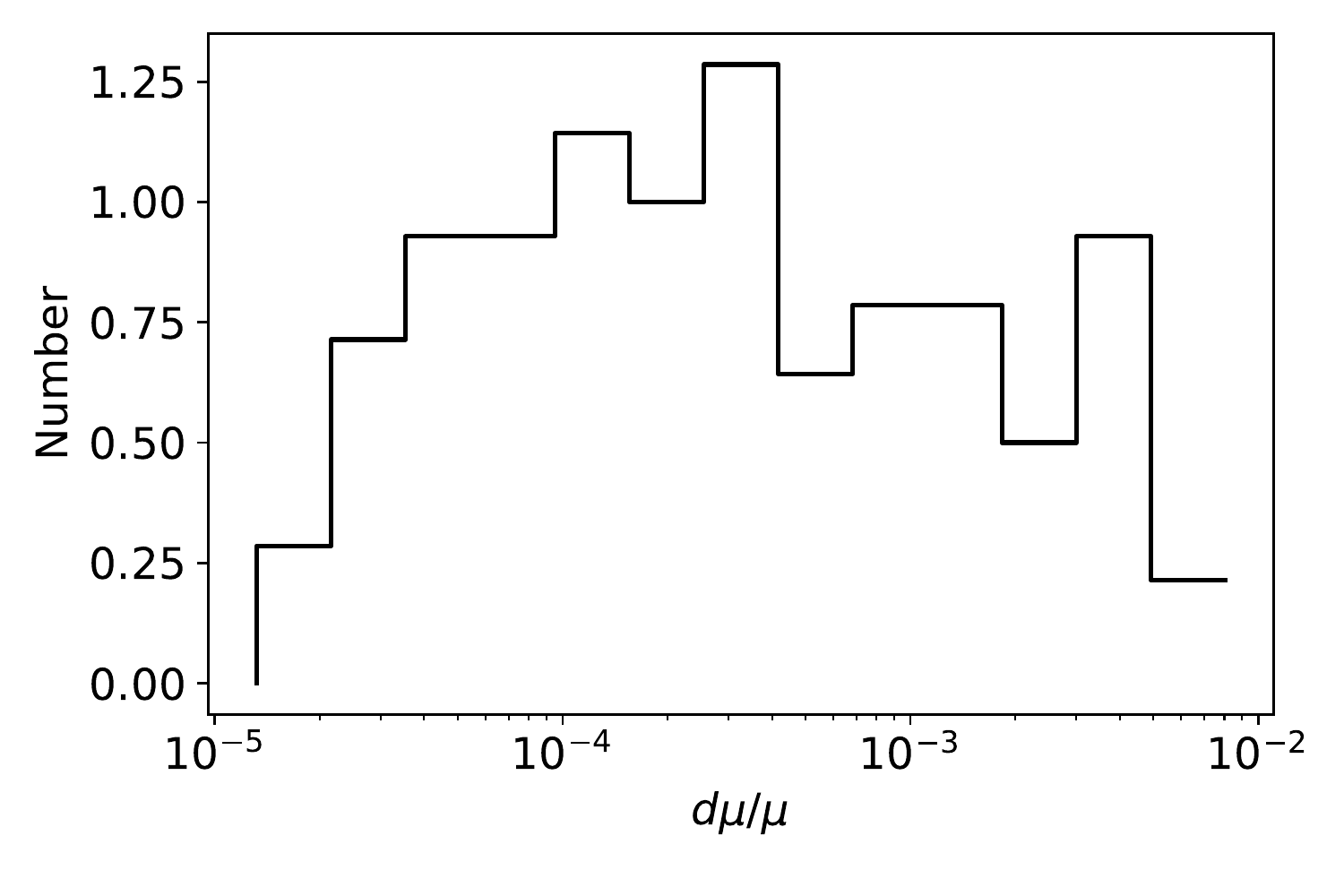}
    \includegraphics[width=0.4\textwidth]{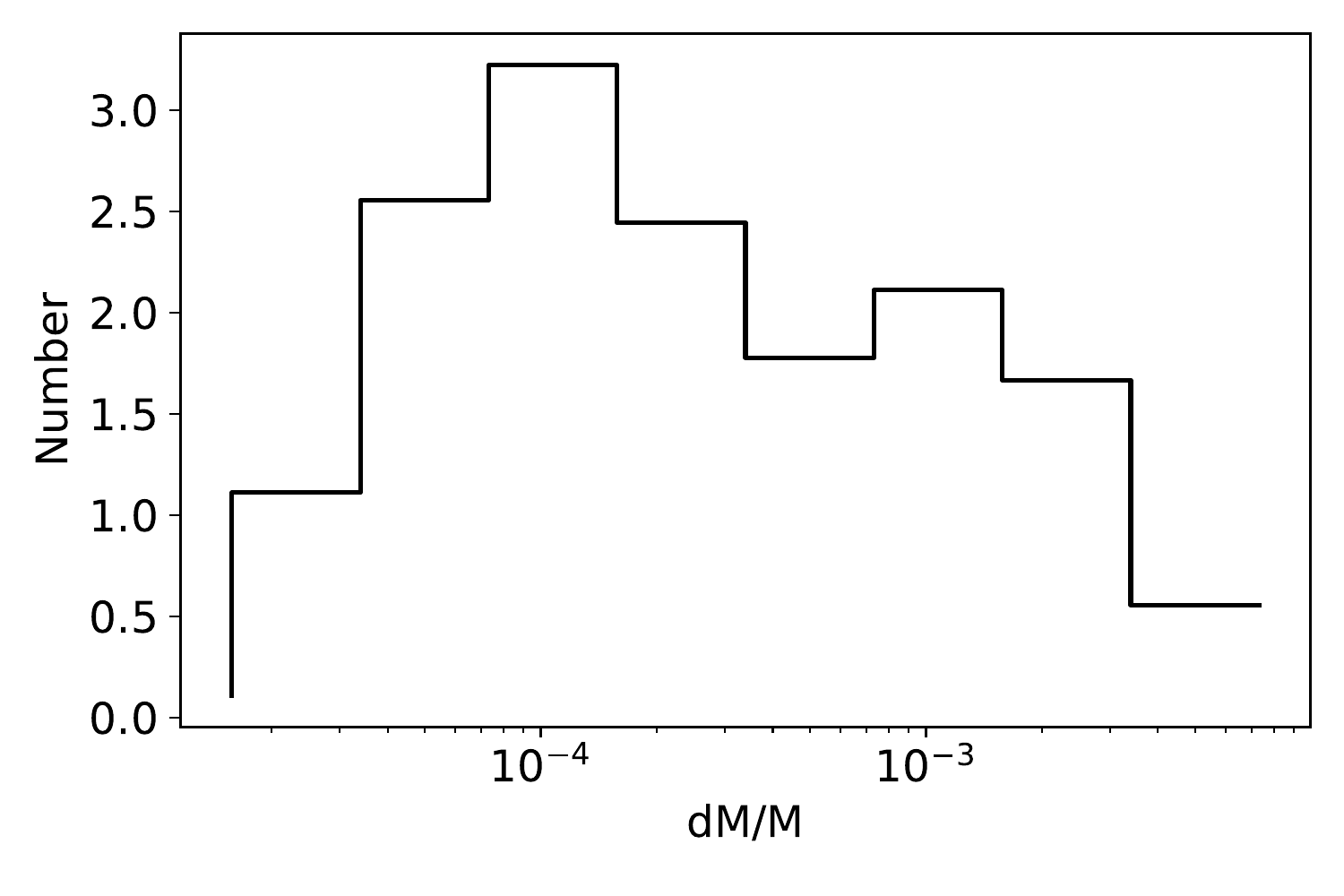}
    \includegraphics[width=0.4\textwidth]{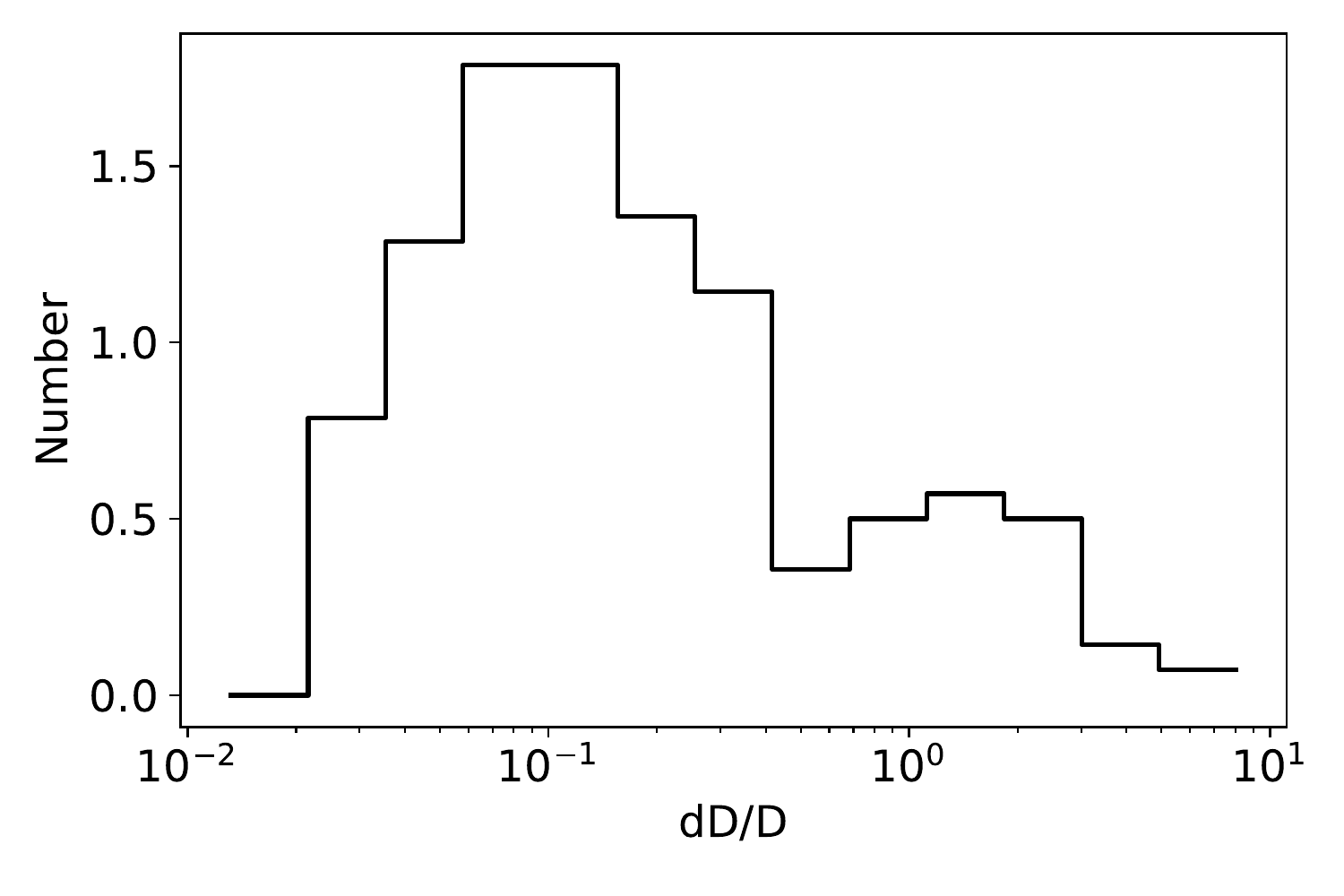}
    \includegraphics[width=0.4\textwidth]{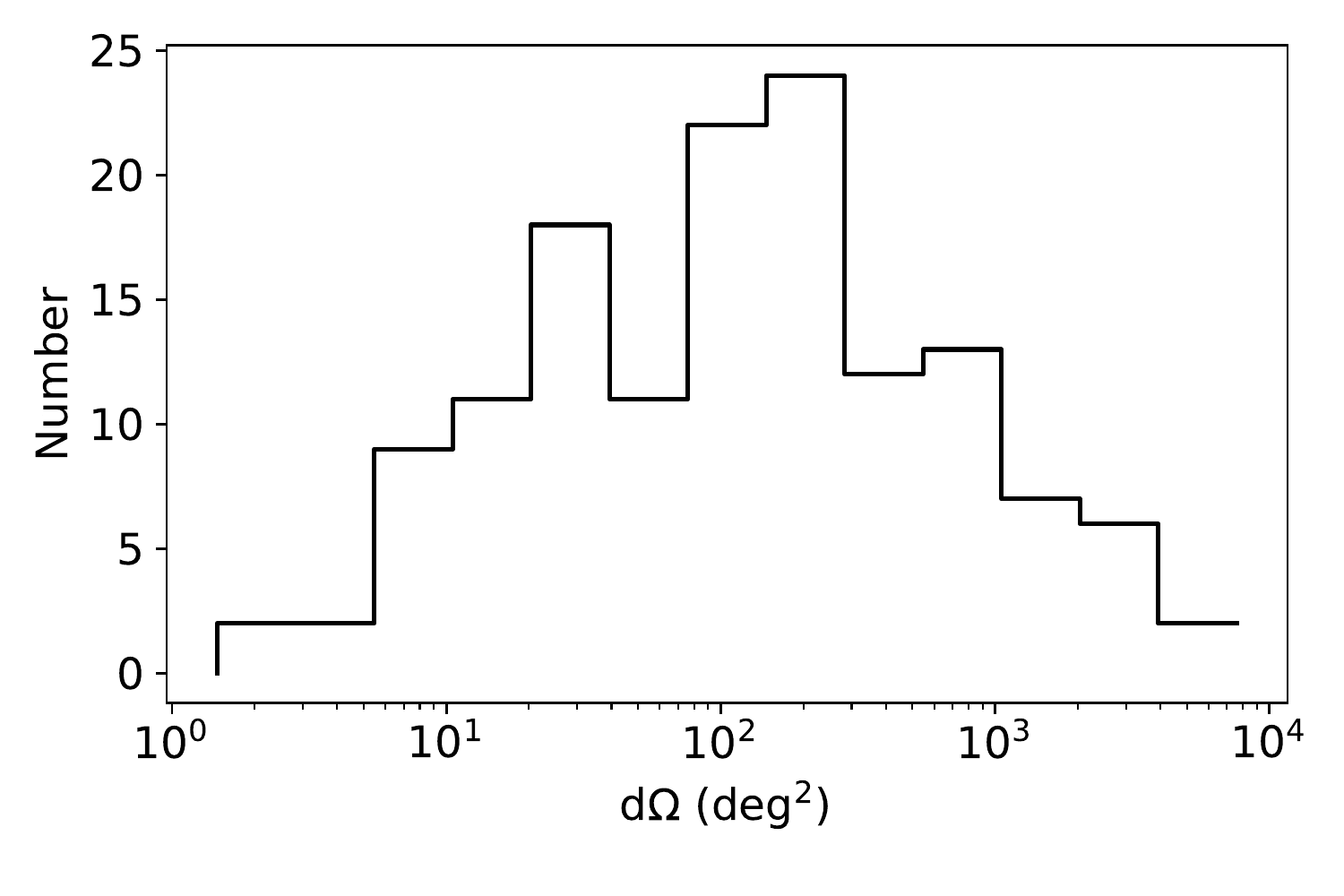}
    \caption{Histogram of relative uncertainties on the mass of the lighter component $\mu$, the mass of the heavier component $M$, the distance $D$ and the uncertainties of the sky location $d\Omega$ in a catalogue of EMRIs detected by the default LISA for a two year observation. The SNR cut-off is set to 20.}
    \label{fig:emris_errors}
\end{figure}

The signal from an EMRI can fall into the detectable frequency range of LISA from an early phase untill the final plunge, which can span quite a long duration from months to years. In order to guarantee the speed of simulation, we use a frequency domain waveform (modulus) generated in advance and stored in files. In each simulation, we go through all the candidate events in the catalogue, read in the corresponding TDI waveform modulus and calculate the SNR against the noise curve. The detected events are then selected against an SNR-threshold. In the EMRIs catalogues that we are using, there are also pre-calculated SNR for each system, which corresponds to a slightly different LISA setup and waveform (see Babak17). In figure \ref{fig:ratio}, we compare their pre-calculated SNRs with ours of the same catalogue for one year observation of the M1 population with our default LISA. Our calculated SNR values disperse within a factor two around the values of Babak17. We attribute this dispersion to the slightly different LISA noise and waveform (AK Schwarzchild versus AK Kerr, see Babak17). 
In figure \ref{fig:snrstars}, we plot the histogram of the SNR of the bright EMRIs in the catalogue of Babak17 and that calculated from this work. The underlying population model is M1. 
In panels of figure \ref{fig:emris_errors}, we plot histograms of relative uncertainties of the masses $\mu$, $M$, distance $D$ and sky location $\Omega$ (in units of square degrees) in a catalogue detected by the default LISA. The observation duration is two years, and the SNR cut-off is set to 20 and the population is M1. The uncertainties are estimated with FIM (section \ref{sec:SNRandFIM}). In the calculation of FIM, derivatives of the complex waveform relative to all the relevant parameters are needed (equation \ref{eq:derivatives}). Therefore, if we were to use the pre-calculated waveform for the uncertainty estimation, the storing files would be $\sim20$ times larger in size than those used for the SNR calculation. On that account, we calculate the late stage waveform in real time and use them for the uncertainty evaluation. In general the parameters of EMRIs are very well determined, expect in some cases the sky position. The estimated level of uncertainties are in agreement with Babak17.


\subsubsection{Results for PTAs}
\begin{figure}
    \centering
    \includegraphics[width=0.45\textwidth]{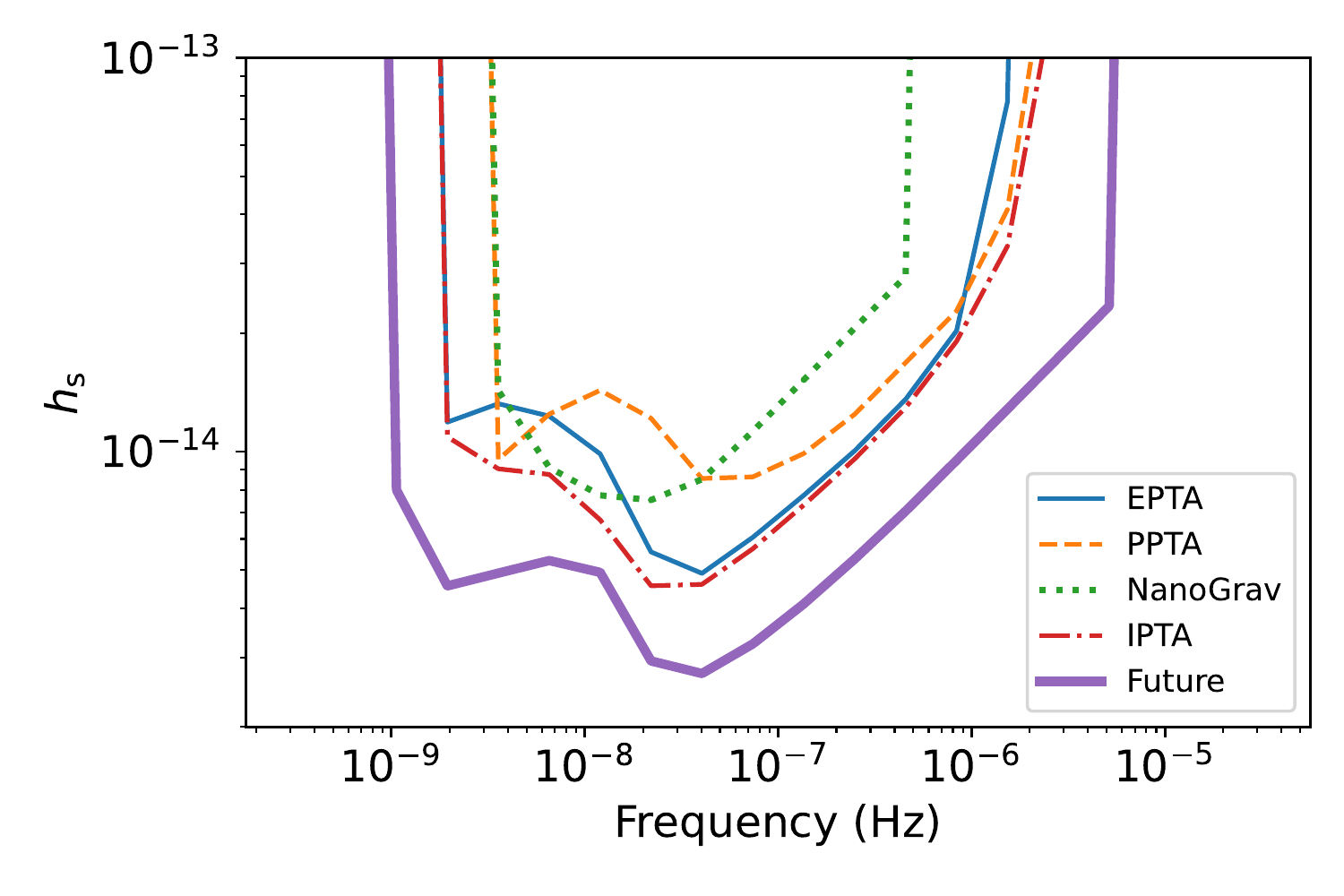}
    \caption{Sensitivity plot for an individual SMBBH as function of frequency, for $\rho_{\rm cri}=10$ and averaged over the celestial sphere for different PTAs.}
    \label{fig:PTAsens}
\end{figure}

For PTAs, we calculate the detection limits for individual SMBBHs and a stochastic background based on the different PTA configurations. Given a certain PTA, a SNR cut-off and the coordinates of the source, we can give a sensitivity curve for GW emitted by an individual SMBBH, as a function of frequency. In figure \ref{fig:PTAsens}, we plot the sky-averaged sensitivity curve to individual sources of EPTA, PPTA, NANOGrav, IPTA and a simulated future PTA (labeled ``future"). For the future PTA, we assume daily observation on the IPTA pulsars for 10 more years, with two more new pulsars being added to the PTA per year (the setting of the future PTA can be customized by users). The corresponding $\rho_{\rm cri}=10$. The results are in agreement with published ones \citep{2016MNRAS.455.1665B,2016MNRAS.459.1737S,2019ApJ...880..116A}. 

In Table \ref{tab:sgwb}, we list the upper limits on the SGWB of different origins, the corresponding $\rho_{\rm cri}=100$. They are broadly in agreement with published results (to within a factor $\sim2$), which are also listed in the table in the parentheses.

\begin{table*}[]
\centering
\small
\begin{tabular}{l|c|c|c}
\hline
         & SBHBH ($A_{\rm yr^{-1}}$) & Cosmic Strings ($\Omega_{\rm gw, yr^{-1}}h^2$) & Relic ($\Omega_{\rm gw, yr^{-1}}h^2$) \\
\hline
EPTA \citep{2015MNRAS.453.2576L} & \eptabh ($3\times10^{-15}$)                         &        \eptacs                                               &               \eptarl ($1.2\times10^{-9}$)                           \\
PPTA \citep{2015Sci...349.1522S} &      \pptabh ($1\times10^{-15}$)                    &        \pptacs                                               &            \pptarl                                  \\
NANOGrav \citep{2018ApJ...859...47A} &      \nanobh ($1.45\times10^{-15}$)                     &       \nanocs                                              &                \nanorl ($3.4\times10^{-10}$)                           \\
IPTA     &        \iptabh                   &     \iptacs   &       \iptarl \\
\hline
\end{tabular}
\caption{The upper limits set with different PTAs to SGWB of different origins, $\rho_{\rm cri}=100$. Numbers in parentheses are values in literature. The most recent reported upper limit on the GW originates from Cosmic String are in terms of the cosmic-string tension  ($G\mu$). The conversion from $G\mu$ to $\Omega_{\rm gw, yr^{-1}}h^2$ depends on models and the reconnection probability $p$, so we cannot give literature values.}\label{tab:sgwb}
\end{table*}

\section{Caveats \& Discussion}\label{sec:discussion}

We have implemented and described a first version of the \GWT, which still has a few caveats and is missing some ingredients that we plan to implement in the (near) future. The most important caveats of the current version are:
\begin{itemize}
    \item Higher order multipole modes of the waveform: Theoretically, GW radiation are treated with spin-weighed spherical decomposition \citep{1980RvMP...52..299T}. The leading term is the $\ell=2$, $|m|=2$ (quadrupole) term. In the current \GWT, we apply a waveform that neglects the higher order modes beyond quadrupole. High order modes will extend the spectrum of a GW to high frequencies \citep{2018PhRvL.120p1102L}, and their significance increase with higher mass inequality, higher total mass and higher orbital inclination angle \citep{2021PhRvD.103b4042M}. Including higher order modes in the waveform can break the degeneracy between the inclination angle and the distance. thus results in more accurate estimate of the parameters of the source \citep{2018PhRvL.120p1102L}. In the O3 observation, there are already some GW events with evidence of higher order modes (GW170729: \citealt{2019PhRvD.100j4015C}; GW190412: \citealt{2020PhRvD.102d3015A}; GW190814: \citealt{2020ApJ...896L..44A}). \cite{2021PhRvD.103b4042M} found that a few percent of binaries will be detected with higher order modes by aLIGO with designed sensitivity. For 3G GW detectors, \cite{2021PhRvD.104h4080D} found that the SNR contribution from higher order modes in some GWTC-3 systems would surpass 10\% of the total SNR. For larger total mass and mass ratio binaries, the higher order modes are expecting to contribute a significant fraction of the SNR. Therefore, neglecting higher order modes in the waveform will results in an underestimate of the detection rate of sources, especially for those with large total mas and mass ratio. In an updated \GWT, the waveform will be replaced with one that includes the higher order modes, {\it e.g.,} \texttt{PhenomHM} \citep{2018PhRvL.120p1102L}.
    \item Uncertainties estimation for Earth-based detectors: the uncertainties are estimated in a hybrid way: those of the intrinsic parameters ($m_{i,z}$ and $\chi$) are calculated with FIM, with all of the amplitude parameters treated as one. In addition, the uncertainty on the red-shift ($\delta z$) is estimated with empirical relations. Then the source-frame masses uncertainties are calculated with error propagation equation (Equation \ref{eq:error_propagation}). Equation (\ref{eq:error_propagation}) is valid only when the covariance between $dm_{i,z}$ and $dz$ is vanishing. Since $m_{i,z}$ are largely determined from matching the phases of the waveform, while the constraint on $z$ comes from the fitting amplitude of the waveform, we expect that the dependence between $dm_{i,z}$ and $dz$ are weak. A more self-consistent and accurate non-Bayesian treatment would be: to treat all extrinsic parameters separately in the FIM (the dimensional of FIM will thus be enlarged), and calculate the covaraince among all parameters explicitly. Then impose the constraints from the triangulation localization and obtain the uncertainties on all parameters.
    
    \item Populations and Waveforms of EMRIs: In order to return the simulated catalogue of detection in a tolerable time for a website user, we use catalogues of EMRIs in which only bright ones are included (pre-calculated SNR$_{\rm{tot}}>20$). Therefore, the user should not set an SNR cut-off lower than $\sim15$, otherwise the returned synthetic catalogue is incomplete. In order to compare with previous results of Babak17, the waveform we employed is analytic kludge, which is known to be fast but less realistic. In the \GWT, it can be replaced with a more accurate waveform augmented analytic kludge (AAK) easily, since the latter can also be simulated with the same package \texttt{EMRI\_Kludge\_Suite}. For the sake of the speed of simulation, we use the pre-calculated frequency domain TDI waveform for the SNR calculation. When estimating the uncertainties using FIM methods of the detected sources, we only include the late stage of their waveform. As shown in the above sections, we will miss some of the low frequency section corresponding to the early inspiral stage, which is in fact detectable by LISA, and results in underestimation of the accuracy of parameters inference. As shown in examples in the above section, such underestimation is not severe. 
    \item PTAs: Our sensitivity curves and upper limits are given according to an SNR threshold and the SNR of GW are calculated based on a simplified parameterised noise spectra. On the other hand, upper limits are reported in literature with confidence levels. Due to the very different nature between the methods, the correspondence between the confidence level in literature and our SNR threshold is difficult to explore. In our examples, we set the $\rho_{\rm{cri}}=10$ for continuous GW in plotting the sensitivity curves, and $\rho_{\rm{cri}}=100$ for SGWB, in order to obtain results which are in order of magnitude in accordance with literature. For continuous GW, the sensitivity scales with the $\rho_{\rm{cri}}$; while for SGWB, upper limits on the characteristic strain scale with the square root of $\rho_{\rm{cri}}$.
\end{itemize}
We plan to include a number of additions to the \GWT in the future. The first ones involve several proposed space-borne detectors, in particular DECIGO, Taiji and Tianqin. In the ground-based and space-borne modules, we will integrate SNR calculators for individual sources that are specified by the user. On a longer time scale, we plan to include more GW sources, \textit{e.g.,} supernovae explosions, single spinning and recycling neutron stars, multiple black holes encounters and catalogs of SMBBH for PTAs. We are also working to extend the \GWT with electro-magnetic counterparts, \textit{e.g., }to return the fluence of short GRBs and peak fluxes of kilonovae. Triangulation of a network of ground-based detectors will also be developed.

In the next step, the \GWT will have the ability to simulate observations of different evolutionary phases of the same population in different GW frequency ranges. For instance, each population of compact object mergers corresponds to a population of persistent GW source from the earlier phase. The former are targets of ground-based interferometers, while the latter are targets of space-borne interferometers. Another instance is the close orbit and inspiral-merger phases of SMBBH, which can be observed with PTA and LISA respectively. We will also include simulated observation on SGWB with ground-based and LISA-like detectors. 

\section{summary}\label{sec:summary}
In this paper, we introduce the \GWT (\url{www.gw-universe.org}), a set of tools that quickly simulates a Universe with kHz/mHz/nHz GW source populations, and observations with different GW detectors, i.e., ground-based interferometers, space-borne interferometers and pulsar timing arrays. We hereby summarize the functionalities and methodologies of the \GWT for each module: 
\begin{itemize}
    \item The module for ground-based interferometers simulates observation of mergers of compact objects, including binary black holes (BBH), double neutron stars (DNS) and black hole-neutron stars (BHNS). The detectors include default aLIGO, Virgo, KAGRA, Einstein Telescope and Cosmic Explorer, and a user customised LIGO/Virgo-like and ET-like detectors. The noise curves of the default detectors are taken from the literature, and those of user customised detectors are simulated with the \texttt{FINESSE} software. After the noise curve and antenna patterns are determined, we calculate the optimal SNR for all sources in the selected source class. A simplified \texttt{IMRPhenomD} waveform that assumes zero effective spin is employed in this step. With a certain SNR-threshold of detection, we marginalize the geometrical parameters and obtain the detectability $\mathcal{D}(m_1, m_2, z, \chi)$ as function of the source's masses, redshift (luminosity distance) and effective spin. The product of $\mathcal{D}(m_1, m_2, z, \chi)$ and the user-selected probability density function (p.d.f) of the source population defines the p.d.f of the detectable sources, $N_d(m_1, m_2, z, \chi)$. A synthetic catalogue of observations is obtained with a MCMC sampling from the $N_d(m_1, m_2, z, \chi)$. We use Fisher Information Matrix (FIM) method to estimate the uncertainties of the parameters of events. In the process of calculating the FIM, we apply the complete \texttt{IMRPhenomD} waveform phases (for aligned spins). 
    \item The module for space-borne interferometers simulates observations with LISA or a customised LISA-like configuration. The noise power density in the TDI-$X$ response channel is calculated with an analytical formula, which includes acceleration noise, laser shot noise, other optical Meteorology noises and confusion noise due to Galactic double white dwarf (DWD) foreground. The targets we include are the inspiral of Supermassive Binary Black Holes (SMBBH), individual resolvable Galactic DWD and Extreme Mass Ratio Insprials (EMRIs). For SMBBH, we calculate the TDI-$X$ LISA responses of a GW source with \texttt{LDC} codes. The optimal SNR is subsequently calculated. There are three population models being considered, namely \texttt{Pop3}, \texttt{Q3\_nodelays} and \texttt{Q3\_delays}, taken from Babak17. There are ten realizations of the simulated catalogues of SMBBH mergers in the Universe within five years for each population model. The \GWT will re-sample from the catalogue according to user specified observation duration, and calculate the SNR for each source in the sample. A synthetic detection catalogue is thus returned based on a SNR threshold of detection set by the user. The uncertainties are estimated with FIM. For DWDs, we use an analytic equation to calculate the modulus TDI-$X$ LISA response to a series of sinusoidal GWs, and therefore the SNR. We consider two samples of DWDs, namely the verification DWDs and a simulated entire Galactic population. For the former sample, the SNR is calculated individually in the catalogue, and a detected catalogue is returned according to a SNR threshold. For the latter sample, due to its huge number, we randomly draw a sub-sample from it, and find the catalogue of detections in the sub-sample. The total expected number of detection is obtained by rescaling the number of events in the returned catalogue. The uncertainties are also estimated with FIM, where we use \texttt{LDC} codes for the complex TDI LISA response, instead of using the analytical equation as in the SNR computation. For EMRIs, we make use of the \texttt{EMRI\_Kludge\_Suite} for the TDI LISA response, and therefore the SNR. We calculate the SNR for each source in pre-simulated catalogues of EMRIs of different populations, and select those with SNR surpassing the SNR threshold. The uncertainties are again computed with FIM. 
    \item In the PTA module, we include four currently operating PTAs: EPTA, PPTA, NANOGrav and IPTA. For the pulsars in these PTAs, we use the following parameters to represent their noises properties and observation campaigns: the levels of white noise, the level of red noise, the red noise spectrum index, total observation duration and averaged interval between observations. We allow users to include new pulsars which will be discovered in the course of future observations. The sky-locations of the new pulsars and their noise properties are randomly assigned according to the distributions of  those of the known pulsars. In this module, the \GWT computes the SNR of a series of monochromatic GWs with given frequency and amplitude, which corresponds to a GW from the orbital motion of SMBBH. Another function of this module is to evaluate the upper limit that a PTA can set to Stochastic GW Background (SGWB) from different origins. 
\end{itemize}

In the (near) future, the \GWT will be extended with new standard detectors, triangulation of a network of ground-based detectors, new source classes and electro-magnetic counterparts and the ability to "observe" the same source model with different detectors. In this way, the \GWT will provide even more functionality to give users quick idea of the power of different GW detectors for their favourite source population.

\section*{Acknowledgements}
    We thank the many researchers that provided results for the different source populations and detectors that made it possible to collect all these together in the \GWT. We also thank the referee for comments and detailed testing that greatly improved the paper and the \GWT. We thank in particular the (Mock) LISA Data Challenge team and the participating groups that made it possible to include so many results in the space module. Special thanks to Stas Babak and Antoine Petiteau for help and support. We thank our colleagues Paul Groot, Samaya Nissanke, Sarah Caudill, Chris van den Broeck, Gemma Janssen, Antonia Rowlinson, Peter Jonker, Selma de Mink, Marc Klein-Wolt and Jess Broderick for the initial discussions that led to this project. This research was made possible by support from the Dutch National Science Agenda, NWA Startimpuls – 400.17.608

\appendix
\section{Conversion among different LISA responses}
\begin{figure}[h]
\includegraphics[width=.45\textwidth]{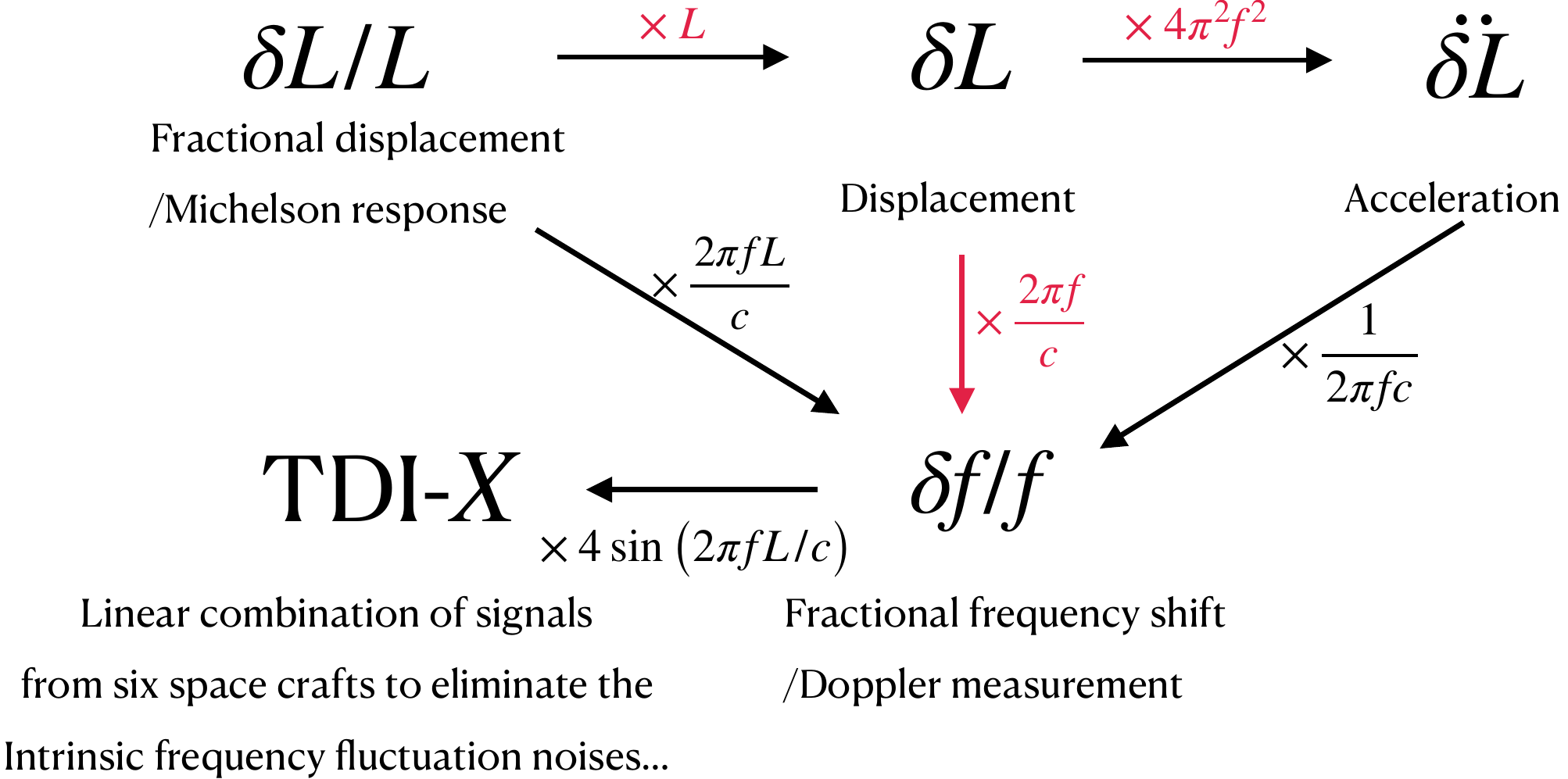}
\caption{Conversion between different kinds of LISA responses}
\label{fig:convert_LISA}
\end{figure}

In above sections, when working with LISA responses to gravitational wave signal and noises, we often need to convert among different kinds of LISA responses. We summarise the conversion relationship in figure \ref{fig:convert_LISA}.
\end{document}